\documentclass[iop]{emulateapj}
\usepackage[]{natbib}
\usepackage{graphicx}
\usepackage{hyperref}
\usepackage[figure]{hypcap}
\newcommand{\FIRST}{{FIRST}}
\newcommand{\SDSS}{{SDSS}}

\slugcomment{Accepted by \aj.}
\shorttitle{CLUSTER ENVIRONMENTS OF RADIO SOURCES}
\shortauthors{WING \& BLANTON}

\begin{document}

\title{Galaxy Cluster Environments of Radio Sources}

\author{Joshua D.\ Wing and
		Elizabeth L.\ Blanton}		
\affil{Astronomy Department and Institute for Astrophysical Research}
\affil{Boston University, Boston, MA 02215}
\email{jwing@bu.edu}

\begin{abstract}
Using the Sloan Digital Sky Survey (\SDSS) and the \FIRST\/ (Faint Images of the Radio Sky at Twenty Centimeters) catalogs, we examined the optical environments around double-lobed radio sources.  Previous studies have shown that multi-component radio sources exhibiting some degree of bending between components are likely to be found in galaxy clusters.  Often this radio emission is associated with a cD-type galaxy at the center of a cluster.  We cross-correlated the SDSS and FIRST catalogs and measured the richness of the cluster environments surrounding both bent and straight multi-component radio sources.  This led to the discovery and classification of a large number of galaxy clusters out to a redshift of $z \sim 0.5$.  We divided our sample into smaller subgroups based on their optical and radio properties.  We find that FR I radio sources are more likely to be found in galaxy clusters than FR II sources.  Further, we find that bent radio sources are more often found in galaxy clusters than non-bent radio sources.  We also examined the environments around single-component radio sources and find that single-component radio sources are less likely to be associated with galaxy clusters than extended, multi-component radio sources.  Bent, visually-selected sources are found in clusters or rich groups $\sim 78\%$ of the time.  Those without optical hosts in \SDSS\/ are likely associated with clusters at even higher redshifts, most with redshifts of $z > 0.7$.
\end{abstract}

\section{Introduction} \label{introduction} \index{Introduction}
Galaxy clusters are the largest gravitationally bound objects in the universe.  They are typically composed of $\sim30-1000$'s of galaxies, diffuse, hot gas, and dark matter.  As such, they can be used as tracers for studying large-scale structure \citep{bahcall1988,postman1992,carlberg1996,ikebe1996,jee2007,bradac2008,kartaltepe2008}.  Clusters also provide an excellent laboratory for the study of galaxy formation and evolution in dense environments \citep{butcher1978,butcher1984,garilli1999,thomas2010}.  Currently there are thousands of known, spectroscopically confirmed, clusters with redshifts $z<0.3$, but few spectroscopically confirmed with redshifts above $z\sim1.0$.  

A common method for detecting galaxy clusters is through optical surveys \citep{abell1958, zwicky1968, abell1989,lumsden1992,dalton1994,gal2003, miller2005, gladders2007, koester2007}.  Earlier methods of detecting galaxy clusters involved visual selection from photographic plates, but more recent techniques involve automated methods such as identifying a red sequence and determining cluster members based on their colors and their proximity to other galaxies that lie on the red sequence \citep{bower1992,ostrander1998, gladders2000, bahcall2003, gladders2005,koester2007}.  This method can work without having to know the spectroscopic redshift of the cluster members.

Identifying galaxy clusters through optical selection can become increasingly difficult with increasing redshift.  At high redshifts, the galaxies composing the cluster become too faint to detect with an exposure time typical of that of a large area survey.  In addition, the peak of the rest-frame optical emission of these high-redshift galaxies will be redshifted into the near-IR.  It is possible to search for, and identify, clusters out to much higher ($z\sim2$) redshifts using near-IR imaging \citep{daddi2004,blanc2008}, but the trade-off is in the small area coverage.  New IR detectors with larger collecting areas (such as the NOAO\footnote{The National Optical Astronomy Observatory is managed by the Association of Universities for Research in Astronomy under a Cooperative Agreement with the National Science Foundation.} Extremely Wide Field Infrared Imager, NEWFIRM) should greatly help the effort of detecting high-redshift galaxy clusters using this technique.  In addition, recent cluster searches in the mid-IR conducted using {\it Spitzer} have been successful in finding clusters, including some with $z > 1$ \citep{brodwin2006,krick2009}.

Galaxy clusters are also identified by detecting the X-ray emission of the hot gas of the intracluster medium (ICM).  Due to the long exposure times needed to detect galaxy clusters in the X-ray, as well as the small field-of-view of most X-ray detectors, it is difficult to conduct a large-scale, systematic survey of X-ray detected galaxy clusters at high redshifts.  X-ray detected clusters are also biased towards the most luminous, and therefore the most massive, galaxy clusters at high redshift, although samples at low redshift are complete \citep{reiprich2002}.

Other methods of detecting galaxy clusters such as the Sunyaev-Zel'dovich (SZ) effect \citep{sunyaev1970} (see \citet{carlstrom2002} for a review) also have inherent biases.  Clusters detected through the SZ signal should have no redshift bias out to a redshift of $z\sim2$.  However, clusters with bright radio point sources (from galaxies with an active nucleus, AGN) make it difficult to detect an SZ signature \citep{lin2007,martini2009,lin2009}.  As such, surveys searching for galaxy clusters using the SZ effect may possibly exclude clusters with radio sources that are bright at high frequencies.  Further, \citet{martini2009} and \citet{krick2009} find that the fraction of AGN in clusters is higher for clusters at high redshift suggesting AGN evolution in clusters similar to that of star-forming galaxies.  Therefore, a higher fraction of high-redshift clusters, as compared to low-redshift, may be excluded from SZ surveys because they contain radio sources.  Another limitation of cluster detection using the SZ effect is the size of the signal from the SZ effect of a galaxy cluster is small and directly related to the mass of the cluster.  This implies that clusters detected through their SZ signal are biased towards finding the most massive clusters \citep{menanteau2010,vanderlinde2010}, similar to X-ray detected clusters.

A method of detecting clusters that involves a more targeted approach would be useful, especially if that method allows for the possible identification of high-redshift clusters.  If we are able to efficiently select the most promising high redshift cluster candidates and examine these sources more closely, we can find a large number of high redshift galaxy clusters with cluster masses possibly more varied than clusters detected using optical/IR, X-ray, or SZ techniques.  Having a wide range of redshifts can help to measure cosmological parameters such as $\Omega_X$, $\omega_X$ and $\Omega_\Lambda$ \citep{allen2004}.

Previous studies \citep{zhao1989,hill1991, allington-smith1993, dickinson1997, deltorn1997, zirbel1997, blanton2000, blanton2001, blanton2003,belsole2007,venturi2007} have show that radio sources are often found in galaxy clusters.  In particular, bent, double-lobed sources are frequently associated with clusters.  A likely explanation for the bending of the radio lobes is the relative motion between the host galaxy and the intracluster medium (ICM).  There are several explanations for this relative motion.  One scenario involves a galaxy with a high peculiar velocity relative to the cluster moving through the ICM.   If the density of the ICM is high enough, and the velocity of the radio galaxy fast enough, it is possible to create enough ram pressure to bend the lobes of the radio source \citep{owen1976,eilek1984, burns1990, ball1993, bliton1998}.  In another scenario, the ICM is set in motion due to a recent cluster-cluster merger.  The radio galaxy, with a small peculiar velocity relative to the surrounding cluster, and the moving ICM experience enough relative motion that the resulting ram pressure can produce the observed bending in the lobes \citep{burns1990,roettiger1996,burns1996}.  Evidence supporting the second scenario is observed with the alignment of the bending of the lobes and the X-ray emission \citep{gomez1997}.  Bent radio sources may also be found in clusters that are relaxed on large scales, e.g.\ Abell 2029 \citep{clarke2004}.  In these cases, the bending of the lobes may be related to gas motions induced by ``sloshing'' \citep{ascasibar2006} of the central cD.  Using bent-lobed radio sources as tracers of clusters may lead to the discovery of a large number of previously undetected high redshift clusters, without having to perform an extremely deep all-sky survey.  This technique may preferentially select clusters undergoing mergers, and/or those that exhibit sloshing.

Double-lobed radio sources are generally divided into two different morphological classes.  \citet{fanaroff1974} discovered a connection between the radio power and morphology of a given extended radio source.  They found that weaker radio sources, those with powers less than $5\times10^{25}$ W Hz$^{-1}$ at $1440$ MHz (assuming a power-law spectrum with $\alpha=0.8$), typically have bright radio cores and lobes fading toward the edges.  These have since become known as FR I sources.  The more powerful radio sources ($P_{1440\: \rm{MHz}} >5\times10^{25}$ W Hz$^{-1}$) have dim or absent cores and edge-brightened lobes and are now referred to as FR II sources.  More recently, \citet{owen1991} and \citet{ledlow1996} showed that the dividing line between FR I and FR II sources is related to both the radio power and the optical luminosity of the host galaxy.

Differences between FR I and II sources may be related to the power output from the central engine as well as the environment of the source.  \citet{zirbel1997} found that the cluster environments surrounding FR I and FR II sources are different.  They found that on average, FR I galaxies are located in richer groups than FR II galaxies.  As redshift increased, the likelihood of an FR II galaxy existing in a rich group increased.  They also found that high-redshift FR I and II galaxies belong to different subsets of galaxy groups than low-redshift FR I and II galaxies.  These galaxy groups, selected based on the presence of a radio source and including both low- and high-redshift FRI and II sources, differ from optically selected galaxy groups.

\citet{croft2007} looked for radio sources associated with sources detected using the maxBCG algorithm from \citet{koester2007}.  The maxBCG algorithm (as described in \citet{koester2007a}) detects galaxy clusters by taking advantage of the spatial clustering, the tight sequence that most cluster galaxies are located on in a color-magnitude diagram (the "E/S0 ridgeline" or red sequence as described above), and the fact that there is often a brightest cluster galaxy (BCG) that is often coincident with the center of the cluster.  \citet{croft2007} found that $\sim20\%$ of the time, the BCG is associated with a radio source at the same location.  This percentage was dependent on the stellar mass of the host galaxy, rising as the mass of the host rises.  Further, they detected that within a radius of $1.4$ Mpc around the BCG, there is an over-abundance of radio sources compared to the field.  This implies that galaxy clusters are likely to be hosts to radio sources.  Also, there is an association between the presence of a cooling flow in a cluster and the central giant elliptical being radio loud.  At least $70\%$ of cooling flows host radio sources in their centers \citep{burns1990,mittal2009,dunn2010}, while fewer than $\sim30\%$ of non-cooling flows have central radio sources.

We present our investigation into the cluster environments surrounding radio galaxies of different morphologies.  Specifically, we examine the optical environments surrounding multi-component bent and straight radio sources of type FR I and FR II.  We also compare these environments to the cluster environments around single-component radio sources.  Throughout this paper we assume $H_0=70$ km s$^{-1}$ Mpc$^{-1}$,  $\Omega_\Lambda=0.7$, and $\Omega_M=0.3$.  In \S\ref{selecting_radio_sources}, we discuss how we select our radio sources from the VLA Faint Images of the Radio Sky at Twenty-cm (\FIRST\/) survey for further examination, in \S\ref{finding_optical_counterparts_in_the_sdss}, we discuss how we identify optical sources in the Sloan Digital Sky Survey (\SDSS\/) catalog associated with our radio sources, and in \S\ref{determining_cluster_richness}, we measure the richness of the optical environments around the radio-selected sources.  In \S\ref{physical_properties_of_radio_selected_galaxy_clusters}, we examine the optical and radio properties of these sources, in \S\ref{correlation_with_x-ray_catalogs}, we discuss possible X-ray emission associated with our samples, and in \S\ref{conclusions}, we present our conclusions and discuss our findings.

\section{Selecting Radio Sources} \label{selecting_radio_sources} \index{Selecting Radio Sources}
Previous studies indicated that bent-double radio sources are associated with galaxy clusters \citep{blanton2000, blanton2001, blanton2003, smolcic2007, kantharia2009, giacintucci2009, oklopcic2010}.  Identifying these sources in radio catalogs and then searching for optical counterparts can define a sample of radio-selected galaxy clusters.  We made use of the \FIRST catalog \citep{becker1995} to identify possible galaxy clusters based on radio emission.  The \FIRST\/ survey covers $\sim25\%$ of the sky and is mostly contained in the northern Galactic cap.  The \FIRST\/ survey has a flux density threshold of 1 mJy, systematic astrometric errors $<0.05\arcsec$ and total positional errors on the order of $\sim1\arcsec$.

Possible galaxy clusters identified through this characteristic radio emission can then be used for studies of cosmology as well as galaxy and cluster formation and evolution.  In this paper, we examine the optical environments around both FR I and FR II extended radio sources (bent and straight).  We also study a sample of single-component radio sources for comparison.

\subsection{The Visually-Selected Bent Sample} \label{the_visually_selected_bent_sample} \index{Selecting Radio Sources!The Visually-Selected Bent Sample}
Visual inspection of a radio source is one way to determine if it is a true bent-double source.  \citet{blanton2000a} visually examined a sample of $\sim32,000$ multiple-component sources from the April 1997 release of the \FIRST\/ catalog (covering $\sim3000$ deg$^2$ of the sky).  The maximum separation between components was $60\arcsec$.  A total of $384$ sources were identified as bent, double-lobed sources.  Many of these sources were studied previously in the optical in an effort to measure the richness of their environments \citep{blanton2000,blanton2001,blanton2003}.  These sources each contain two or more components listed in the \FIRST\/ catalog.

\subsection{The Automatically-Selected Bent Sample} \label{the_automatically_selected_bent_sample} \index{Selecting Radio Sources!The Automatically-Selected Bent Sample}
In order to utilize the entire \FIRST\/ catalog (as of April 2003, covering $\sim9000$ deg$^2$ of the sky), we employed an automated program to identify the bent-double sources within the $ >800000$ sources in the catalog.  \citet{proctor2006} created a pattern recognition program that used a random training set of $2823$ sources from the $\sim16000$ \FIRST\/ three-component sources to assign a score to each three-component source as to its bent-ness or straight-ness. The maximum separation between any two components was $57\farcs6$, similar, although not identical, to that for the visually-selected sources.  \citet{proctor2006} visually inspected each of the training set sources and classified them as either bent, non-bent, or ambiguous.  Of the $2823$ training set sources, \citet{proctor2006} classified $147$ as bent-double sources, $1395$ as non-bent-double sources, and $1281$ as ambiguous sources.  Using these classifications, the program was able to search the entire \FIRST\/ catalog for three-component sources and identify bent and straight double-lobed radio sources.

\citet{proctor2006} gave each of the $\sim16000$ three-component sources in the entire \FIRST\/ catalog a score that related to the probability that the source was a true bent-double source.  For example, eight out of ten sources with a score of $0.80$ are expected to be true bent-double sources.  In order to limit the sample to only sources most likely to be true bent-double sources, we limited the auto-bent sample to only those sources with a score of $0.50$ or higher.  Sources with lower scores might also be true bent-double sources, but the number of such sources is overwhelmed by serendipitous sources that are not related.  This limits the sample to $1546$ sources.  Using only the sources with the highest scores eliminates many of the "three-component" sources that only appear related due to projection effects.  \citet{proctor2006} defined the central source for these automatically-selected bent sources as the source opposite the longest side when making a triangle of the three sources.

\begin{figure}
\begin{center}
\capstart
\includegraphics{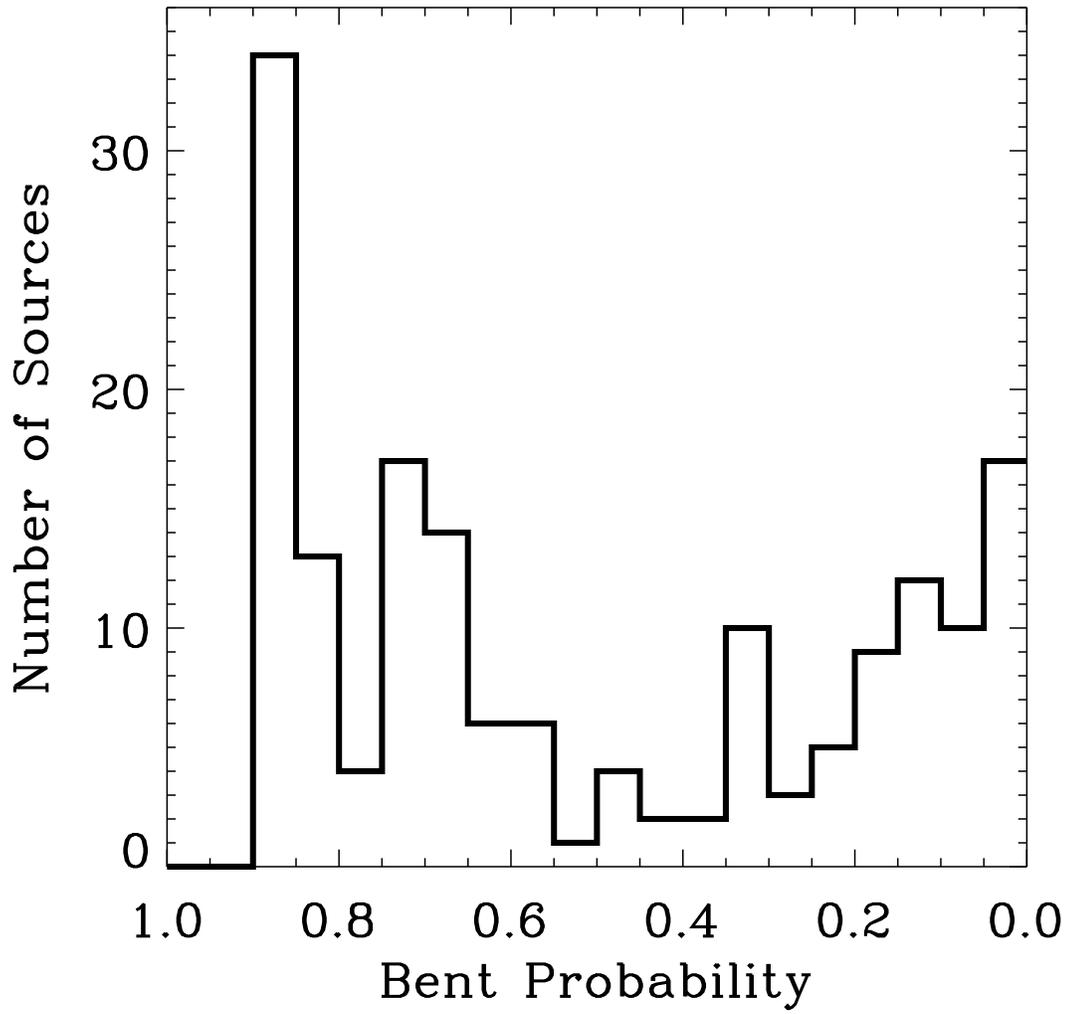}
\caption{A histogram of the probability scores given to the 167 three-component sources in the visual-bent sample.  A score of 1.0 corresponds to a probability of 100\% that the source is a bent double-lobed source.  The auto-bent sample is composed of sources with scores higher than 0.5.  Some of the visual-bent sources have low probabilities as determined by the computer algorithm.} \label{bentscore}
\end{center}
\end{figure}

Since both the visual- and auto-bent samples are selecting for similar sources from different versions of the same catalog, we expect overlap between the two samples.  Visual inspection of bent-lobed sources allows for the selection of sources that the automatic selection technique may classify as ambiguous.  There are 167 sources (of 384) in the visual-bent sample that are three-component, double-lobed sources (the others have only two, or greater than three, components).  Searching the entire $\sim16000$ source auto-bent sample for these 167 double-lobed sources, we find that there are 24 sources located within $1\arcsec$ of the central location of sources in the visual-bent sample. An additional 126 sources from the entire $\sim16000$ source auto-bent sample have central locations within $10\arcsec$ of the location of the corresponding visual-bent source.  The other 17 double-lobed sources in the visual-bent sample are located within an area with a maximum separation of $24\arcsec$ from the corresponding source in the entire $\sim16000$ source auto-bent sample.  Thus, we recovered all 167 visually-selected three-component double-lobed sources in the entire $\sim16000$ source auto-bent sample.  The separation between the visual-bent and the automatically selected sample positions is likely due to the automatic selection of the center of the bent-lobed radio source being different than the center of the source as determined visually.  When there was a known optical counterpart for a visually-selected bent-lobed radio source, the position of the optical source was used as the position of the center of the bent-lobed radio source.  Not all of the visual-bent three-component double-lobed sources have scores high enough to be included in our auto-bent sample.  Only 94 of the 167 sources identified in the visual-bent sample as three-component, double-lobed sources have bent probability scores higher than the $0.50$ score that we set as a baseline.  Figure~\ref{bentscore} shows the probability scores of the visual-bent, three-component bent-lobed sources.  The sources with scores below $0.50$ were not included in the auto-bent sample.  Therefore, the total overlap between the visual-bent and auto-bent samples is 94 sources.

\subsection{The Straight Sample} \label{the_straight_sample} \index{Selecting Radio Sources!The Straight Sample}
We also examined a sample of $3232$ sources from the \FIRST\/ catalog which represent all of the aligned (straight) three-component sources (with the distance between components limited to $57\farcs6$) in the entire \FIRST\/ survey region.  We selected these sources using the same code from \citet{proctor2006}.  The straight-lobed sources are those sources identified by the code of \citet{proctor2006} as having $(d_{min}+d_{mid}) / d_{max} < 1.01$ and $d_{min}/d_{mid} \ge 0.7$ where $d_{min}$ is the minimum pair-wise distance between the three radio sources, $d_{mid}$ is intermediate pair-wise distance, and $d_{max}$ is maximum pair-wise distance.  In practice, the dividing line between the auto-bent and straight samples is an opening angle of $\sim160^{\circ}$.  It is possible that the angle at which the observer views a source (classified as straight-lobed) is such that the lobes appear to be aligned even though the source is bent.  In this case, sources that appear to be aligned and are included in the straight sample are actually bent-lobed sources.  If bent-lobed radio sources are more often associated with galaxy clusters, then the inclusion of these apparently straight bent-lobed sources in the straight sample serves to increase the number of straight-lobed sources located in rich clusters.

\subsection{The Single-Component Sample} \label{the_single_component_sample} \index{Selecting Radio Sources!The Single-Component Sample}
If galaxy clusters are more likely to be associated with extended, multi-component radio sources, there should be a corresponding lack of galaxy clusters when looking at single-component radio sources.  In order to compare extended sources to single-component sources, we randomly picked out $3356$ sources in the \FIRST\/ catalog that had no other source within $60\arcsec$.  The number of single-component sources in the \FIRST\/ catalog totaled $576,267$, but we chose to work with a smaller sample size to better correspond with the sizes of the other samples.

One of the questions we wish to address with the single-component sample is how often are radio sources in general associated with cluster environments?  If we find single-component radio sources are located in cluster environments with the same incidence as bent-double radio sources, there is no evidence to suggest that bent-double radio sources are any more likely to be tracers for galaxy clusters than any other radio source.  This makes our sample of single-component radio sources a good baseline to compare typical radio sources to our samples of multi-component, extended radio sources.

\subsection{Determining Accurate Radio Flux Densities} \label{determining_accurate_radio_flux_densities} \index{Selecting Radio Sources!Determining Accurate Radio Flux Densities}
The \FIRST\/ B-array observations sometimes resolve out extended source emission.  This can result in an underestimate of the total flux of the object.  The NRAO VLA Sky Survey ({\it NVSS}; \citet{condon1998}) surveyed the sky north of $\delta = -40^{\circ}$ at a lower resolution (the D- and DnC-array) than \FIRST\/ (the B-array) at the VLA.  If a source is in both catalogs we can use the {\it NVSS} flux and compare it to the \FIRST\/ flux to determine if the source has extended emission that has been resolved out.  Going by the convention set forth by \citet{ivezic2002}, we searched the {\it NVSS}\/ within $15\arcsec$ of the location of each component of the \FIRST\/ source.  If the location of the \FIRST\/ source is within $15\arcsec$ of the location of a source in the {\it NVSS}\/ catalog, the probability is less than $1\%$ that the two sources are random associations \citep{ivezic2002}.

The extended sources in our sample are often comprised of sources located more than $15\arcsec$ from each other, thus we searched for an {\it NVSS} counterpart around not only the central radio source but also the two lobes.  There were cases where a single {\it NVSS} source corresponded to two or more \FIRST\/ components.  In these instances, we assumed that the single {\it NVSS} source was an under-resolved version of the multiple-component \FIRST\/ source and thus singularly represented the multiple \FIRST\/ components.  For these sources we used the higher of the {\it NVSS} flux or the sum of the flux of the \FIRST\/ components with which the {\it NVSS} source was associated.

\section{Finding Optical Counterparts in the \SDSS} \label{finding_optical_counterparts_in_the_sdss} \index{Finding Optical Counterparts in the SDSS}
The Sloan Digital Sky Survey (\SDSS\/) \citep{york2000} is a drift-scan survey of the northern Galactic cap.  The \SDSS\/ cataloged photometry in five bands ({\it ugriz}) for over 350 million unique sources and spectra of over a million sources as of data release 7 (DR7, \citet{abazajian2009}).  DR7 contains photometric redshifts calculated for every extragalactic source in the \SDSS\/ catalog with a new hybrid technique that gives much more accurate results than previous data releases.  DR7 also includes updated and more accurate magnitudes for all of the sources in the {\tt photo} catalog that better accounts for sky background subtraction issues for crowded fields.  The astrometry of the sources in DR7 was updated as well, yielding more accurate source positions.

The \FIRST\/ survey was designed to cover the same area of the sky as the \SDSS\/.  This makes it convenient to search the \SDSS\/ for optical counterparts for all of the \FIRST\/ sources from the samples we have described above.  The positional accuracy of both \SDSS\/ and \FIRST\/ are less than $1\arcsec$ so we were able to search in a very small area around each radio source for an associated optical source.  Because of the large number of sources in the \SDSS\/ catalog, it is necessary to identify sources which are likely to be chance-coincidences, instances when a physically unrelated optical source appears close in projection to a radio source.  To achieve this, we searched the \SDSS\/ for all optical sources within $10\arcsec$ of the core of the multi-component radio sources as discussed below.  We found that a radius of $10\arcsec$ allowed us to maximize the possibility of finding an optical counterpart while keeping the number of chance-coincidences to a minimum.

\subsection{Correlating \FIRST\/ Sources With \SDSS\/ Sources} \label{correlating_first_sources_with_sdss_sources} 
\begin{figure}
\begin{center}
\capstart
\includegraphics{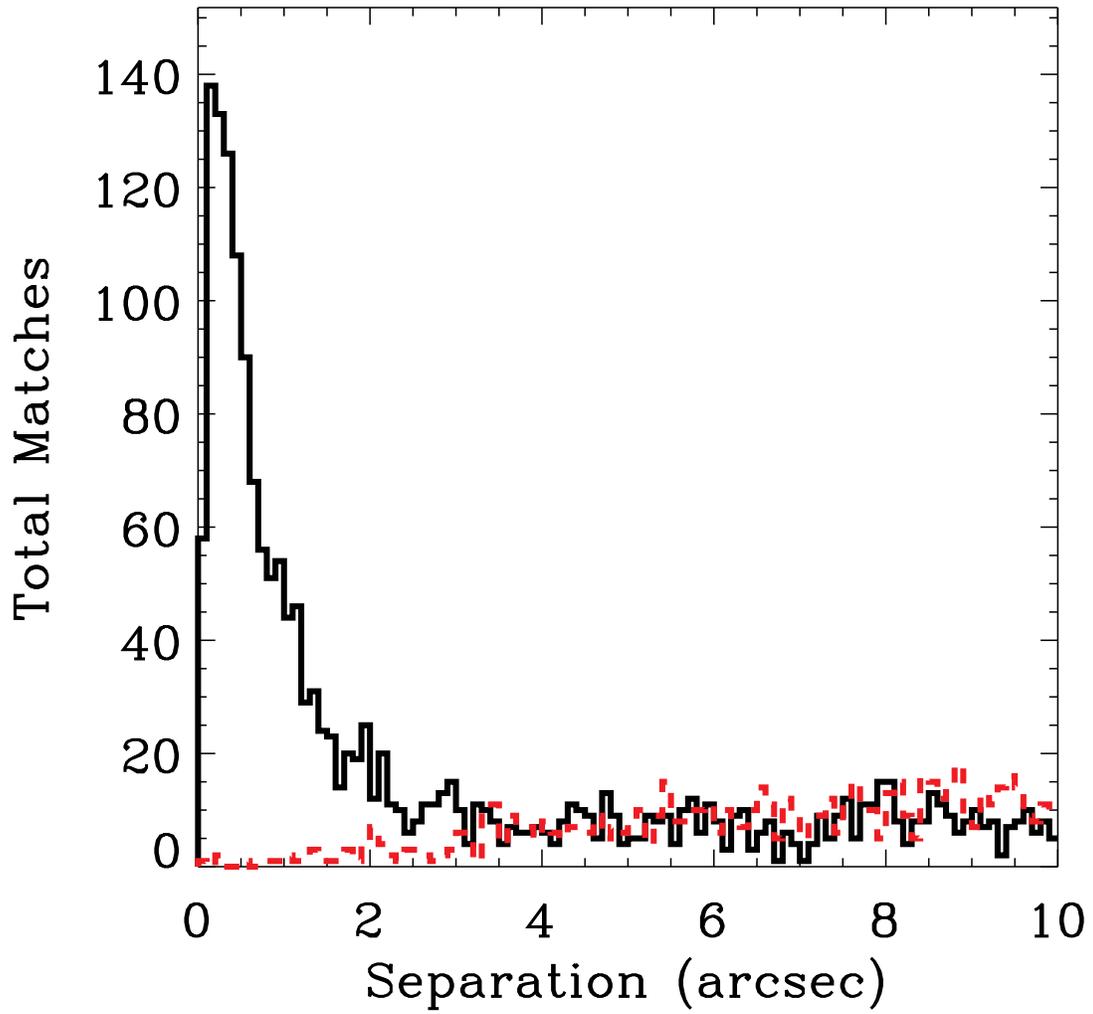}
\caption{The black solid-line histogram represents the number of optical sources associated with radio sources (in this case from the straight sample) as a function of separation from the radio source center.  The peak in the distribution at small separations is a result of real radio/optical counterparts.  The red dashed-line histogram plots the distribution of optical matches as a function of separation for the shifted positions of the radio sources, i.e. a random sample of points on the sky.  There is no peak at small separations but as the separation increases, the number of matches increases.  At large separations the distribution of the real sample matches that of the random sample.} \label{matchsep}
\end{center}
\end{figure}

\index{Finding Optical Counterparts in the SDSS!Correlating FIRST Sources With SDSS Sources}
Following the procedure set forth by \citet{moran1996}, we matched the original central radio source positions with the \SDSS\/ catalog.  We then shifted both the RA and Dec of every radio source in each of our samples by $30\arcmin$ and matched the two catalogs again.  The optical sources matched with these random positions are not real radio/optical counterparts but instead give an idea as to the number of chance-coincidences to expect as a function of separation between the radio and optical positions.  We compared the number of \SDSS\/ matches as a function of increasing distance from the center of the radio source to the number of \SDSS\/ matches that the shifted positions give as a function of increasing distance from the location of these positions.  Looking at the number of matched sources as a function of separation between the radio source and the optical source (keeping only the closest optical source to each radio source) should show a peak at small separations for real radio/optical counterparts.  For the random sample, there is no peak at small separations but instead a steadily rising number of matches as the separation between the radio and optical source is increased.  This rising number of matches is a result of the increased area that is being searched as separation is increased.  We see in Figure~\ref{matchsep} that the number of random matches closely follows the real matches starting at a separation of around $2\arcsec$.  This is for the straight sample, but it is similar for all of the samples.

\begin{figure}
\begin{center}
\capstart
\includegraphics{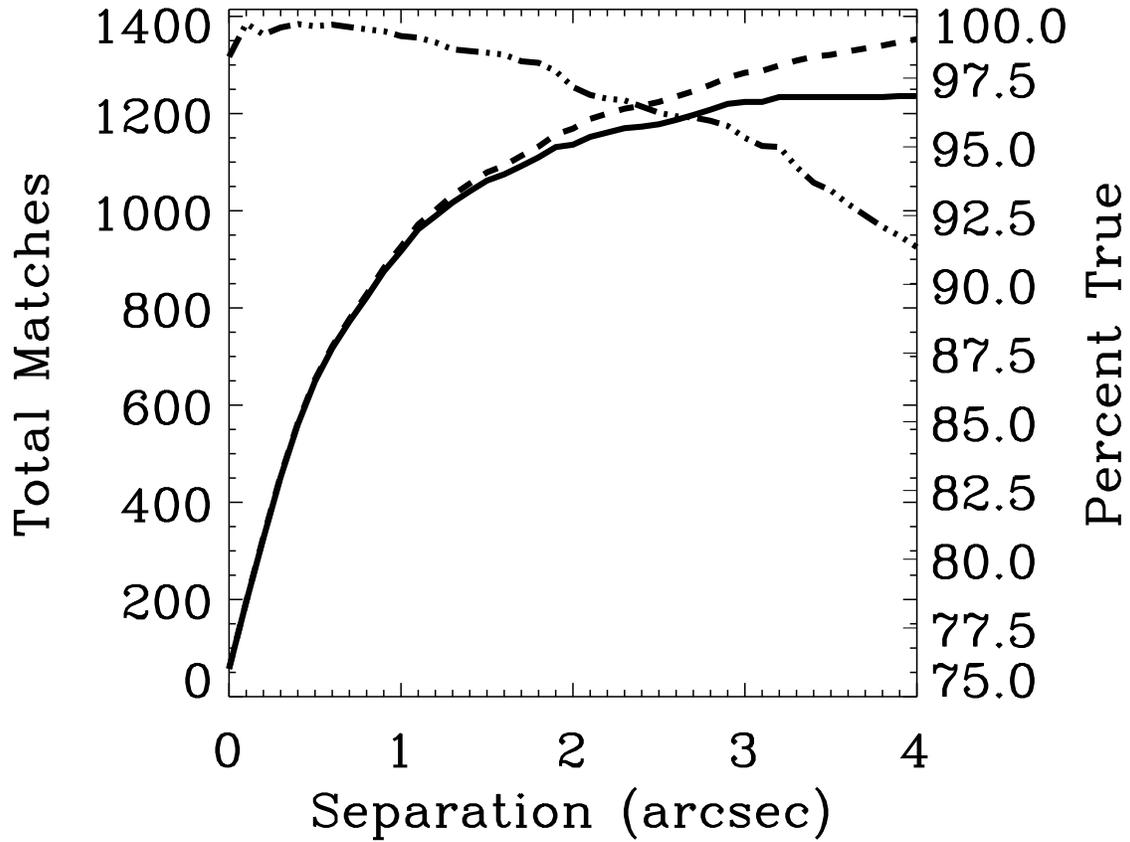}
\caption{The solid-line represents the cumulative number of good, non-chance-coincidence, matches as the separation between the radio and optical sources is increased.  The dashed-line shows the cumulative total number of matches with increasing separation.  At a separation of around $3\arcsec$ the number of good matches starts to level off and the total number of matches continues to rise.  This is illustrated by the dash-dotted-line which shows the percentage of matches that can be expected to be true matches as the separation is increased.  We aimed for a $95\%$ reliability cut-off, hence the $3\farcs2$ separation limit for considering an \SDSS\/ source a match with a \FIRST\/ source for the straight sample shown here.} \label{matchfrac}
\end{center}
\end{figure}

Figure~\ref{matchfrac} shows how increasing the allowed separation between the radio source and the optical source increases the number of chance-coincidences.  We see that the total number of \SDSS\/ matches continues to increase upwards, but the number of good matches that are expected, those that are not chance-coincidences, does not increase as quickly.  There are true, non-chance-coincidence, matches at larger separations.  However, if we allow matches to come from larger separations, while recovering a higher fraction of the true radio/optical counterparts expected, we increase the number of chance-coincidences in our sample much more quickly than the number of true radio/optical counterparts.

We aimed to make our samples $95\%$ reliable, meaning that we expect $5\%$ of the sources to be serendipitous chance-coincidences.  The dash-dotted line in Figure~\ref{matchfrac} shows how the percentage of reliable sources in our sample decreases as the separation is increased.  We settled on $95\%$ for a number of reasons.  First, we want our sample to be as free as possible of chance-coincidences and we have achieved this by requiring a $95\%$ reliability cut-off.  We would also like our sample to be as complete as possible.  That is, recovering a large percentage of the true radio/optical counterparts.  We define the completeness of the sample as the fraction of radio/optical counterparts identified when limiting matches to only those contained within the radius set by our $95\%$ reliability criteria compared to the total number of radio/optical matches within $10\arcsec$.  Constraining our samples to $95\%$ reliability maximizes the completeness of the samples while simultaneously limiting the number of chance-coincidences.

It is difficult to determine the center of a bent-lobed radio source automatically.  We find that as a result of the possible ambiguity as to which of the three components of the doubled-lobed radio source is the central component, the maximum allowable separation between the optical source and the presumed center of the radio source for the visual-bent and auto-bent samples is larger than for the other samples.  This is similar to results in \citet{mcmahon2002}, that for multiple component radio sources the separation between the center of an extended radio source and its optical counterpart is larger than for radio point sources.  Limiting the auto-bent sample to a $95\%$ reliability cut-off limits the sample completeness to below $60\%$.  As a compromise, to increase the completeness of the auto-bent sample we have lowered the reliability cut-off to $90\%$.  This brings the completeness of the auto-bent sample more in line with the other samples.  We find that the samples were $95\%$ reliable (or $90\%$ for the auto-bent sample) out to radii between $1\farcs9$ and $6\farcs8$ from the radio source, depending on the sample.

The results of these selection criteria are shown in Table~\ref{sample_breakdown}.  Column 1 of Table~\ref{sample_breakdown} identifies the name of each sample, column 2 lists the number of radio sources that each sample contains, column 3 lists the number of FIRST radio sources in each sample that are contained within the SDSS footprint, column 4 gives the total number of radio sources in each sample with a unique match within the \SDSS\/ database located $10\arcsec$ or less from the center of the radio source, column 5 gives the radius, in arcseconds, out to which an optical match in the \SDSS\/ database located within that distance from the center of the radio source is $95\%$ likely to be coincident with the radio source and column 6 gives the number of sources remaining in each sample that have an optical counterpart in the \SDSS\/ within a radius corresponding to that $95\%$ confidence.  Column 7 gives the completeness of each sample (see \S\ref{identifying_possible_clusters}), column 8 lists the number of radio/optical sources that meet our color criteria (see \S\ref{identifying_possible_clusters}), column 9 gives the total number of sources in each sample with redshift $z \ge 0.01$ and a Schechter correction factor (see \S\ref{schechter_correction}) less than 3, and columns 10 and 11 list the number of FR I and FR II sources in column 9 (see \S\ref{determining_radio_power_and_fr_morphology}) in each sample, respectively.  

\subsection{Determining Redshift} \label{determining_redshift} \index{Finding Optical Counterparts in the SDSS!Determining Redshift}
A source in the \SDSS\/ can have up to four different redshift values associated with it, one measured spectroscopically and the others based on the color and shape of the spectrum of the source.  In DR7, over $1.2$ million sources have associated spectra and thus a spectroscopically measured redshift.  If one of the radio/optical sources has a spectroscopically measured redshift in \SDSS\/, we use that value as the redshift of the source for the rest of our calculations.  Unfortunately, that still leaves a vast majority of the $\sim357$ million sources in the \SDSS\/ without spectroscopically measured redshifts.  For these sources, redshifts were inferred photometrically.

The \SDSS\/ {\tt photoz2} table was generated by a neural network program which used a training set of known galaxy colors and redshifts to determine a more accurate photometric redshift \citep{oyaizu2008}.  The {\tt photoz2} table provides photometric redshift estimates for $\sim77$ million sources in DR7 which have been classified as galaxies and have a magnitude of $m_r < 22$.  There are two different estimations of the photometric redshift contained within the {\tt photoz2} table.  The {\tt photozcc2} photometric redshift is to be used for faint sources which have $m_r > 20$.  The {\tt photozd1} photometric redshift has a smaller error and is to be used for brighter sources with $m_r < 20$.  Unfortunately, not all of the sources in the \SDSS\/ catalog have {\tt photoz2} redshifts.

\begin{figure}
\begin{center}
\capstart
\includegraphics{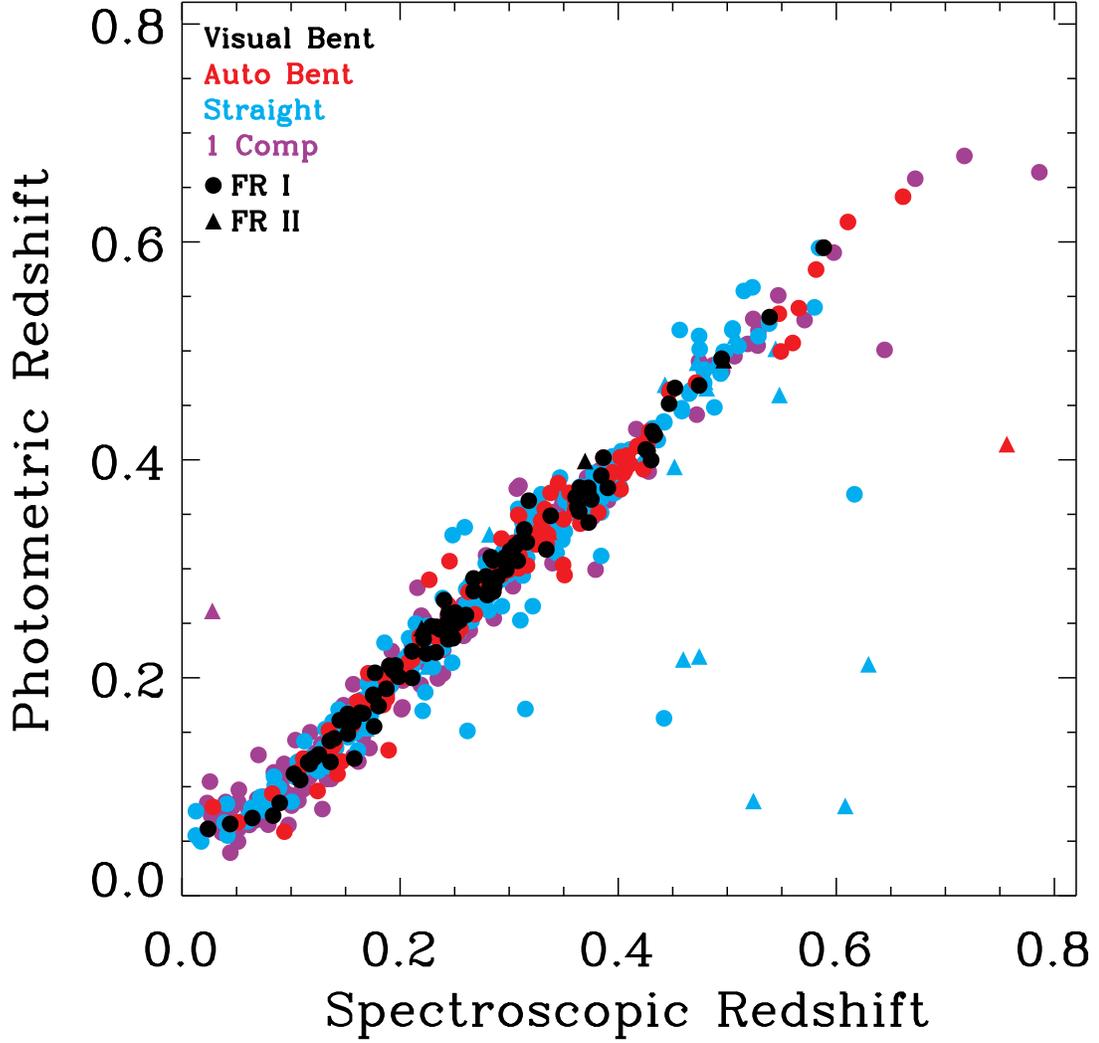}
\caption{A plot of the spectroscopic redshifts versus the photometric redshifts for the sources in each sample.  The photometric redshift here is from the {\tt photoz} table.  Not all of the sources with measured spectroscopic redshifts have calculated photometric redshifts.  This mostly applies to the sources identified by \SDSS\/ as point sources.  In general, the spectroscopic and photometric redshifts agree well.  Purple filled symbols represent the single-component sample, blue filled symbols represent the straight sample, red filled symbols represent the auto-bent sample, and black filled symbols represent the visual-bent sample.  The FR I sources are represented by circles and the FR II sources are represented by triangles.} \label{specvsphotoz}
\end{center}
\end{figure}

Nearly every extragalactic source in DR7 has a redshift in the {\tt photoz} table.  The photometric redshifts in the {\tt photoz} table were obtained with use of the template fitting method \citep{budavari2000,csabai2003}.  This method compares the expected colors of a given galaxy type with those observed for an individual galaxy.  Template fitting involves taking a small number of spectral templates for discrete galaxy types and then choosing the best fit redshift by comparing the galaxy morphology, colors, and apparent magnitude and determining the redshift value that gives an appropriate luminosity.  This basic photometric redshift measurement is much improved in DR7 and gives results consistent with spectroscopically measured redshifts \citep{abazajian2009}.  Figure~\ref{specvsphotoz} shows the correlation between the spectroscopic and photometric redshifts for the different samples in the instances where there exists a spectroscopic redshift.  Examination of Figure~\ref{specvsphotoz} shows that the photometric redshifts are very similar in most cases to the spectroscopic redshifts.

\begin{figure}
\begin{center}
\capstart
\includegraphics{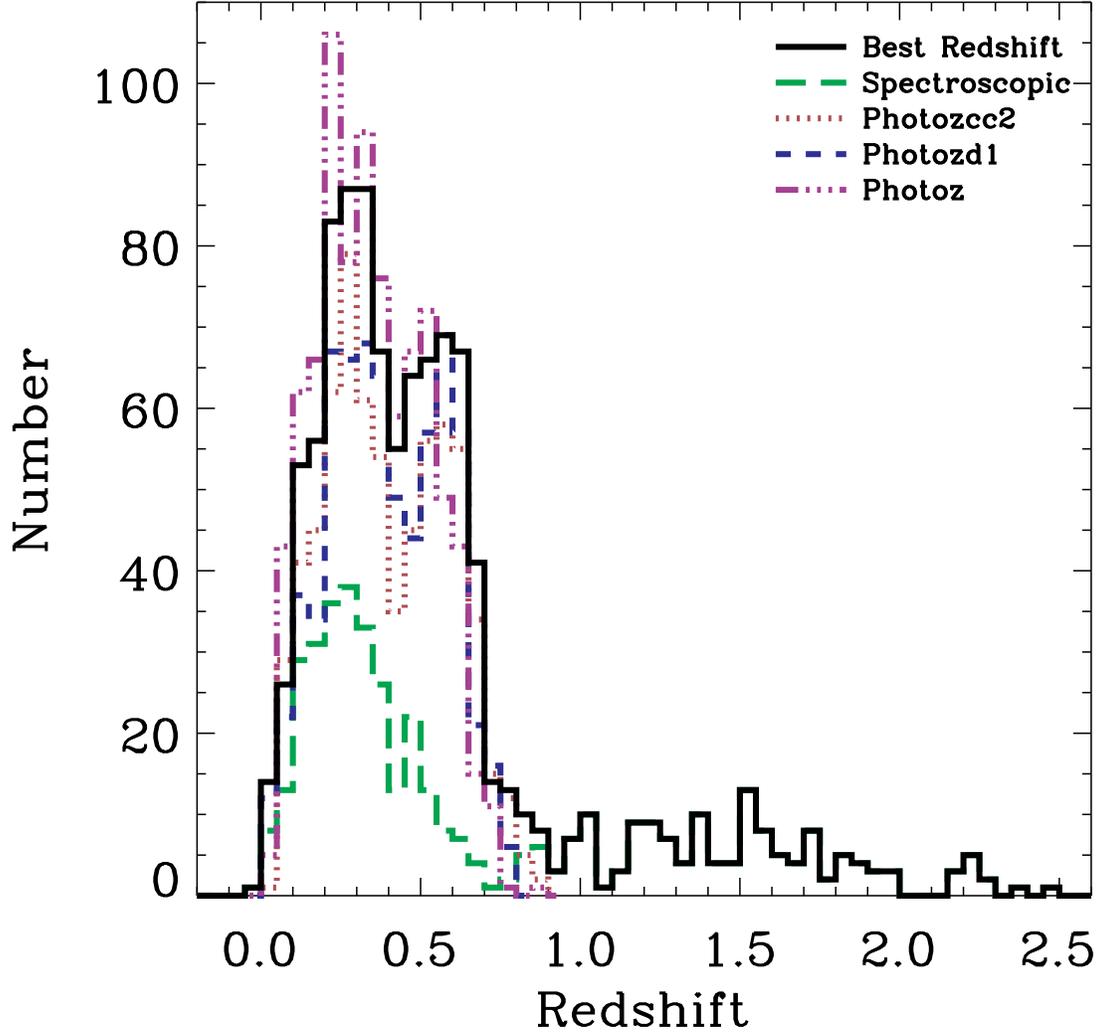}
\caption{The (black) solid-line is representative of the overall distribution of redshifts, the best redshift for each source as described in \S\ref{determining_redshift}.  The green long-dashed-line represents the spectroscopically measured redshifts, and the maroon dotted- and blue short-dashed-lines represent the redshifts given by the adaptive learning decision tree program that calculated the {\tt photoz2} catalog (see \S\ref{determining_redshift}).  The purple dash-dotted line represents the redshifts as calculated by the {\tt photoz} catalog.  This plot is for the straight sample.  The other samples have similar distributions for the various redshift measurement techniques.  The distributions of the best redshifts for each sample can be seen in Figure~\ref{zhist}.} \label{compact_zhist}
\end{center}
\end{figure}

The spectroscopic redshifts are the most reliable, so if a matching optical source has an associated spectroscopic redshift, we use that value as the redshift of the source.  If the source does not have a spectroscopic redshift, but does have an associated {\tt photoz2} redshift, we used the specific {\tt photoz2} redshift depending on the apparent magnitude of the source, as described above.  The errors for the {\tt photoz2} photometric redshifts are much smaller than the errors associated with the more simply calculated {\tt photoz} photometric redshifts.  In other cases, we used the redshift given by the {\tt photoz} table as the redshift of the source.  Figure~\ref{compact_zhist} shows the distribution of redshifts for each of the different methods of calculating redshift for the straight sample.  The {\tt Best Redshift} line on the histogram refers to the single redshift given to each radio/optical source with the priority as described above.  The other samples have similar distributions.  Overall there are a small number of sources with spectroscopic redshifts, but all sources with redshifts above $z=1.0$ have been obtained spectroscopically.  The distribution of redshifts from the {\tt photoz2} table are similar for both ({\tt photozcc2} and {\tt photozd1}) estimations.  Redshifts obtained from the {\tt photoz} table have a peak at lower redshifts, similar to the {\tt spectroscopic} table and lower than that of the {\tt photoz2} table.

\begin{figure}
\begin{center}
\capstart
\includegraphics{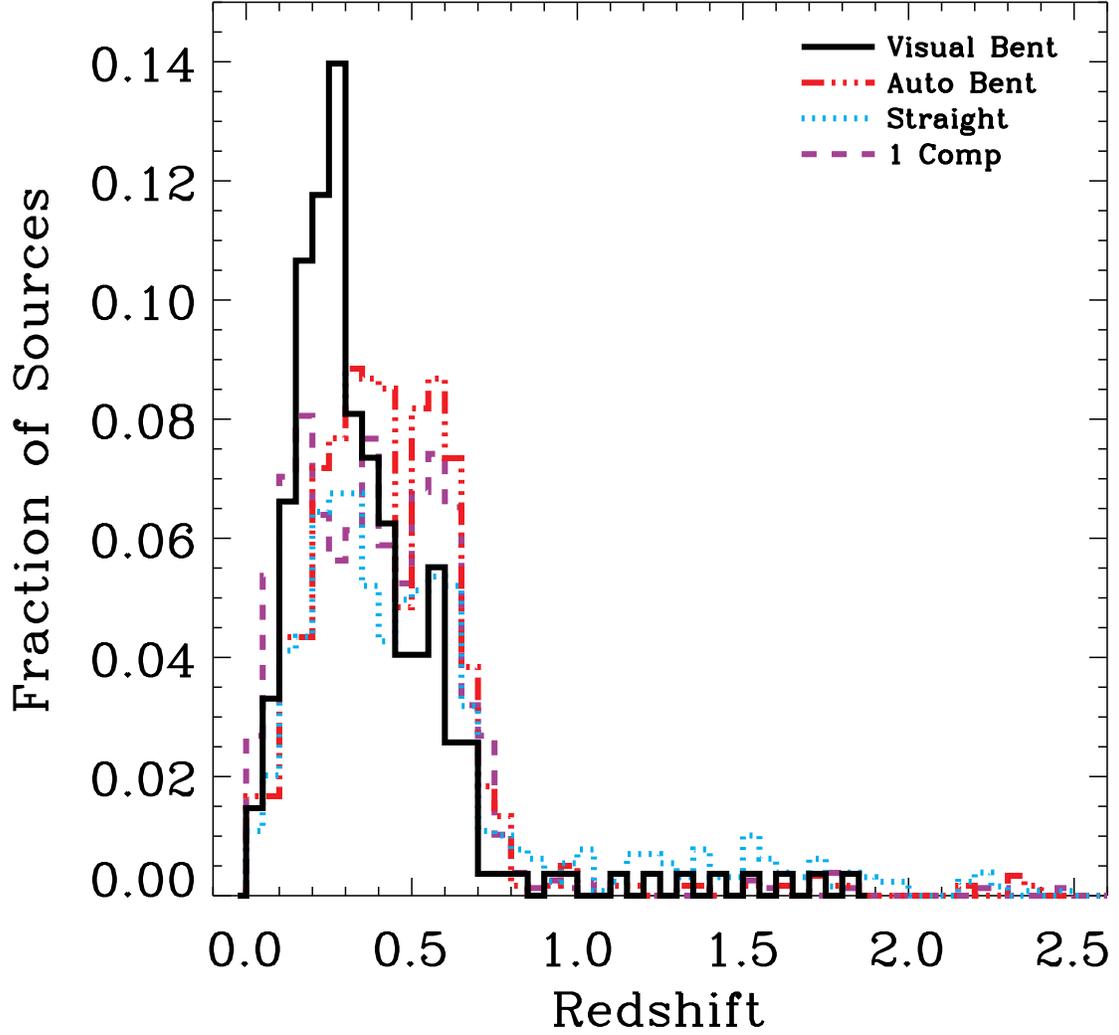}
\caption{The distribution of redshifts for each sample.  The black solid-line represents the visual-bent sample, the red dash-dotted line represents the auto-bent sample, the blue dotted line represents the straight sample, and the purple dashed line represents the single-component sample.  These are the best redshifts (see \S\ref{determining_redshift}) for each optically identified radio source.  The visual-bent sample peaks at a lower redshift than the other samples.  This is most likely due to the nature of the selection.  It is easier to classify a source as a bent-double source if it is at lower redshift and angularly larger on the sky.} \label{zhist}
\end{center}
\end{figure}

Figure~\ref{zhist} shows the redshift distribution of each radio source sample using the best redshift available as described above.  The visual-bent sample peaks at a lower redshift than the other samples, most likely a result of the visual selection of each source.  Low redshift sources are easier to identify by eye as bent-double sources.  The straight sample has a higher fraction of high redshift sources than the other samples.

\subsection{Identifying Possible Clusters} \label{identifying_possible_clusters} 
\begin{figure}
\begin{center}
\capstart
\includegraphics{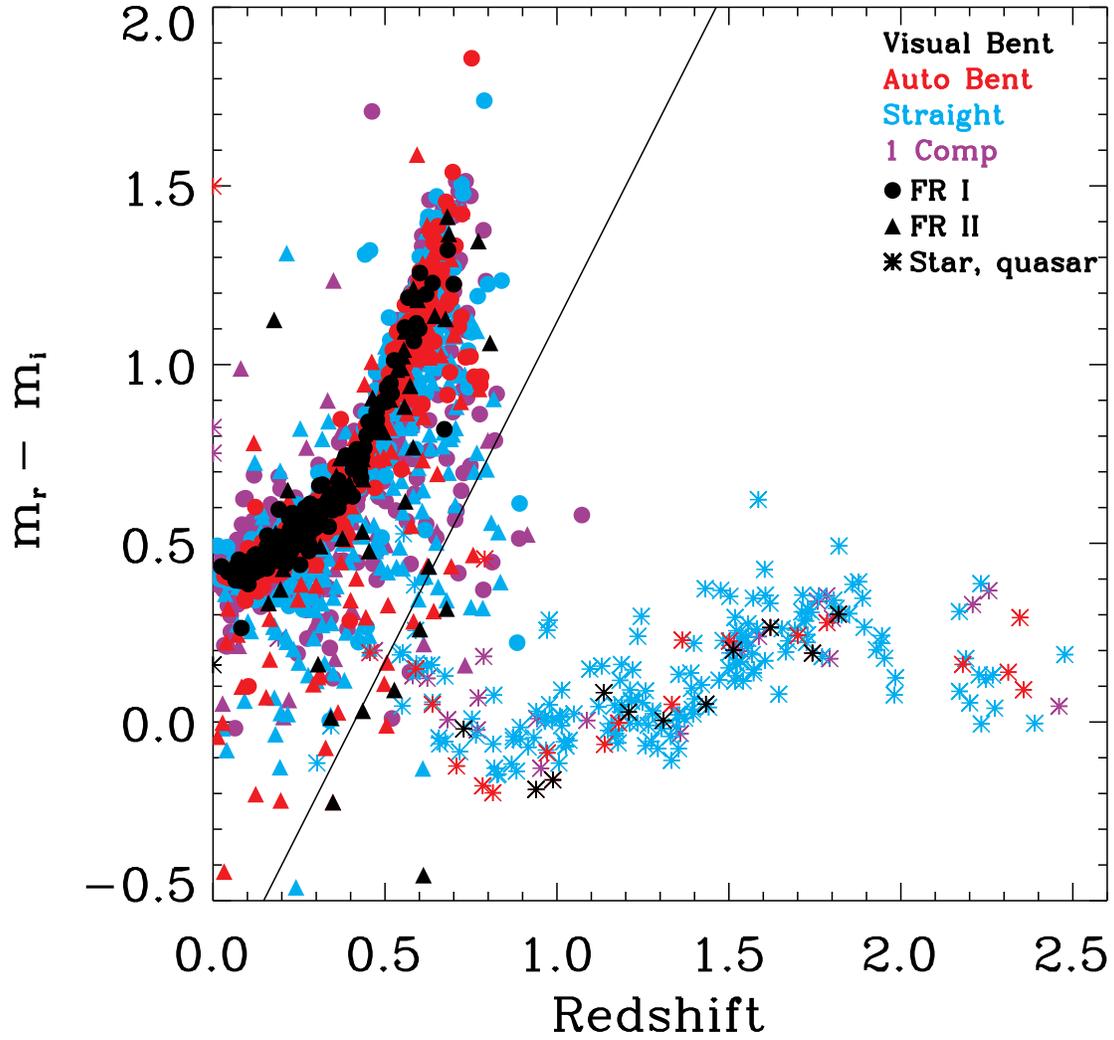}
\caption{Redshift vs. color for our radio source hosts.  The symbols are the same as in Figure~\ref{specvsphotoz}.  The point sources, as identified by \SDSS\/ (either misidentified stars or quasars), are represented by asterisks.  For the most part, the elliptical galaxies at the centers of galaxies clusters should have similar colors.  Thus, as redshift increases, the color of the galaxy should get redder.  The solid line represents the cut-off between sources that are likely to be red elliptical galaxies and those that are likely quasars.} \label{zvcolor}
\end{center}
\end{figure}

\index{Finding Optical Counterparts in the SDSS!Identifying Possible Clusters}
We expect that optical hosts associated with radio galaxies in clusters detected to the limit of the \SDSS\/ will mostly be red elliptical galaxies.  If we assume that the galaxies hosting the radio sources are ellipticals with similar colors, they will form a recognizable distribution on a plot of color versus redshift.  Figure~\ref{zvcolor} shows the results of comparing a source's redshift with its $r-i$ color.  Plotting redshift versus $r-i$ color allows us to differentiate between red elliptical galaxies and higher redshift blue quasars.  We drew a line in Figure~\ref{zvcolor} to set a boundary between ellipticals and quasars.  The sources below the line are likely associated with quasars and we excluded them from our samples to avoid attempting to search for galaxy clusters around quasars in the \SDSS..  Any galaxies potentially associated with a cluster around these high-redshift quasars would be too faint to be detected with the \SDSS.  These objects will be explored in more detail in a future paper.  In this paper we are more interested in the cluster environments surrounding elliptical galaxies.  We removed these blue sources from our samples as our ''color-cut'', along with any sources that are without a detection in any of the five \SDSS\/ color bands.  The removal of these sources accounts for less than $10\%$ of the sample for all of the samples except the straight sample, where the removal of these high-redshift blue sources accounts for more than $20\%$ of the sample.  We see this as the difference between columns 6 and 8 in Table~\ref{sample_breakdown}.

\subsection{Determining Radio Power and FR Morphology} \label{determining_radio_power_and_fr_morphology} \index{Finding Optical Counterparts in the SDSS!Determining Radio Power and FR Morphology}
With the redshift of the optical source giving us the distance to the source, we were able to calculate an absolute magnitude and a radio power.  Sources were de-reddened using the maps of \citet{schlegel1998} and k-corrected using code written specifically for use with \SDSS\/ sources \citep{blanton2007}.  With apparent magnitudes adjusted for reddening and k-corrected to a reference frame of $z=0$, we calculated rest-frame absolute magnitudes and radio powers for each source.

\begin{figure}
\begin{center}
\capstart
\includegraphics{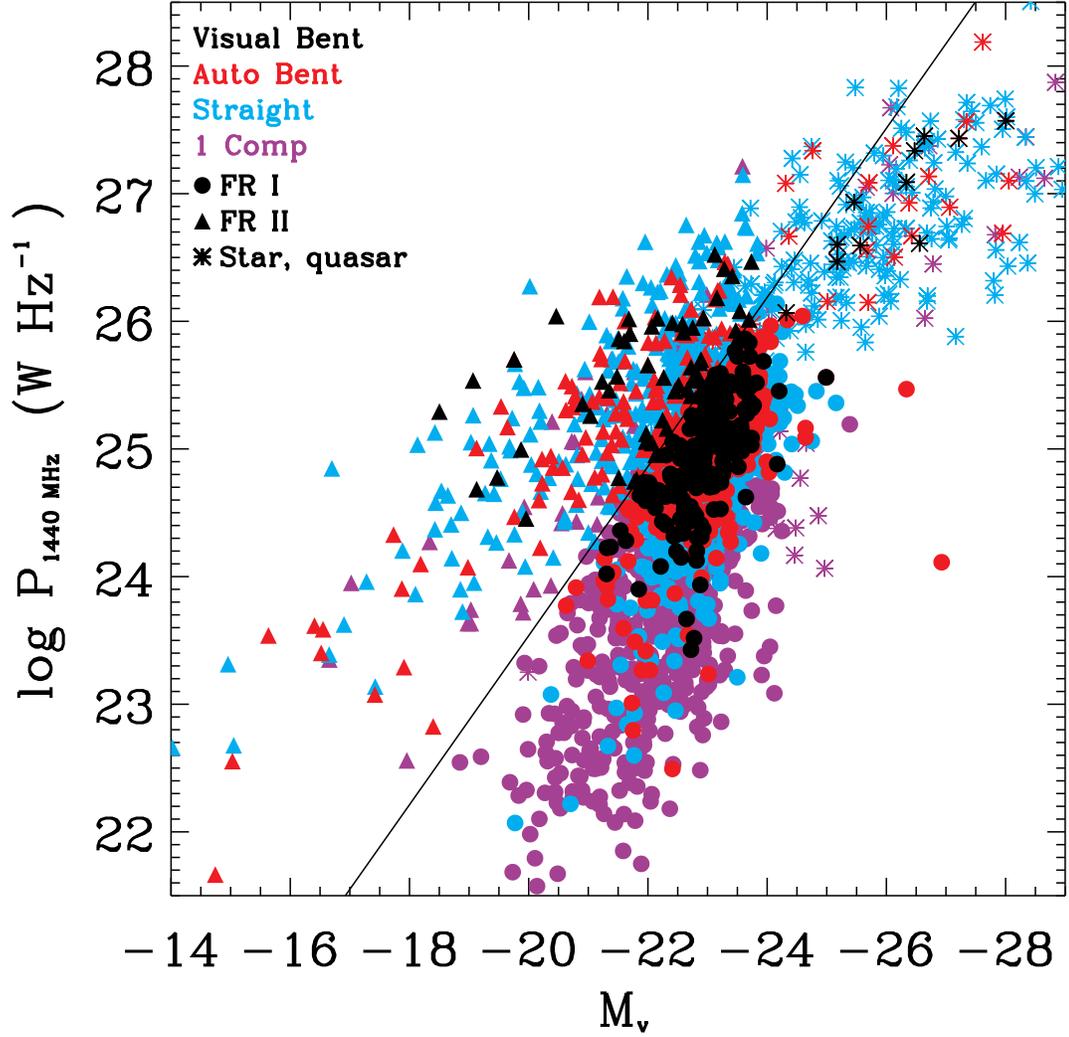}
\caption{Absolute V-magnitude vs. $1440$ MHz radio power for sources in our samples.  The symbols are the same as in Figure~\ref{zvcolor}.  \citet{ledlow1996} showed that based on a comparison of a source's absolute V magnitude and its radio power, a distinction could be made between FR I and FR II sources.  The line represents the division that \citet{ledlow1996} found between FR I and FR II sources.  The FR II's are above and to the left of the line and the FR I's are below and to the right of the line.  Using this, we were able to classify sources as either FR I or FR II and use this classification to make comparisons between cluster richness for each type.} \label{vvpower}
\end{center}
\end{figure}

\citet{ledlow1996} made a distinction between FR I and FR II sources using the radio power of the source as well as its absolute V-band magnitude, $M_V$.  They found that while FR II sources generally are found to have powers greater than $P_{1440\: \rm MHz}=10^{25}$ W Hz$^{-1}$, there was a correlation between the power of the radio source, the V-band absolute magnitude of the optical source associated with the radio source, and the morphology of the radio source.  The line drawn in Figure~\ref{vvpower} replicates the distinction that they found between FR I and FR II sources and we classified our sources based on these criteria.  We converted magnitudes from the \SDSS\/ g and r bands to the Johnson V band using the conversions in \citet{smith2002}, namely that $V=g-0.55(g-r)-0.03$.

Using this distinction between FR I and FR II sources, we compared the optical environments surrounding both types of sources.  Thus, we were able to not only compare bent-double sources and straight sources to single-component sources, we also examined the differences between FR I and FR II sources within these samples.  This allowed us to identify the type of radio sources most likely to be found in rich galaxy clusters.  While the FR I/II criteria are usually only applied to extended sources, we applied them here to our single-component sample as well which contained a mix of extended and non-extended sources using the \citet{ledlow1996} division.

\begin{figure}
\begin{center}
\capstart
\includegraphics[scale=.49]{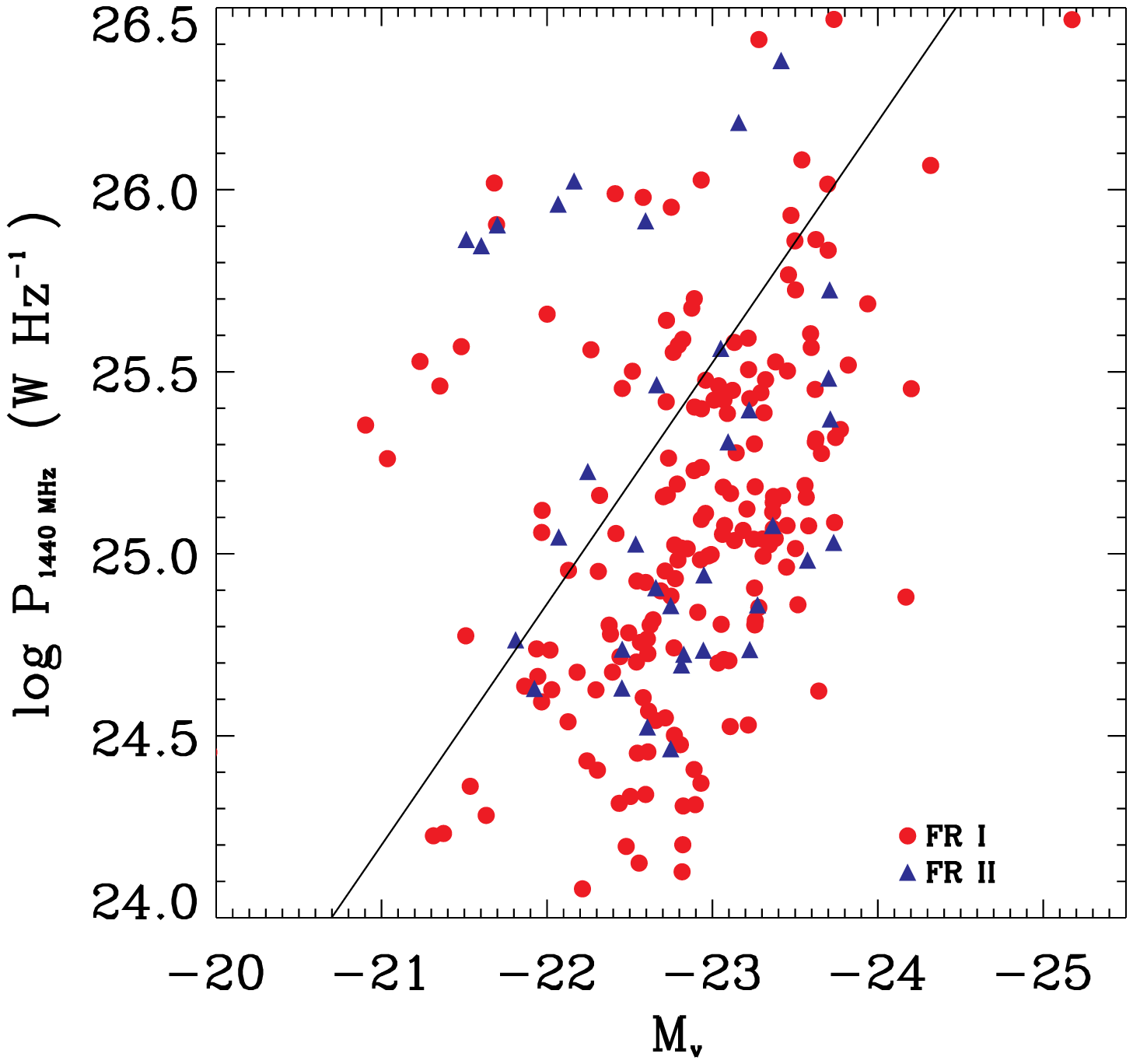}
\includegraphics[scale=.49]{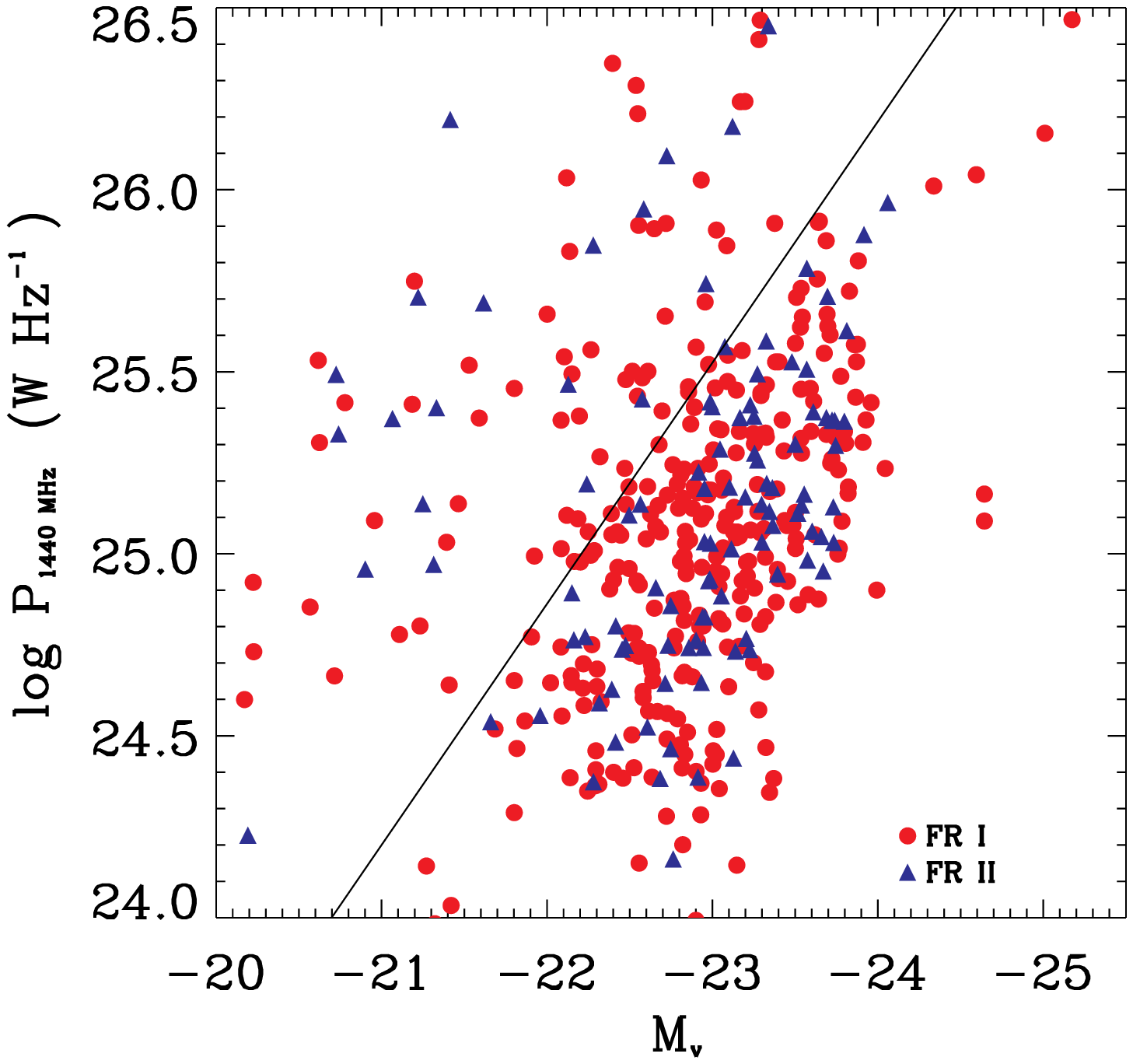}
\includegraphics[scale=.49]{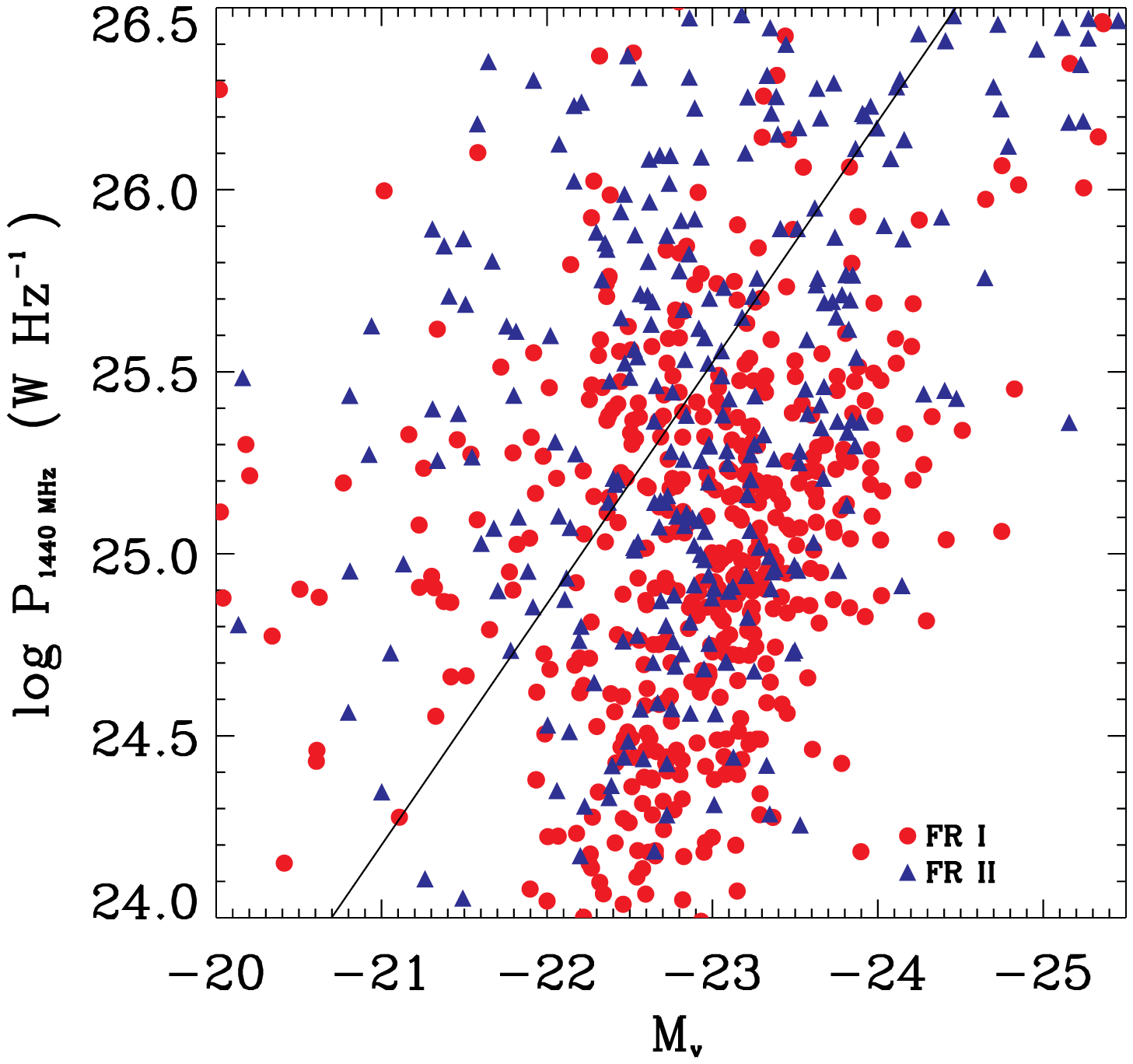}
\caption{Absolute V-magnitude vs. $1440$ MHz radio power for sources in our samples with visual FR classifications.  Filled red circles represent sources we have visually classified as FR I and blue filled triangles represent sources we have visually classified as FR II.  The top-left panel shows the results of the visual-bent sample, the top-right panel shows the results of the auto-bent sample, and the bottom panel shows the results of the straight sample.  We find that while there is a general trend for the FR I type radio sources to be located below the threshold set forth in \citet{ledlow1996}, it does not hold for all sources, in agreement with \citet{best2009}.} \label{fig:vvvpower}
\end{center}
\end{figure}

We also visually determined the FR I and FR II classifications of all radio sources in the bent and straight samples with an associated optical source that met our reliability criteria.  We visually inspected the \FIRST\/ radio image and contour maps of each of these sources.  For the sources that are clear FR I or FR II sources, we identified them as such.  For the sources where the visual classification was less certain, we marked them as possible FR I or FR II sources.  There were also several sources that fit neither the FR I or FR II criteria and we marked those sources as "other".  The plots in Figure~\ref{fig:vvvpower} are similar to Figure~\ref{vvpower}, except we use the FR I  or FR II classification we determined visually for the visual-bent,  auto-bent,  and straight samples.

Similar to the results of \citet{best2009} and \citet{lin2010}, we found that while there may be a general trend for FR I sources to reside below the line proposed by \citet{ledlow1996}, it is by no means an absolute separation.  Specifically, we found that for the visual-bent sample, $81\%$ of the sources we visually identified as FR I sources are also FR I sources according to the criteria of \citet{ledlow1996}.  For the same visual-bent sample, only $40\%$ of the sources we visually classified as FR II would also be FR II sources in the \citet{ledlow1996} scheme.  For the auto-bent sample, $79\%$ of FR I sources and $21\%$ of the FR II sources that we visually-identified follow the \citet{ledlow1996} demarcation.  Likewise, for the straight sample, $76\%$ of the visually-identified FR I and $44\%$ of the visually-identified FR II sources follow the \citet{ledlow1996} criteria.  This is possibly the result of extended emission being resolved out leading to the mis-classification of some FR I sources as FR II sources.  We are more likely to mis-classify an FR I source as an FR II source than the other way around.

Table~\ref{fr_class} gives the numerical breakdown of visually classified FR I and FR II sources in the two samples.  Columns 3 and 4 are the number of clearly identified FR I and FR II sources in each sample, respectively.  Columns 5 and 6 list the number of ambiguous FR I and FR II sources, and column 7 gives the number of sources in each sample with a radio morphology that cannot be classified, even tentatively, as an FR I or FR II source.  Columns 8 and 9 list the total number of FR I and II visually-classified sources, respectively, that remain after redshift and apparent magnitude selection (see \S\ref{schechter_correction} and \S\ref{nearby_clusters}).  There were several sources in the automatically selected samples that appeared to be either two or three unrelated sources.  This visual inspection allowed us to classify those sources as either questionable FR I/II (in the case where it appeared that there was an unrelated radio source near a double-lobed source) or as other (in the case where all three radio sources appeared to be unrelated to each other).  This allows us to state with more certainty that the sources that are securely classified as FR I/II are in fact true three component radio sources, and not spurious chance-coincidences.

\begin{figure}
\begin{center}
\capstart
\includegraphics[scale=.5]{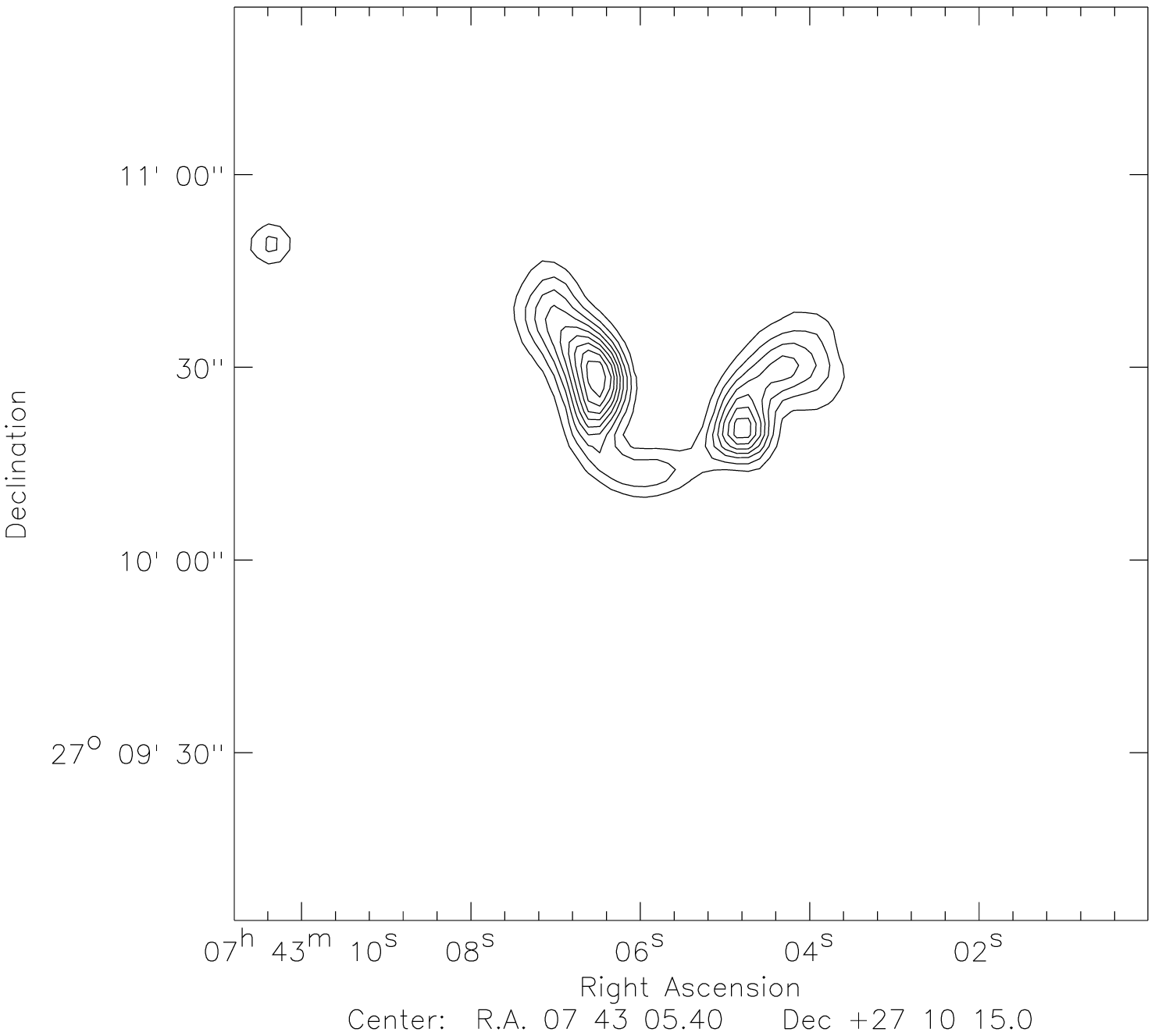}
\includegraphics[scale=.5]{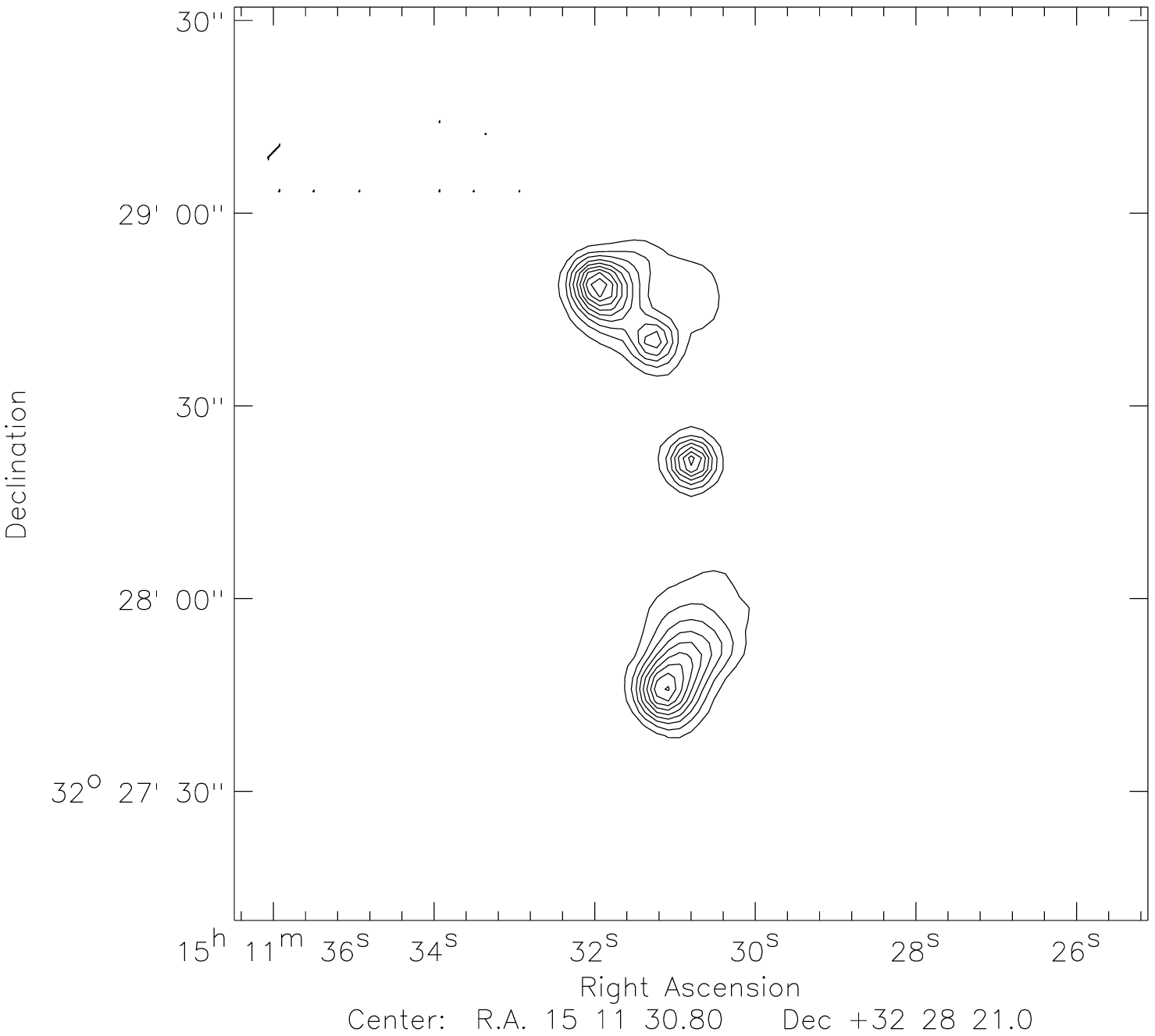}
\includegraphics[scale=.5]{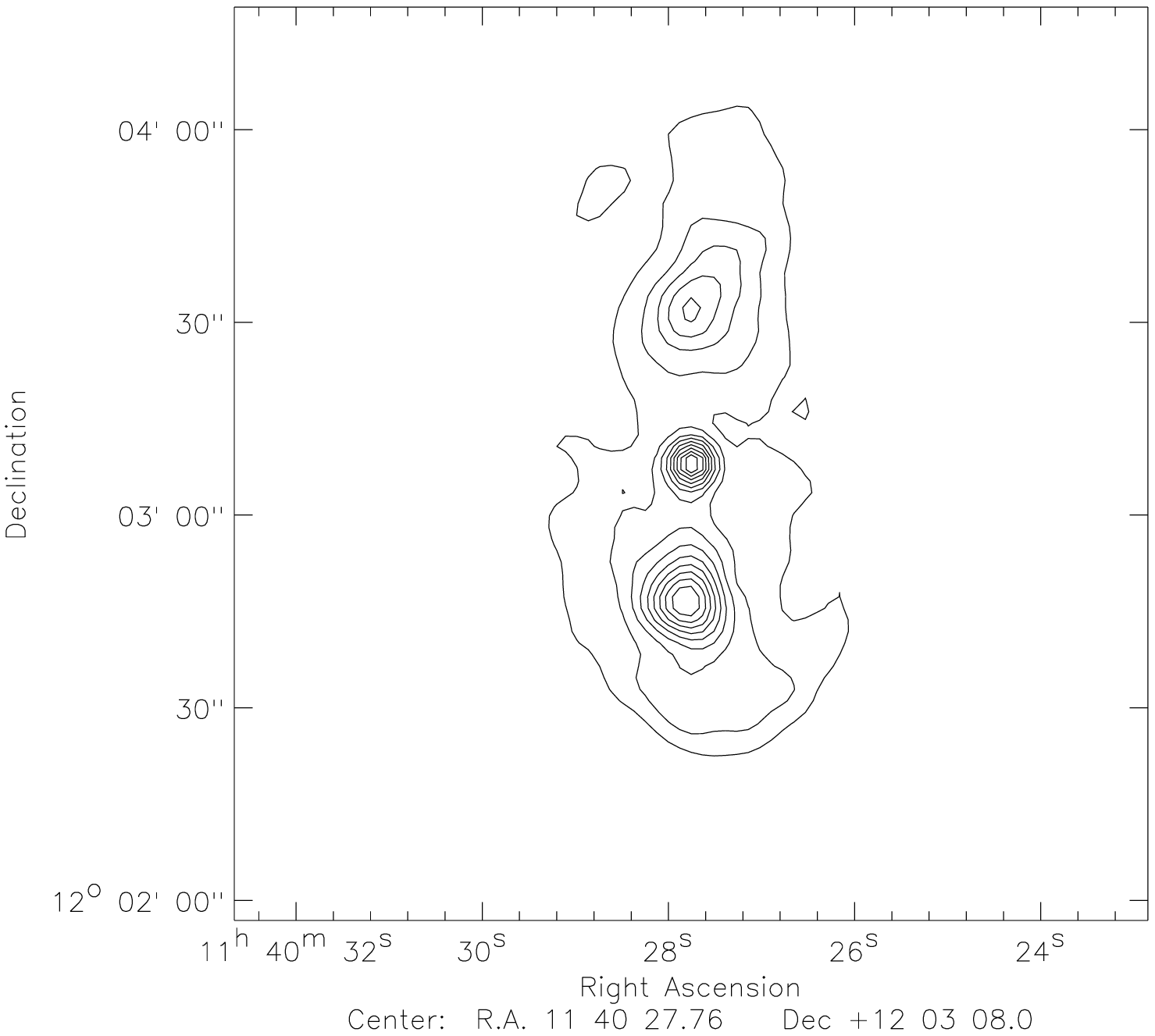}
\includegraphics[scale=.5]{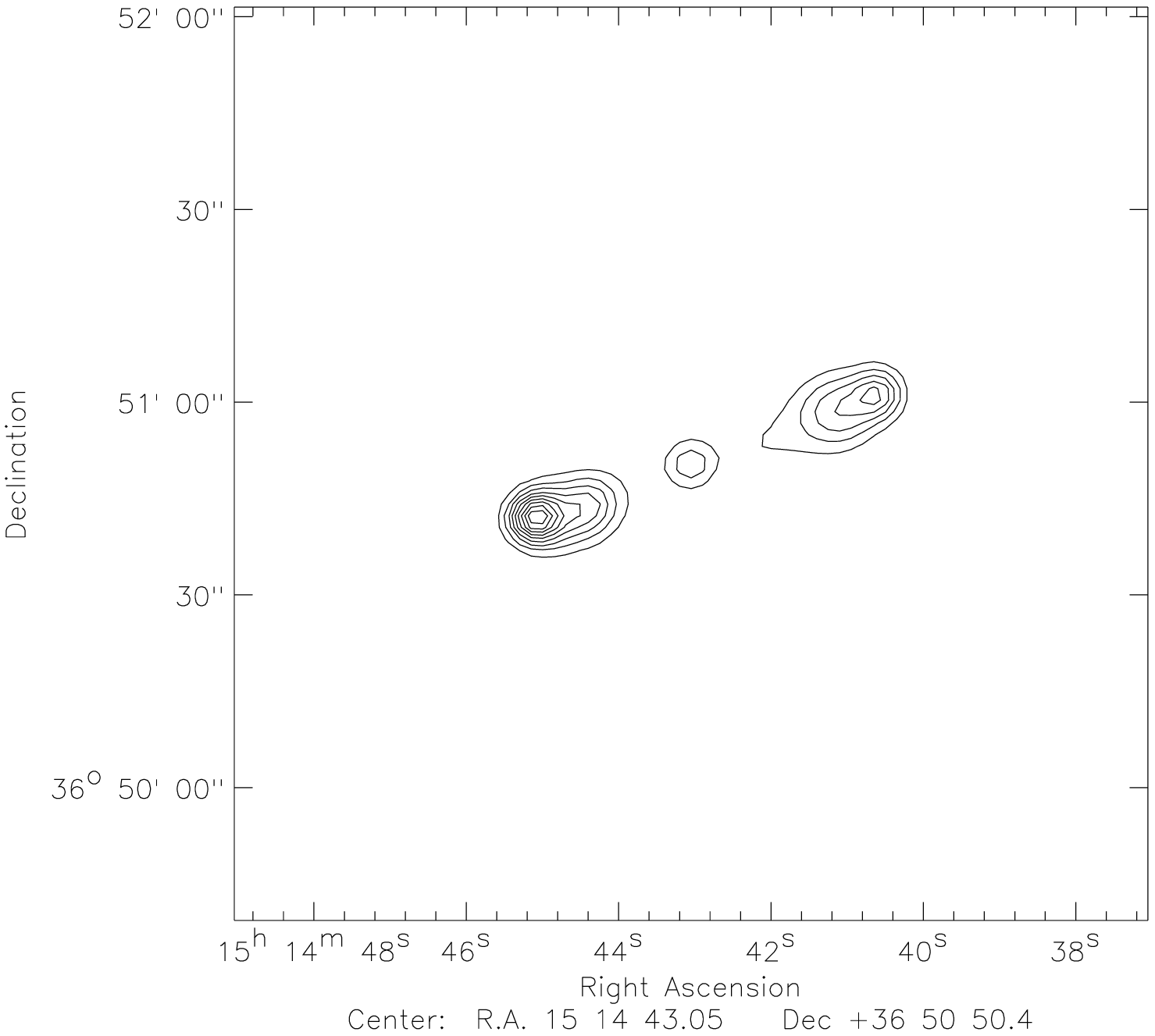}
\caption{Radio contours of characteristic sources from our samples.  The top panels both show bent-lobed sources, the left-hand panel illustrates an example of a bent FR I source, the right-hand panel is an example of a bent FR II source.  The bottom panels both show straight-lobed radio sources, the left-hand panel is an example of a straight FR I source, the right-hand panel is an example of a straight FR II source.  The contours are linearly scaled with ten contours with minimum contours of 0.76 mJy, 1.2 mJy, 3.1 mJy, and 13 mJy and maximum contours of 9.1 mJy, 16 mJy, 31 mJy, and 160 mJy for the upper-left, upper-right, lower-left, and lower-right panels, respectively.} \label{fig:contours}
\end{center}
\end{figure}

The radio sources located at high redshifts present problems for a definitive visual FR I/II classification.  For these sources, the three components are typically located in very close proximity on the sky and thus difficult to determine if the lobes are edge-bright, etc.  The resolution of the \FIRST\/ survey also limits accurate FR I/II determination, especially for the low surface-brightness sources.  For the sources where this becomes an issue, it is up to the discretion of the examiner whether the source is classified as I or II and it can vary even after repeated viewings of the same source.  Thus, visual classification is not necessarily a better solution than classifying using the \citet{ledlow1996} criteria.  We show some typical FR I and FR II sources from the visual-bent and straight samples in Figure~\ref{fig:contours}.

\subsection{Source Extent} \label{source_extent} 
\begin{figure}
\begin{center}
\capstart
\includegraphics{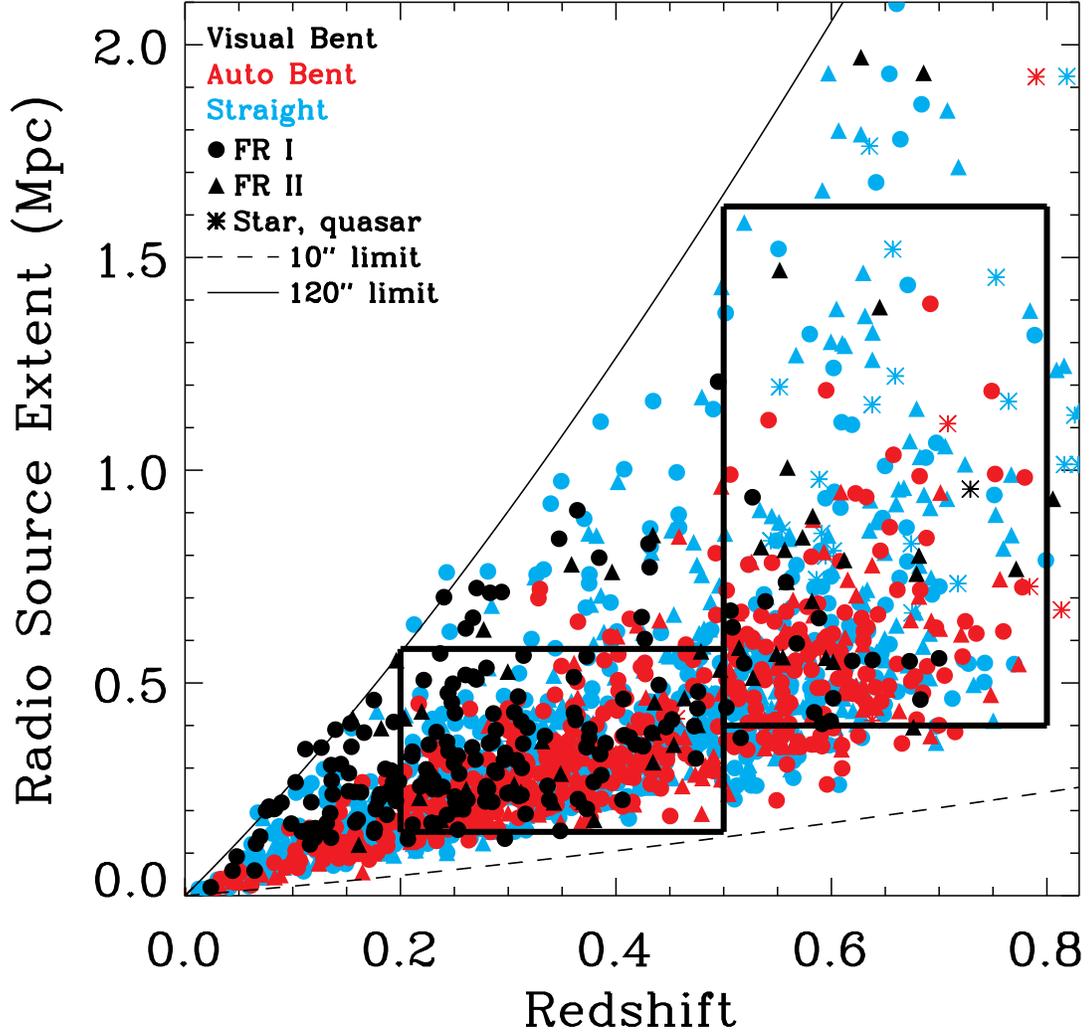}
\caption{Radio source extent vs. redshift.  The symbols are the same as in Figure~\ref{zvcolor}.  The dashed line corresponds to the physical extent at a given redshift for a multi-component source that has a total angular size of $10\arcsec$.  This is the lower limit to the detection and identification of a multi-component source in the \FIRST\/ survey.  The solid line corresponds to an angular size of $120\arcsec$.  This is the approximate upper limit for the physical extent of our sources.  The few sources above this line are slightly larger owing to the extent of the lobes.  The boxes represent samples of sources that are without selection effects for physical size and/or redshift.  The two boxes are used to constrain the samples in order to compare similar distributions of sources.} \label{zvsize}
\end{center}
\end{figure}

\index{Finding Optical Counterparts in the SDSS!Source Extent}
We determined the physical extent of the extended sources.  We measured the angular distance from the center of each lobe to the central source and added these together.  In addition, we included the semi-major axis of each lobe in the total extent of the source, so that our extents are effectively the distance from the edge of one lobe to the other, as if the source was straight.  We converted these angular sizes to physical extents using the redshift of the source as determined by the optical counterpart.  These physical extents are affected by projection effects caused by the inclination of the source as a result of our viewing angle.  Figure~\ref{zvsize} compares the extent of the radio source with the redshift of the source.  We plotted upper and lower limits for the angular extent of the sources.  The center of each component of a source had to be within $60\arcsec$ of another component to be in the visual-bent sample and $57\farcs6$ for the auto-bent and straight samples, but the lobes can have extended emission, giving some sources total extents greater than the presumed $120\arcsec$ (for the visual-bent sample) or $115\farcs2$ (for the auto-bent and straight samples) maxmium angular extent.  An extent of $120\arcsec$ is shown as the solid line in Figure~\ref{zvsize}.  The angular resolution of the \FIRST\/ survey is $5\arcsec$.  In order for a source to be identified as a multi-component source, all of the sources (at least the central source and two lobes) need to be resolved independently.  This implies that there is a minimum angular extent of $10\arcsec$ for all of the three-component sources in our sample.  This is shown as the dashed line in Figure~\ref{zvsize}.

\begin{figure}
\begin{center}
\capstart
\includegraphics[scale=0.49]{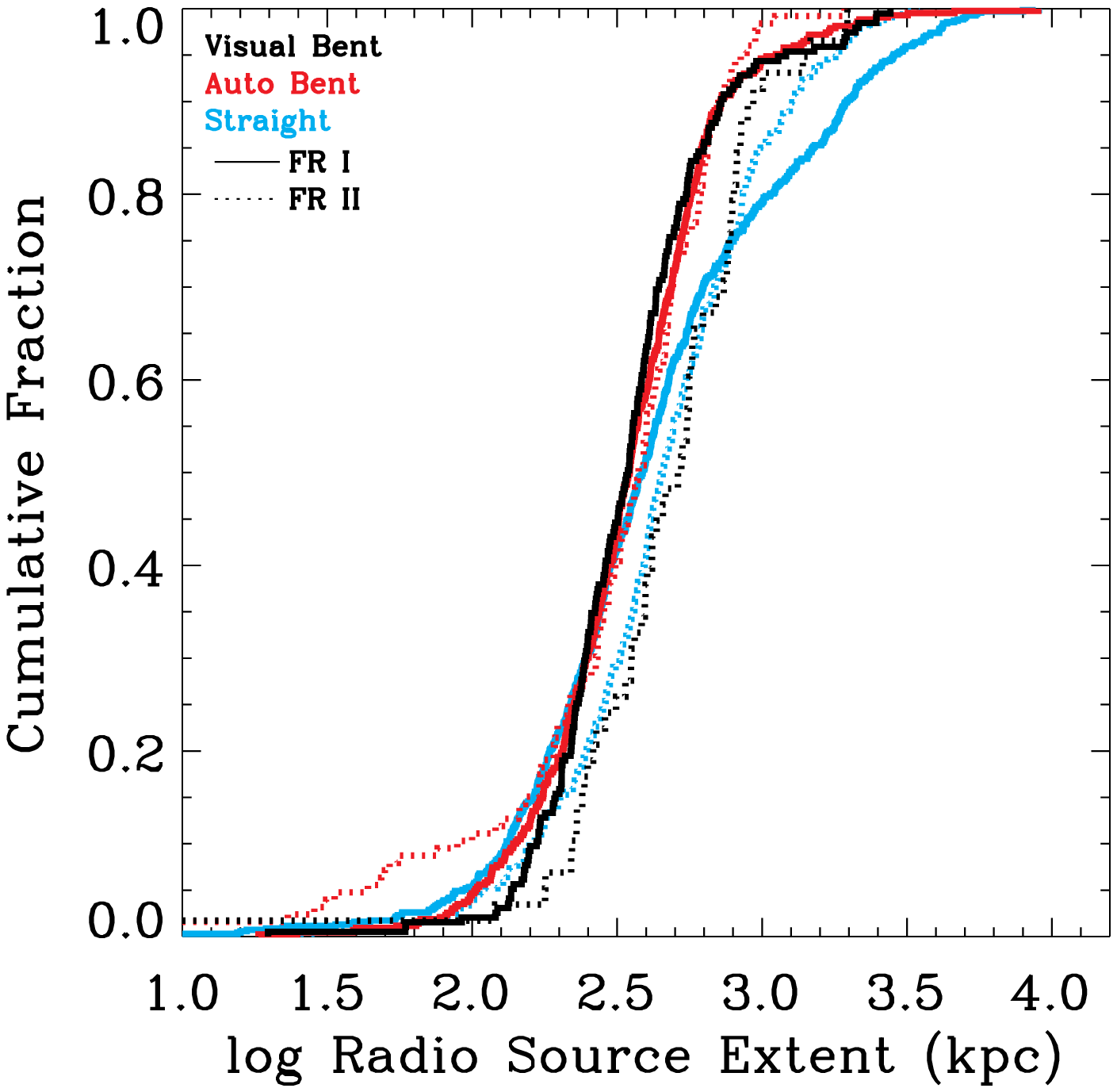}
\includegraphics[scale=0.49]{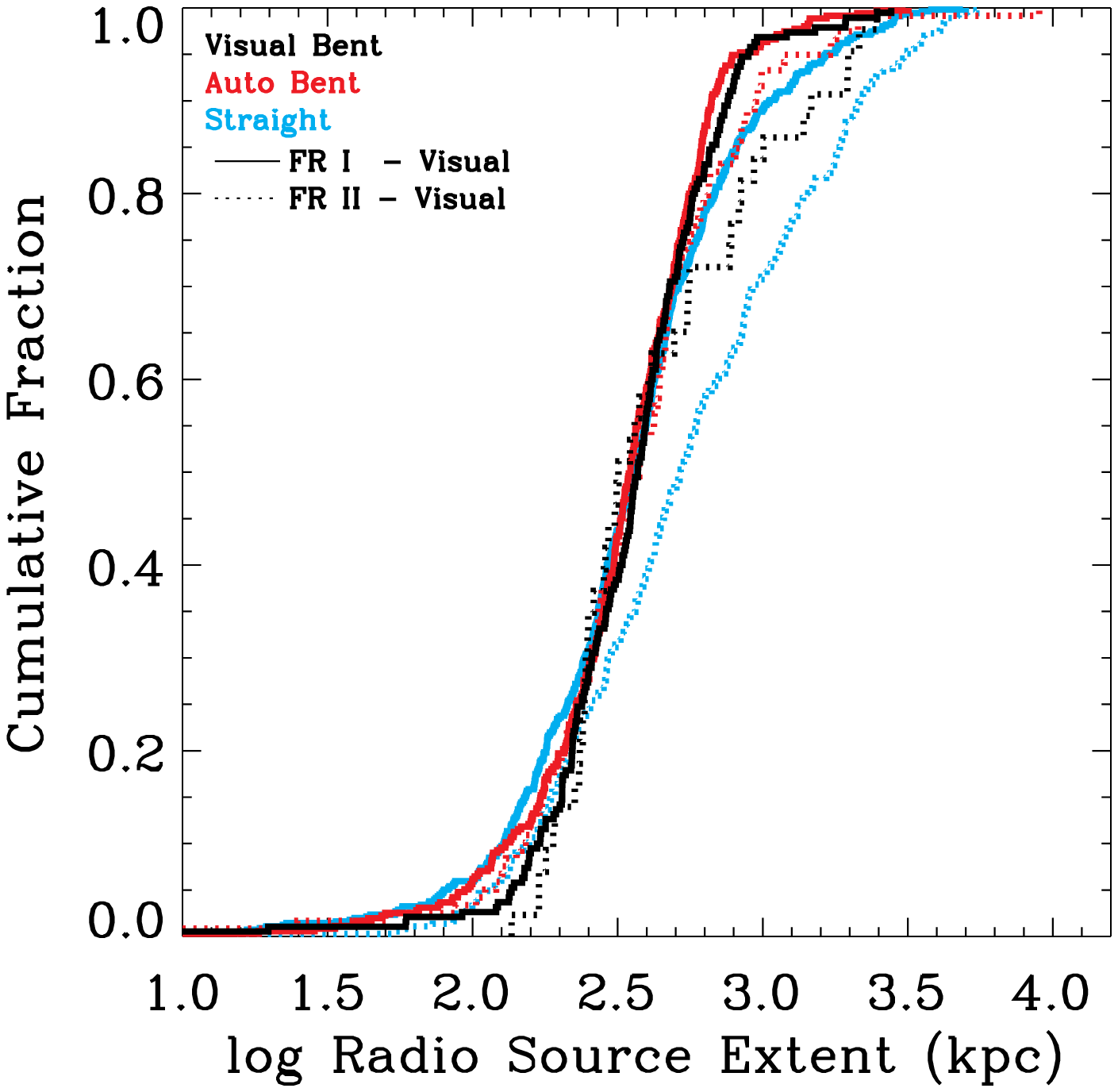}
\caption{The cumulative distribution of the physical extent of FR I and FR II sources in the different samples.  The solid lines represent the distributions of the FR I sources and the dotted lines represent the distributions of the FR II sources.  Black solid and dotted lines represent the visual-bent sample, red for the auto-bent sample, and blue for the straight sample.  The left panel shows the distribution when using the \citet{ledlow1996} criteria for determining FR I/II morphology and the right panel is the distribution when using our visual FR I/II classification.  In general, the FR II sources are larger than the FR I sources.} \label{multicumufrac}
\end{center}
\end{figure}

Previous authors argue that the physical extent of a radio source is analogous to the age of that source:  the older a source is, the more time it has had to expand from the central engine.  Thus the more extended sources are also the oldest sources.  \citet{best2009} finds that for their sample, FR I sources are more extended, and thus older.  We find, for all three of our samples, that the FR I sources are in fact the smaller sources.  This can be seen in Figure~\ref{multicumufrac} where we plotted the cumulative fraction of sources as a function of source extent.  We see that, with the exception of the largest straight-lobed radio sources, FR II sources (short dashed line) are larger than FR I sources (solid line), both when determining FR I/II morphology based on the \citet{ledlow1996} criteria (the left-hand plot in Figure~\ref{multicumufrac}), and determining the morphology visually (right-hand plot).  We determined the K-S test probability that the FR I and FR II populations from each sample were drawn from the same parent population.  We found that when FR morphology was determined by the method of \citet{ledlow1996} the size distributions of the visual-bent sample had a K-S test probability of $0.0730\%$, the auto-bent sample had a probability of $60.0\%$, and the straight sample had a probability of $0.0631\%$ that the FR I and II sources came from the same parent population.  When the FR morphology was determined visually, we found that the K-S test probability was $16.5\%$ for the visual-bent sample, $16.0\%$ for the auto-bent sample, and $5.67\times10^{-10}\%$ for the straight sample that the FR I and II sources came from the same parent population.  These low probability scores (with the exception of the \citet{ledlow1996} classified auto-bent sample) imply that FR I and FR II sources are drawn from different parent populations in terms of size distribution.  This may be a result of FR II sources being generally more powerful than FR I sources.  The more powerful FR II sources will be able to expand to larger extents before the expansion is halted by any intervening ICM.

\begin{figure}
\begin{center}
\capstart
\includegraphics[scale=.49]{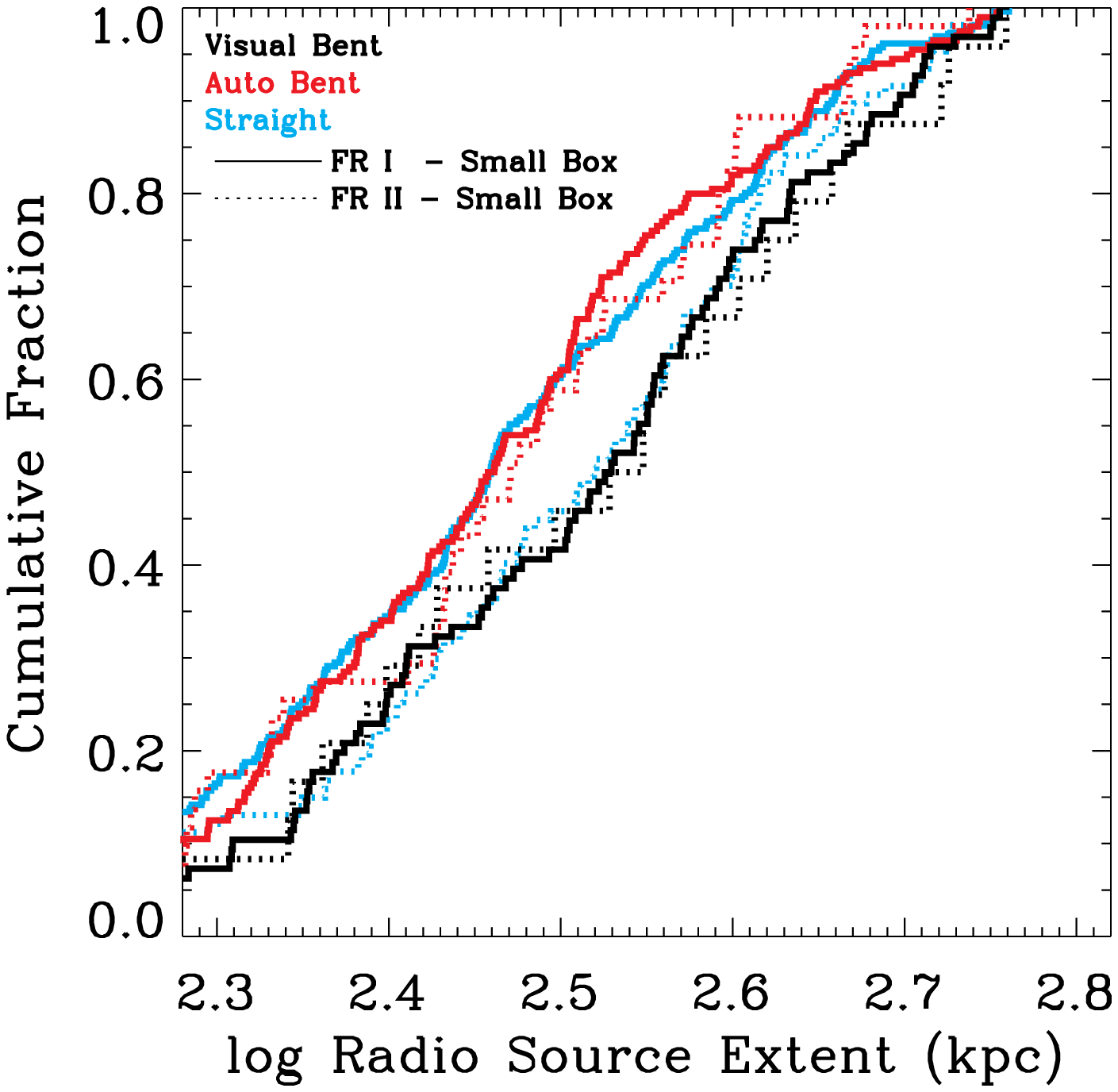}
\includegraphics[scale=.49]{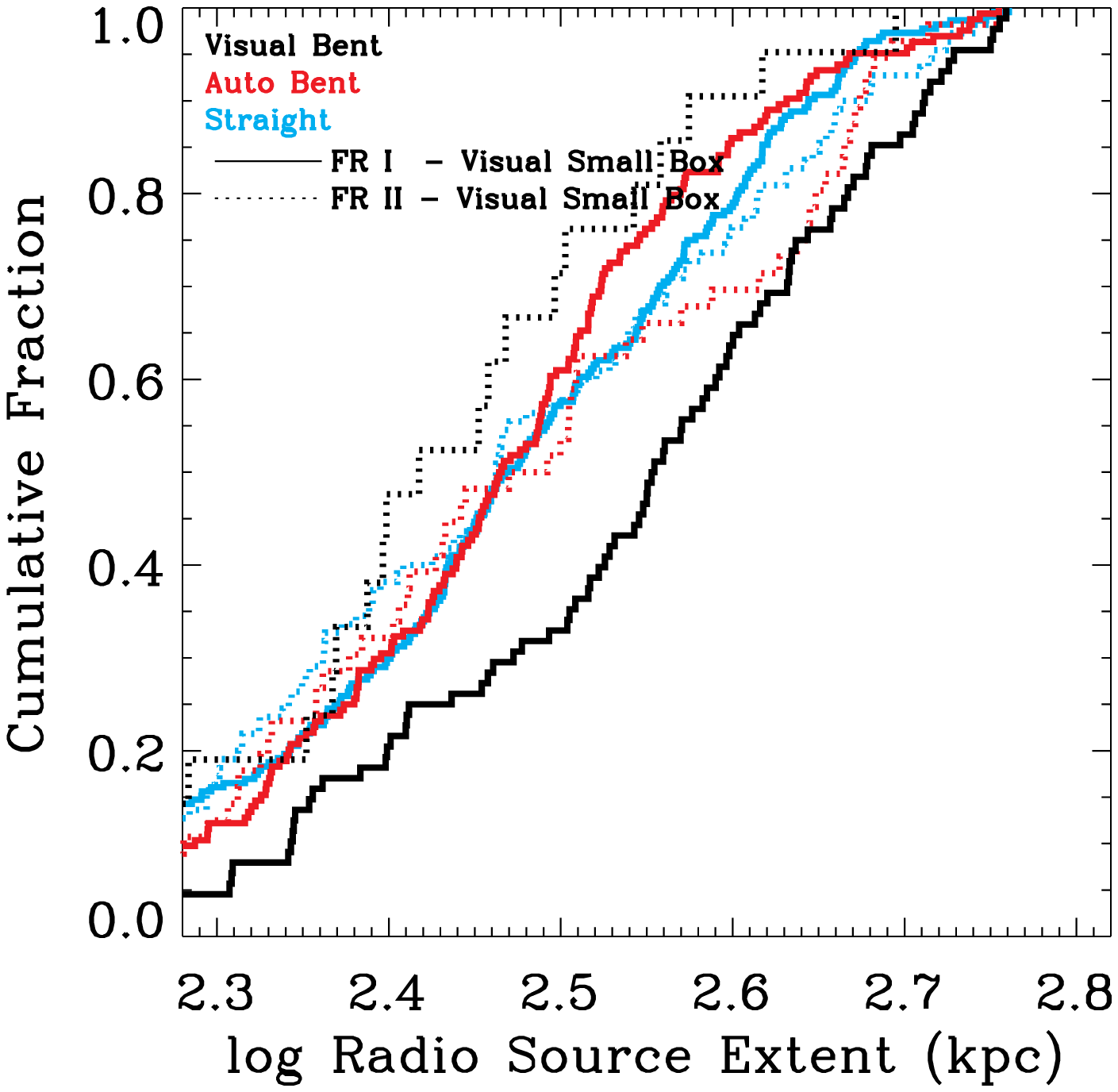}
\includegraphics[scale=.49]{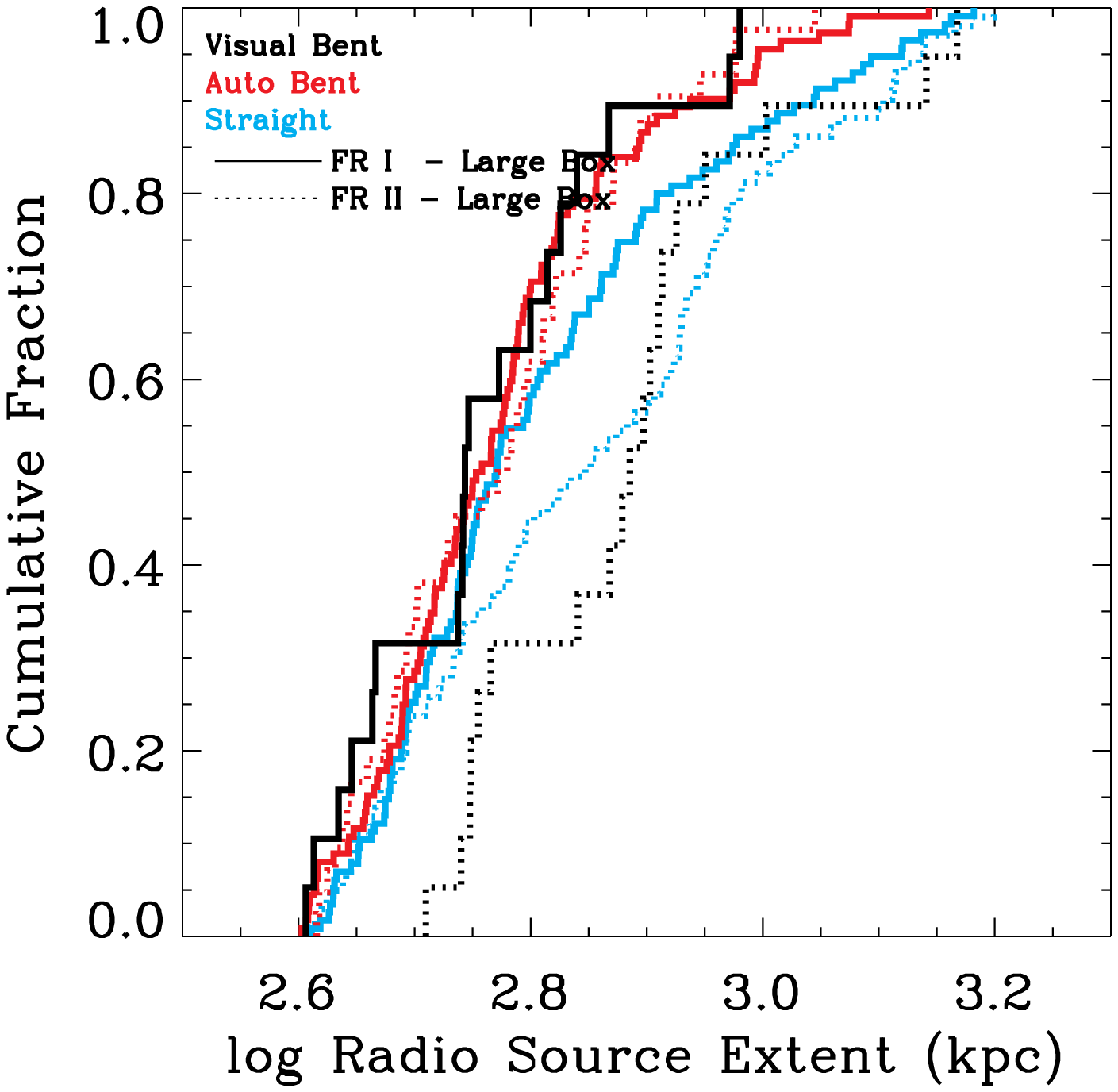}
\includegraphics[scale=.49]{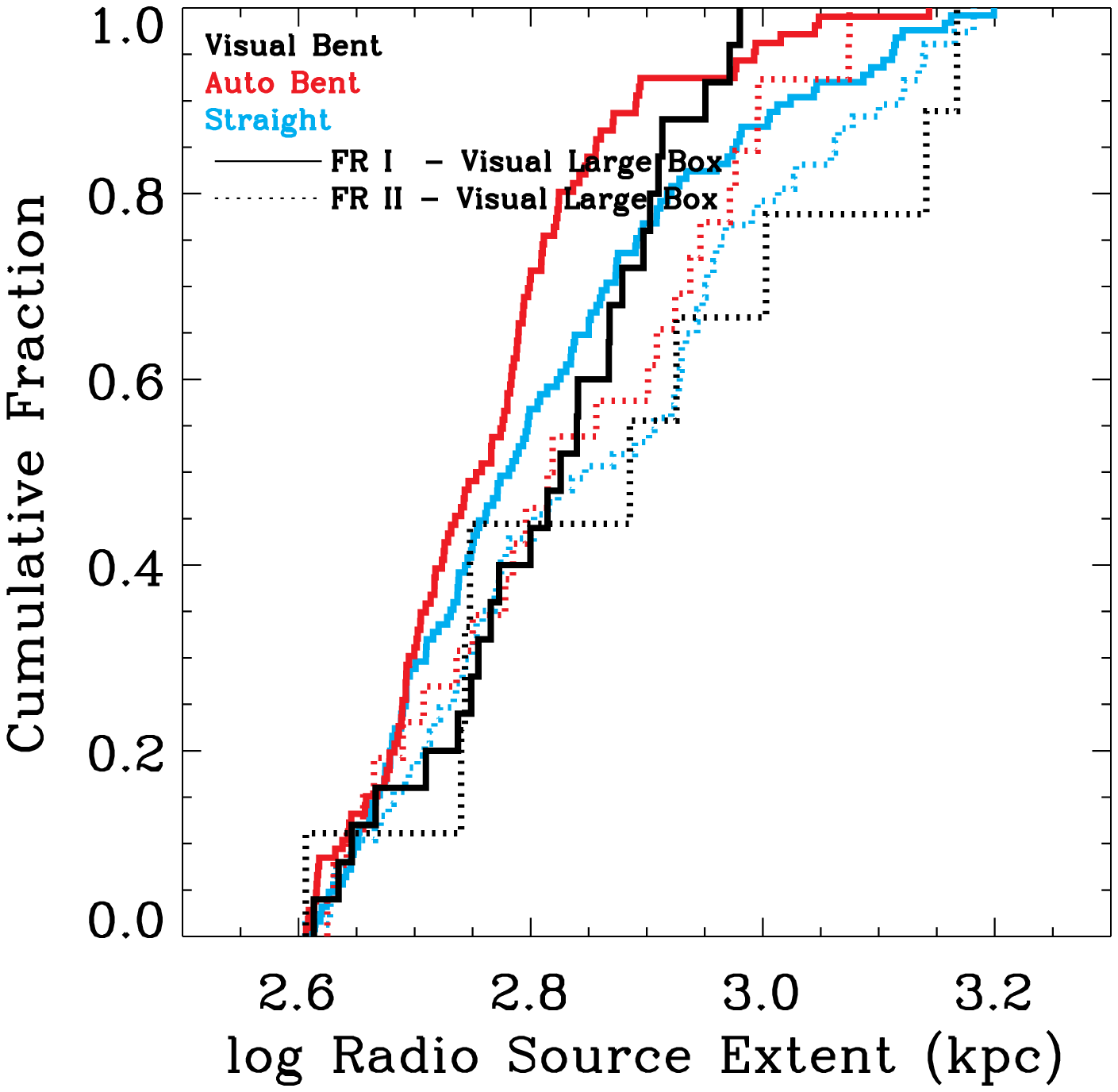}
\caption{A reproduction of Figure~\ref{multicumufrac}, except using only sources contained within the small (top panels) box from Figure~\ref{zvsize} and the large box (bottom panels).  Specifically, the upper-left panel shows the cumulative distribution of the radio source extent for sources with FR classification derived using the \citet{ledlow1996} criteria and located within the small box in Figure~\ref{zvsize}.  The upper-right panel shows the same distribution for sources located within the small box in Figure~\ref{zvsize} and classified as FR I or II based on our visual inspection.  The lower-left panel shows the distribution of sources contained in the large box and classified as FR I or II based on \citet{ledlow1996} and the lower-right panel shows the distribution of sources contained in the large box and classified as FR I or II based on our visual inspection.  In general, we see that FR II sources are larger than FR I sources.} \label{boxedmulticumufrac}
\end{center}
\end{figure}

Further examination of Figure~\ref{zhist} shows that while they are generally similar, our samples do not have identical redshift distributions.  Most notably, the visual-bent sample peaks at a redshift of $z\sim0.2-0.3$, lower than that of the other samples.  We limited our samples to those sources with components separated by a maximum of $\sim60\arcsec$ from each other.  Of course, this is an angular size and not a physical size, so nearby, physically extended sources are likely to have components separated by more than this maximum angular separation.  Since the visual-bent sample is composed of lower-redshift sources in general, this sample may include a lower fraction of sources with large physical extents than the other samples. In order to examine a subset of sources that are free of this bias, we created two groups of subsamples based on Figure~\ref{zvsize}.  We defined two boxes on this plot, one box that is smaller and is composed of lower-redshift, less physically extended radio sources, and one box of higher-redshift sources containing more extended radio sources.  The sources in these boxes (which we refer to as the small box and the large box later in this paper) are free of bias regarding their angular extent as a function of redshift.  Both boxes limit the physical size of the sources such that at a given redshift they would be included in our samples, regardless of their angular extent.  Examining only the sources in these boxes should remove any effect of angular size of the radio source in the results when examining the physical sizes of these sources.  We plotted the size distributions of the FR I and FR II sources within these two boxes in Figure~\ref{boxedmulticumufrac} for our different samples.  Table~\ref{table_size_distribution} lists the number of FR I and II sources (determined both by the \citet{ledlow1996} criteria and visually) in each sample contained within the boxes.  Specifically, columns 2 and 3 list the number of FR I and II sources, respectively, determined using the criteria of \citet{ledlow1996}, contained within the small box.  Columns 4 and 5 list the number of FR I and II sources, respectively, determined visually, contained within the small box.  Columns 6 and 7 list the number of FR I and II sources, respectively, (using \citet{ledlow1996} criteria), contained within the large box.  Columns 8 and 9 list the number of FR I and II sources, respectively, determined visually, contained within the large box.

In general, we find the same result as previously when we limit the sample to sources within the boxes.  As seen in Figure~\ref{boxedmulticumufrac}, the FR II sources are still mostly larger than the FR I sources.  The most obvious exception is for sources in the visual-bent sample that have been classified as FR I or II visually (upper right panel of Figure~\ref{boxedmulticumufrac}).  In this case, the FR I sources are found to be larger.  Specifically, the K-S test probabilities are as follows.  When determining FR morphology using the method of \citet{ledlow1996} for the sources contained within the small box (upper-left plot), we found that the visual-bent FR I and II sources had a $99.9\%$ probability of coming from the same parent population, the auto-bent FR I and II sources had a probability of $56.5\%$, and the straight FR I and II sources had a probability of $2.61\%$ of coming from the same parent population.  When we examined these same sources located within the large box (lower-left plot), we found that the visual-bent sample had a K-S probability of $0.601\%$, the auto-bent sample had a probability of $85.5\%$, and the straight sample had a probability of $0.985\%$.  When we determined the FR morphology visually for sources within the small box (upper-right plot), we found that the visual-bent sample had a K-S test probability of $0.223\%$, the auto-bent sample had a probability of $11.2\%$, and the straight sample had a probability of $49.7\%$.  The same sources located in the large box (lower-left plot) had K-S test probabilities of $37.7\%$ for the visual-bent sample, $0.994\%$ for the auto-bent sample, and $0.617\%$ for the straight sample.  Given the results for the other samples and classification methods, as well as the difficulty in classifying sources as FR I or II visually, we draw the general conclusion that FR I sources appear to be smaller in our samples.

\section{Determining Cluster Richness} \label{determining_cluster_richness} \index{Determining Cluster Richness}
We examine the cluster properties of the radio sources in our samples.  This includes measuring the number of galaxies around each radio source.  This richness measurement will identify clusters as well as groups and even areas with less-than-average galaxy counts.  Characterizing the richness in the optical environment around each radio source will allow us to infer the basic cluster properties of the different samples of radio sources.

\subsection{Choosing A Richness Measurement System} \label{choosing_a_richness_measurement_system} \index{Determining Cluster Richness!Choosing A Richness Measurement System}
We determine the richness of the environments using a method similar to \citet{allington-smith1993} and \citet{zirbel1997}.  In those papers, as well as \citet{blanton2000,blanton2001}, the richness measurement $N^{-19}_{0.5}$ corresponds to counting all galaxies within a $0.5$ Mpc radius of the radio source with absolute magnitudes brighter than $M_V=-19$.  This is an improved metric over that of \citet{hill1991} which involved counting all galaxies within $0.5$ Mpc of the radio source with apparent magnitudes in the range of $m_{rg}$ to $m_{rg}+3$, with $m_{rg}$ being the magnitude of the radio source.  The appealing aspect of using the $N^{-19}_{0.5}$ method is that it measures the same absolute magnitude range for all of the clusters, down to the magnitude limit of the observations.  This allows for a more accurate measure of the cluster richness than through the use of apparent magnitudes alone.  In the case where the radio galaxy is not the brightest galaxy in the cluster, using the apparent magnitude of the optical source associated with the radio source does not account for galaxies in the cluster that might be brighter than this.  Counting all of the galaxies down to an absolute magnitude limit avoids this problem.

When \citet{allington-smith1993} chose $0.5$ Mpc as a radius within which to search for cluster members, they were balancing search area with their smaller detector size.  They acknowledged that they expected to find poor groups instead of rich clusters, so the $0.5$ Mpc radius would be adequate for these means.  However, for richer clusters this radius is too small, i.e. \citet{abell1958} used a radius of $1.5h^{-1}$ Mpc.  We chose a radius of $1.0$ Mpc as a compromise between the two.  This is also a reasonable area in which to search for cluster members in the much deeper \SDSS\/ catalog.  Our richness metric is thus $N^{-19}_{1.0}$, a count of all galaxies brighter than $M_r=-19$ within $1.0$ Mpc of the radio source.

We have used the r-band in the \SDSS.  If we assume that a typical elliptical galaxy has a color of $V-R\approx0.9$ and using the conversion between the standard Johnson bands and the \SDSS\/ bands provided in \citet{smith2002}, we find that $V\approx r+0.8$.  Thus by using the \SDSS\/ r-band, we searched, on average, $0.8$ magnitudes fainter than \citet{allington-smith1993}.  By including galaxies nearly a full absolute magnitude fainter as well as searching out to a larger radius around the radio source allows us a better measurement of the richness of the richer clusters.

\subsection{Accounting for Background Counts} \label{accounting_for_background_counts} \index{Determining Cluster Richness!Accounting for Background Counts}
Because we are using the \SDSS\/, we have the advantage of being able to estimate the background galaxy count locally around each cluster candidate.  We are not limited by the size of the CCD as we would be if we were observing each of these fields individually.  For the local background galaxy estimation we searched within an annulus with inner and outer radii of $2.7$ and $3$ Mpc, respectively, around the radio source for all optical sources brighter than $M_r=-19$.  Most sources located this far from the radio source will be unassociated with the potential cluster.  Of course, the area of this annulus is larger than the area in which we are searching for potential cluster galaxies.  To account for this we normalized the number of galaxies found within the background annulus to match that of the inner $1.0$ Mpc by multiplying by the ratio of the areas.

\begin{figure}
\begin{center}
\capstart
\includegraphics[scale=.42]{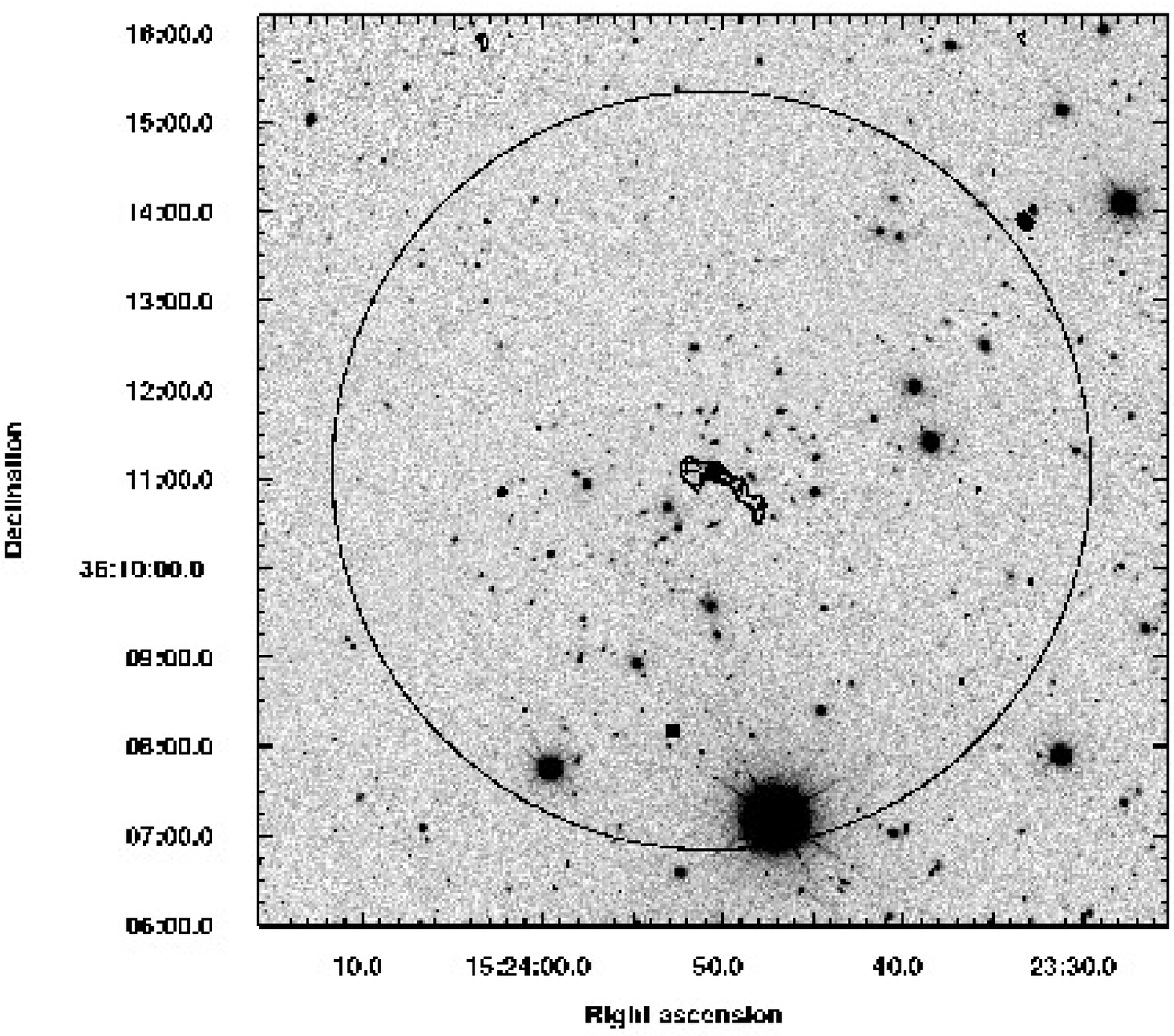}
\includegraphics[scale=.42]{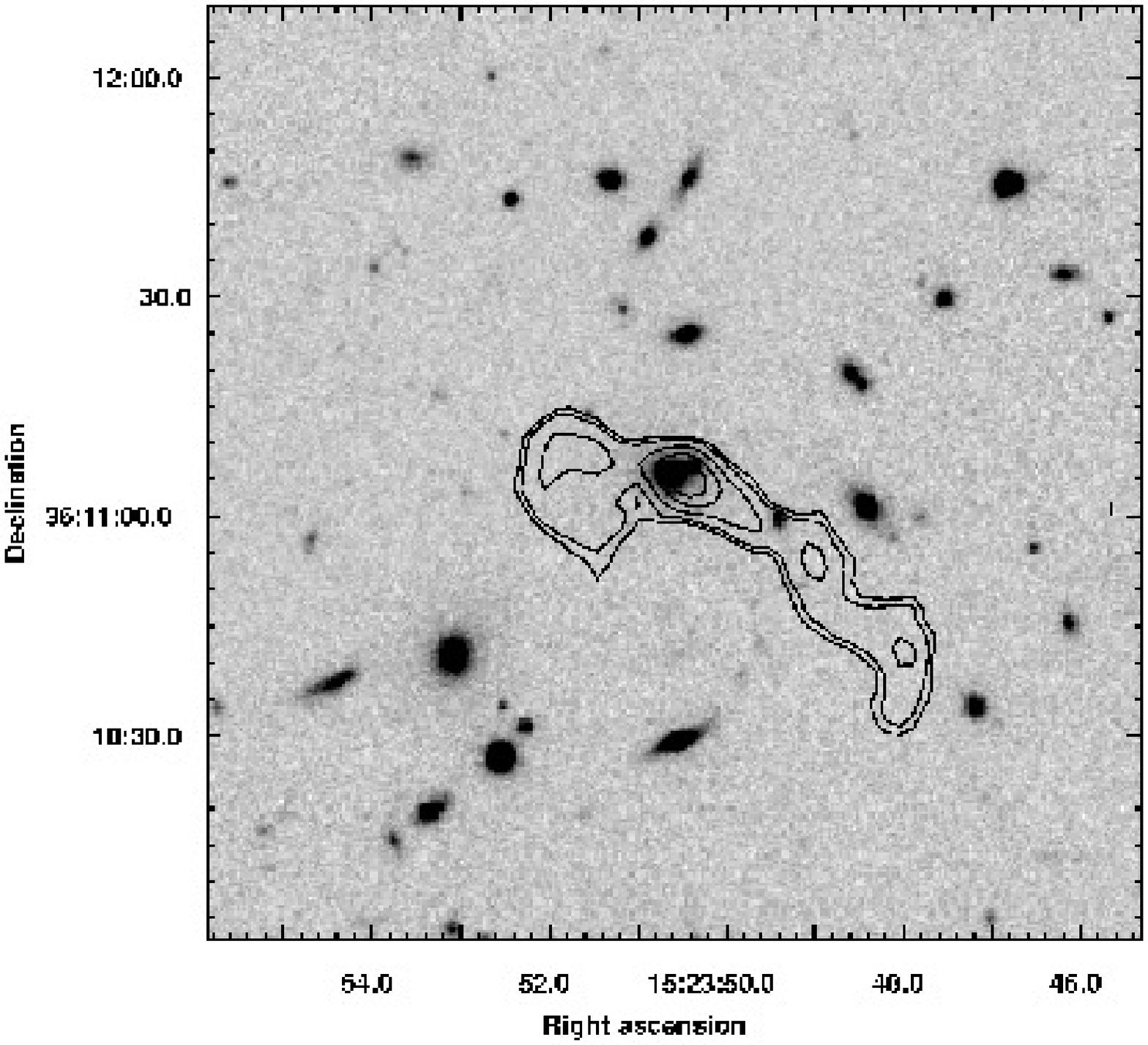}
\includegraphics[scale=.42]{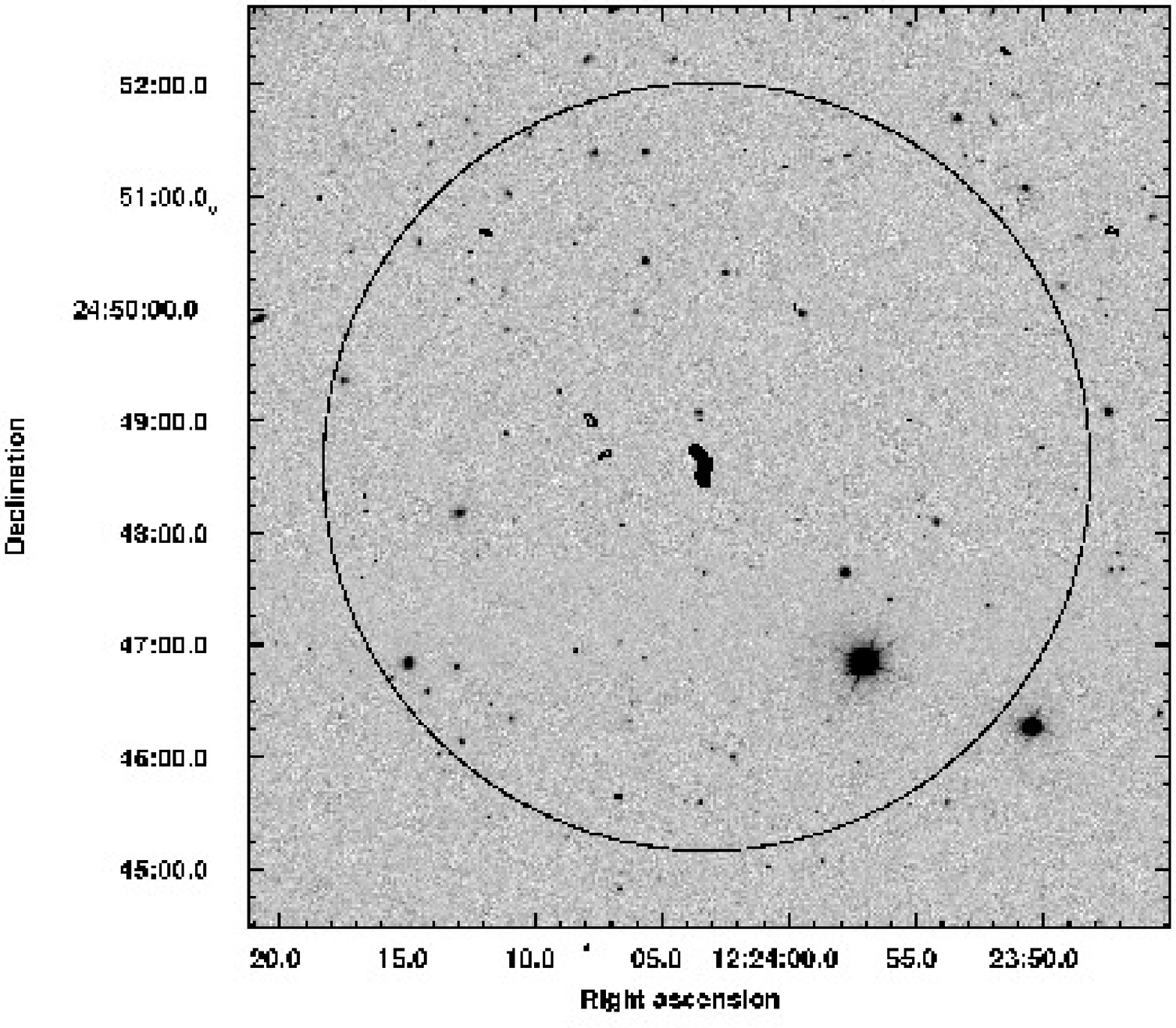}
\includegraphics[scale=.42]{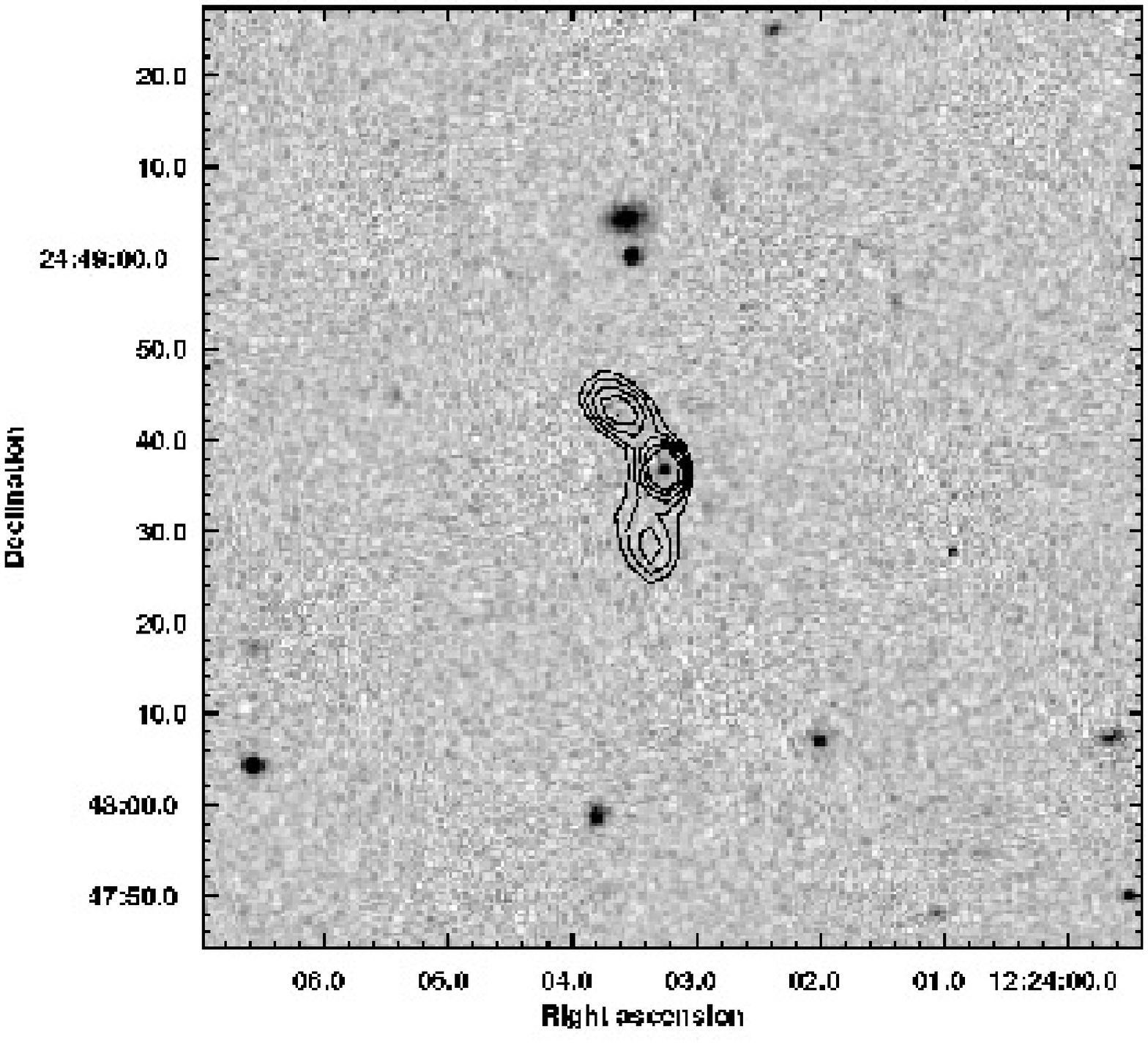}
\caption{\SDSS\/ {\it r}\/-band images with \FIRST\/ contours overlaid for both rich and poor environments.  The left-hand panels show a view of the area in which we looked for cluster members (illustrated by the circle showing an area with a radius of $1.0$ Mpc around the radio source), the right-hand panels show a zoomed in ($250\times250$ kpc) view.  The top panels show a rich cluster located at a redshift of $z=0.25$ with $N^{-19}_{1.0}=102$.  The bottom panels show a non-cluster located at a redshift of $z=0.32$ with $N^{-19}_{1.0}=-9$.  Both sources are visually selected bent sources.  The contours are square root scaled with five contours with minimum contours of 0.55 mJy and maximum contours of 1.1 mJy and 4 mJy for the upper and lower panels, respectively.} \label{fig:bentenvironments}
\end{center}
\end{figure}

\begin{figure}
\begin{center}
\capstart
\includegraphics[scale=.42]{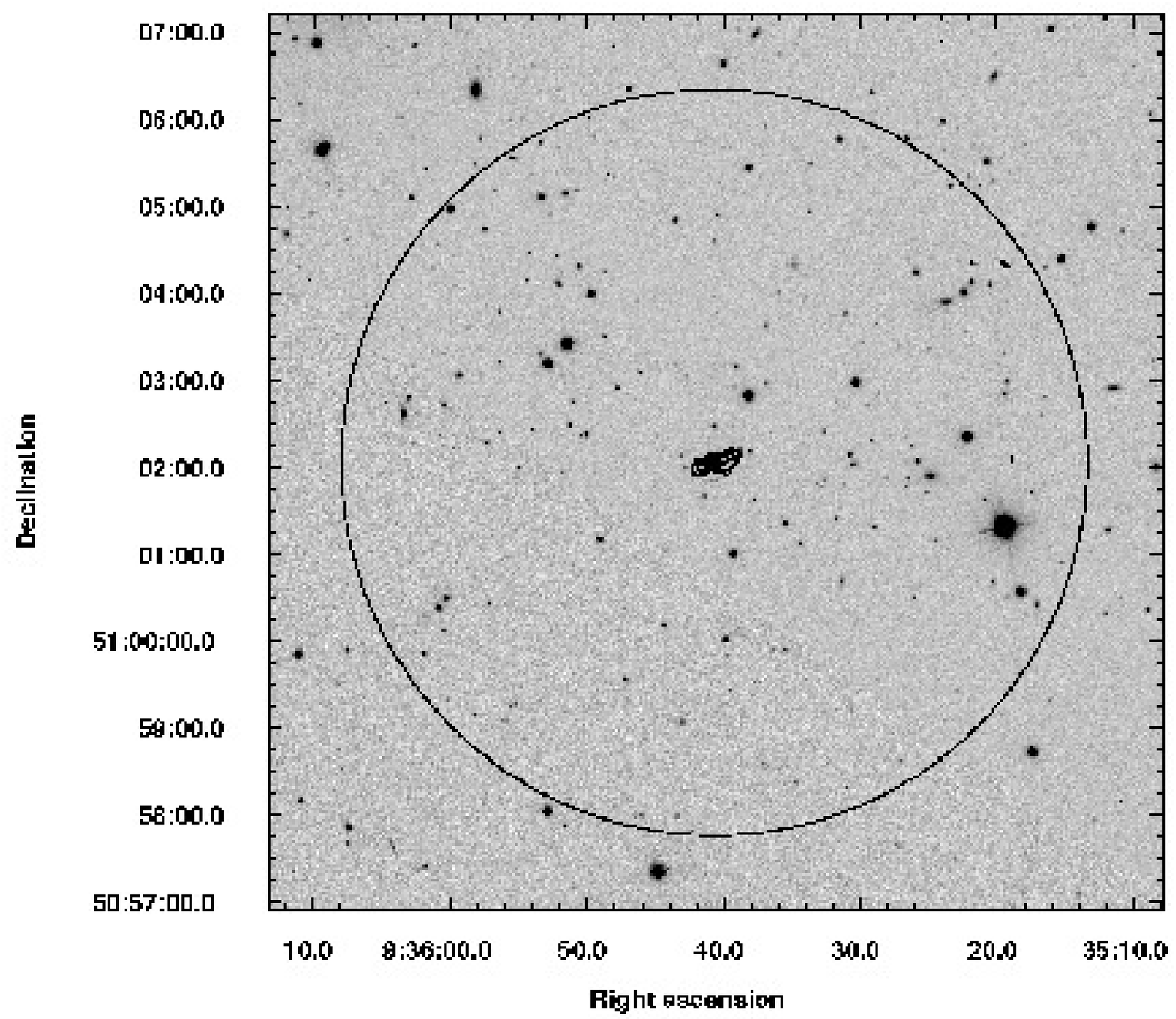}
\includegraphics[scale=.42]{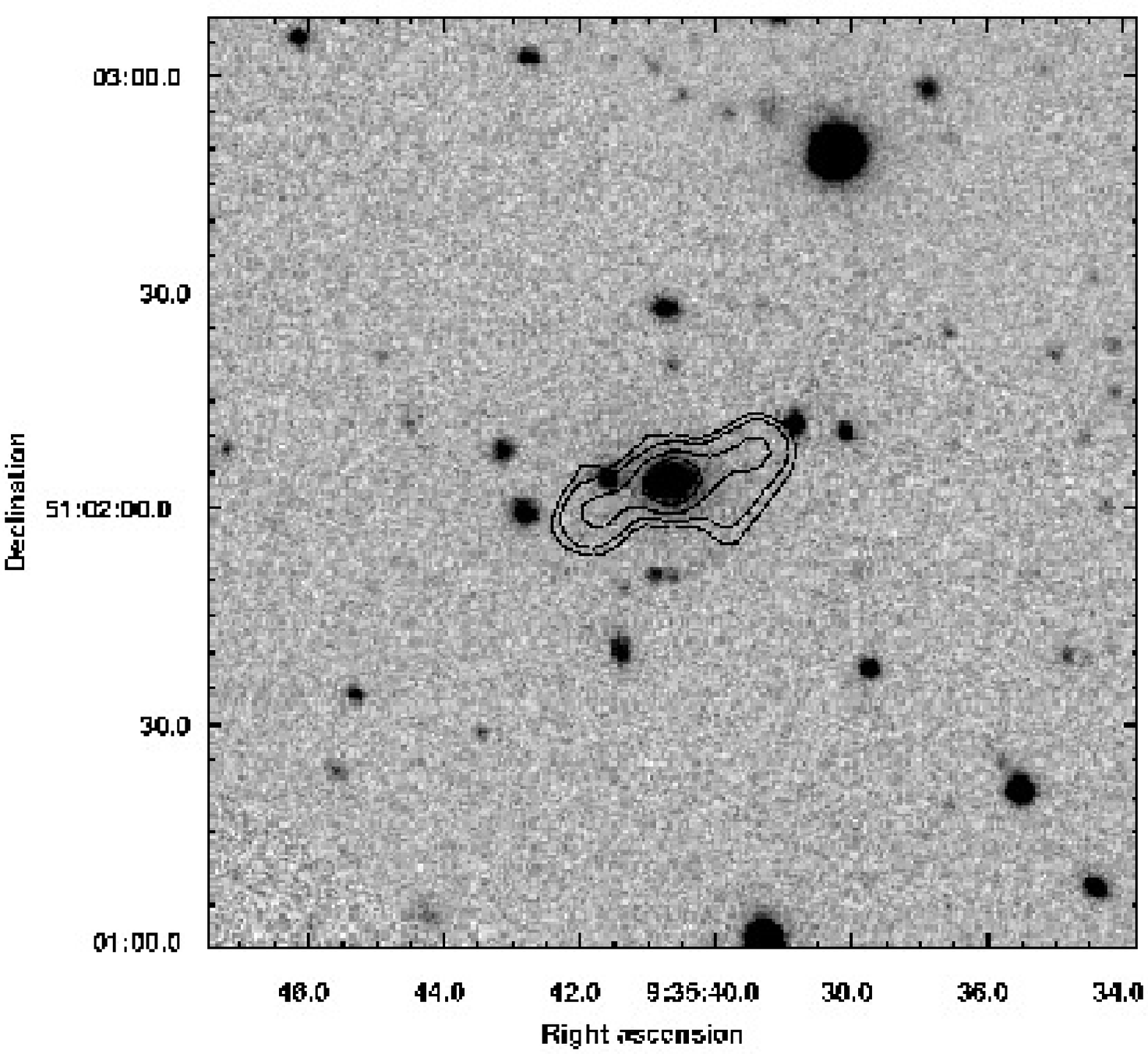}
\includegraphics[scale=.42]{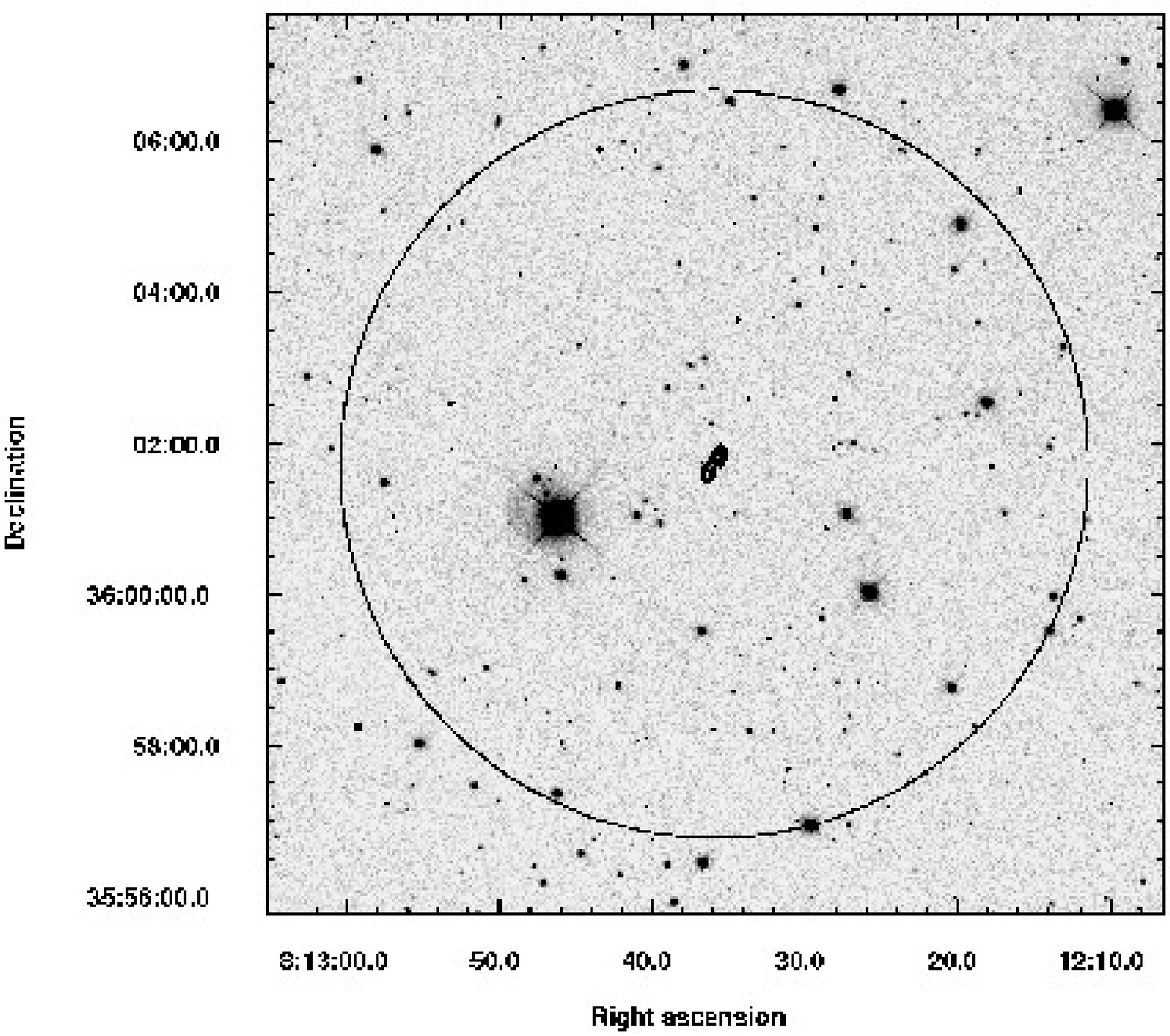}
\includegraphics[scale=.42]{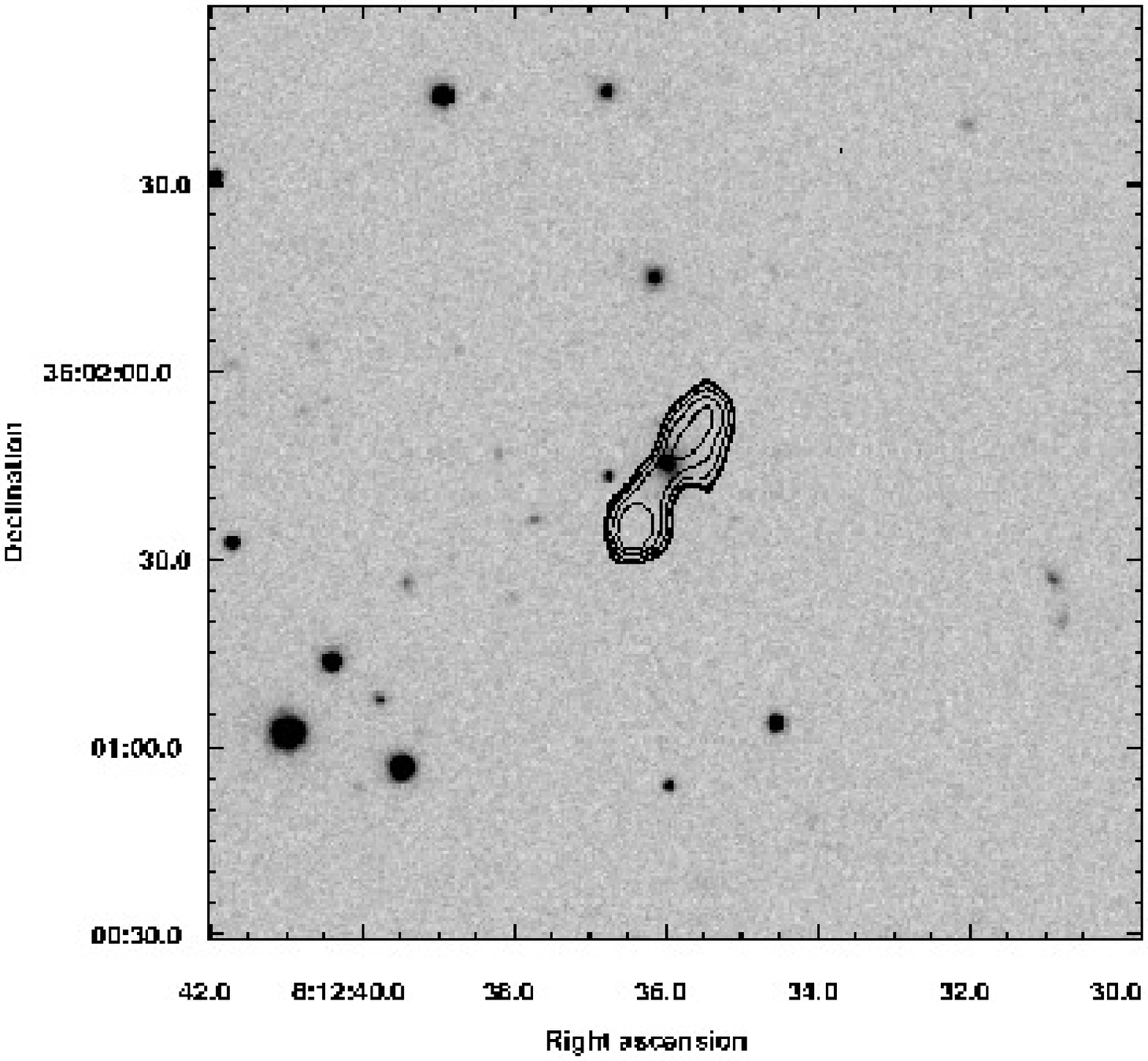}
\caption{\SDSS\/ {\it r}\/-band images with \FIRST\/ contours overlaid for both rich and poor galaxy environments around straight sources.  The left-hand panels show a view of the area in which we looked for cluster members (illustrated by the circle showing an area with a radius of $1.0$ Mpc around the radio source), the right-hand panels show a zoomed in ($250\times250$ kpc) view.  The top panels show a rich cluster located at a redshift of $z=0.25$ with $N^{-19}_{1.0}=72$.  The bottom panels show a poor environment located at a redshift of $z=0.20$ with $N^{-19}_{1.0}=-25$.  Both sources are automatically selected straight sources.  The contours are square root scaled with five contours with minimum contours of 0.55 mJy, and maximum contours of 5 mJy and 2 mJy for the upper and lower panels, respectively.} \label{fig:straightenvironments}
\end{center}
\end{figure}

Then we subtracted the normalized background galaxy counts from the galaxy counts within the inner $1.0$ Mpc to obtain a measurement for the overdensity of galaxies within $1.0$ Mpc of the radio source.  This overdensity allows us to approximate the richness of the potential cluster.  Figures~\ref{fig:bentenvironments} and ~\ref{fig:straightenvironments} show examples of bent-lobed and straight-lobed radio sources in both rich and poor environments.  Other sources look similar to these typical sources.  The sources included in Figures~\ref{fig:bentenvironments} and ~\ref{fig:straightenvironments} are located at typical redshifts ($0.2 < z < 0.3$) for our different samples.

\subsection{Schechter Correction} \label{schechter_correction} \index{Determining Cluster Richness!Schechter Correction}
The limiting magnitude of \SDSS\/ in the $r$ band is $m_r \sim 22$.  For sources at high enough redshifts, the absolute magnitude chosen ($M_r=-19$) will be fainter than the limiting magnitude of the \SDSS\/.  These fields then need to be corrected to account for these unobserved sources.  To do this, we employed a \citet{schechter1976} luminosity function with $M_r^*=-21.21$ and $\alpha=-1.05$ based on the findings of \citet{mblanton2003} specific to \SDSS\/ galaxy observations.  The luminosity function predicts the number of expected galaxies over a given magnitude range.  We calculated a k-correction for each of the sources, as in \citet{blanton2007}.  The k-correction, related to the color of the source, allows for a more accurate determination of the absolute magnitude over a given wavelength band in the source's rest-frame.  Typically, the redshift at which the apparent magnitude of an $M_r=-19$ galaxy becomes fainter than the limiting magnitude of the \SDSS\/ is $z\sim0.26$.  The correction was defined as $f_c=\phi(M_r)/\phi(M_{r,lim})$ with $M_r$ representing the faint end ($M_r=-19$) of the absolute magnitude range we wish to search in and $M_{r,lim}$ representing the absolute magnitude corresponding to the limiting magnitude of the \SDSS\/ at the redshift of the source.  We multiply the correction factor, $f_c$, by the background-adjusted galaxy counts to give a limiting-magnitude corrected richness measurement.

We limited our sample to those sources whose redshifts and colors are such that the value of the applied Schechter correction is less than $3$, to increase the reliability of our richness measurements.  For example, a hypothetical source with a Schechter correction factor of $f_c\approx10$ and a measured over-density (richness) of 3 sources has a Schechter corrected richness of 30, placing it solidly into the rich group/poor cluster category.  This may not necessarily correspond to a cluster but simply a source with a high Schechter correction factor.  Limiting our sample to those with small correction factors ($f_c \le 3$) allows us to more confidently determine if there truly is a galaxy cluster present.  The redshift this correction factor corresponds to is typically $z\sim0.50$.

\subsection{Abell Cluster Comparison} \label{abell_cluster_comparison} 
\begin{figure}
\begin{center}
\capstart
\includegraphics{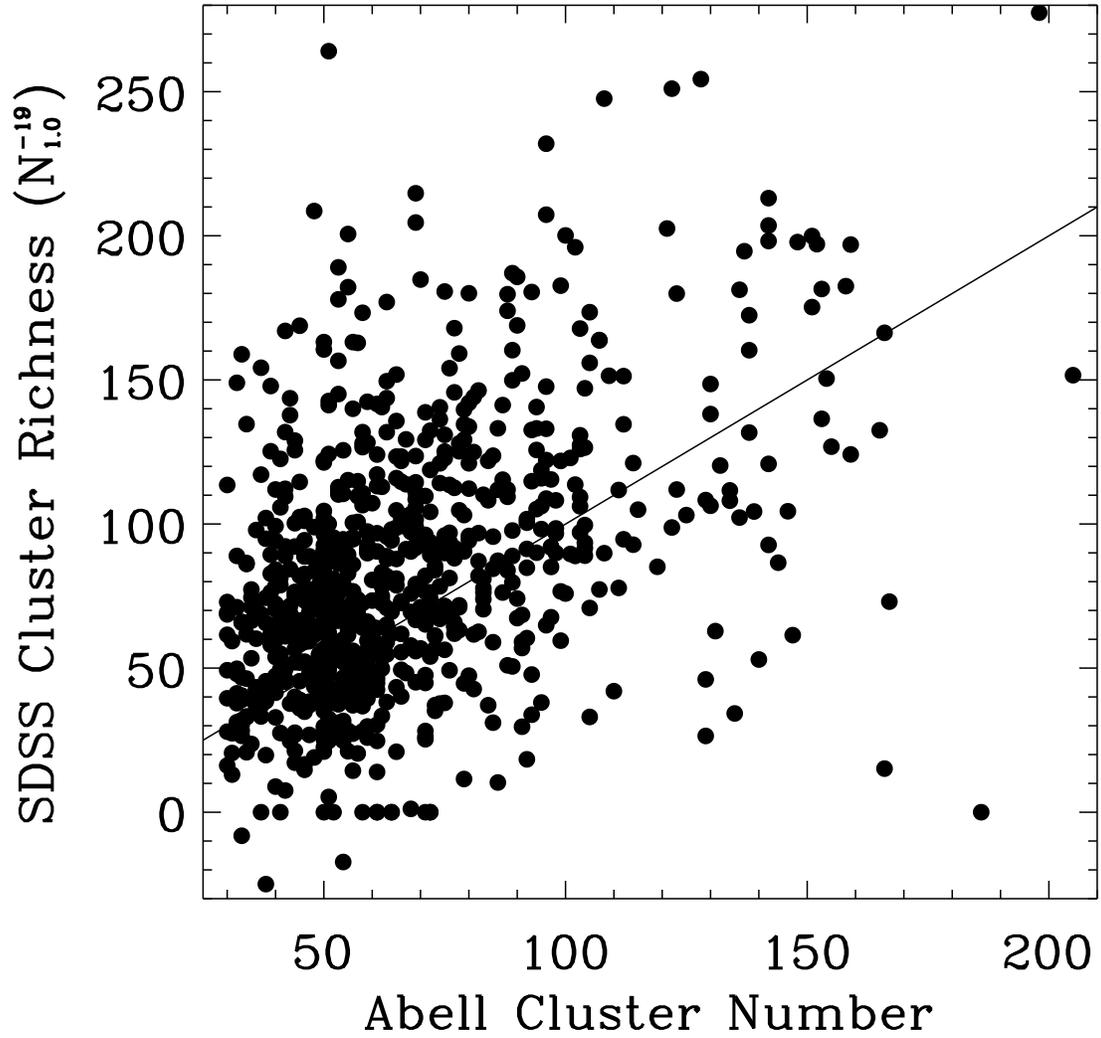}
\caption{The relationship between the number of cluster members cataloged by \citet{abell1958} compared to our $N^{-19}_{1.0}$ method.  The line represents a one-to-one correlation between cluster galaxy counts.  There is a general trend (with a Spearman correlation coefficient of $0.423$) for the richer clusters in the \citet{abell1958} catalog to also have higher $N^{-19}_{1.0}$, but there is considerable scatter.} \label{abellcluster}
\end{center}
\end{figure}

\index{Abell Cluster Comparison}
As a comparison, we used Abell clusters located within the footprint of the \SDSS\/ and measured the richness of those clusters using our $N^{-19}_{1.0}$ metric.  There are $819$ Abell clusters located within the \SDSS\/ footprint with associated redshifts in the NASA/IPACExtragalactic Database (NED) small enough as to not warrant a Schechter correction.  Based on the redshift of each Abell cluster, we calculated the apparent magnitude of an $M_r=-19$ source at the redshift of each of the $819$ clusters that matched our criteria, and searched within $1.0$ Mpc of the center of the cluster for all sources brighter than this magnitude limit.  The apparent magnitude was found using a k-correction from \citet{sarazin1982} and \citet{sandage1973}, assuming that $k(z)=2.5\log(1+z)$ and that $k(z) \approx z$ for sources where $z \leq 0.2$.  We utilized this calculation for the k-correction because there is not necessarily a source located at the center of the cluster as defined by \citet{abell1989}.  Without a source on which to base our k-correction we have applied this estimation.  Almost all of the galaxy clusters are located at redshifts of $z<0.2$ and thus have very small k-corrections.  Figure~\ref{abellcluster} shows the relationship between the number of cluster members that \citet{abell1989} identified and the cluster richness we have calculated using $N^{-19}_{1.0}$.  The relationship between these two numbers is not very clear, similar to the findings of \citet{wen2009} (their Figure~17), although a general trend is apparent (with a Spearman correlation coefficient of $0.423$), albeit with significant scatter.  \citet{wen2009} argue that projection effects (especially for sources with $z \ge 0.1$) and an inconsistent magnitude range for each potential cluster (due to the use of apparent magnitudes by \citet{abell1989} instead of absolute magnitudes) cause these inconsistencies between their cluster member counts and the cluster member counts from \citet{abell1989}.

\subsection{Nearby Clusters} \label{nearby_clusters} \index{Determining Cluster Richness!Nearby Clusters}
A problem presents itself for the lowest redshift sources.  We have set a radius of $2.7$-$3.0$ Mpc for determining the unrelated background sources for each potential cluster.  For sources where the redshift is very low, the area of sky searched corresponding to the annulus from $2.7$-$3.0$ Mpc away from the radio source becomes immense.  Searching this area becomes infeasible at low enough redshifts as the number of sources in the \SDSS\/ in an area that large easily exceeds a few million.  Thus we limit our sample to only sources with a redshift of $z\ge0.01$.  Sources below this limit are nearby and most likely well known, removing them from the samples means that we can not make generalizations about the most nearby galaxy clusters containing radio sources but does not bias the results for sources whose redshift is greater than this lower limit.  Between the four samples, there are a total of only $10$ sources with redshifts lower than this value.  Removing these sources, as well as sources with $f_c>3$, is seen as the difference between columns 8 and 9 in Table~\ref{sample_breakdown}.

\subsection{Cluster Richness Results} \label{cluster_richness_results} \index{Determining Cluster Richness!Cluster Richness Results}
Table~\ref{table_richness} lists the results of the cluster richness measurements.  Column 1 lists the sample name, and column 2 gives the fraction of each sample found in an environment with an over-density of $20$ or more sources calculated as described above.  An over-density of $20$ sources likely corresponds to a poor cluster or a group of galaxies.  For all of the columns, the number in parentheses corresponds to the total number of sources in that sample.  Columns 3 and 4 give the fraction of FR I and FR II sources (with the FR classification determined by the power-magnitude method as in \citet{ledlow1996}) in each sample, respectively, located in environments with an over-density of $20$ or more sources.  Columns 5, 6, and 7 give the fraction of each sample, in the same order as before, located in an environment with an over-density of $40$ or more sources.  This corresponds to a richer cluster.  The first line for each sample contains only those sources which have magnitudes and redshifts such that they do not need to be Schechter corrected.  The second line for each sample also includes all sources whose Schechter correction is less than $3$.

Table~\ref{table_richness_visual} gives the percentages of those sources {\it visually} classified as FR I/II located in environments of $20$ or $40$ or more galaxies, in the same manner as Table~\ref{table_richness}.  It should be noted that although there is overlap between the visual-bent and auto-bent samples, after the cuts on the samples have been made (as described in \S\ref{identifying_possible_clusters}, \S\ref{schechter_correction}, and \S\ref{nearby_clusters}), there is an overlap of only $44$ sources between the two samples (including objects with Schechter correction less than 3).  This represents $\sim16$\% of the auto-bent sample and $\sim27\%$ of the visual-bent sample, so differences in the richness distributions of these two samples are not surprising.

We see that, in general, multi-component sources are more often located in rich cluster environments than single-component sources.  We also see that bent sources are more often located in galaxy clusters than straight sources.  We see a general trend towards FR I sources being more often associated with rich clusters than FR II sources.  This is true for FR class determined visually (Table~\ref{table_richness_visual}) as well as by using the classification scheme proposed by \citet{ledlow1996} (Table~\ref{table_richness}).  An exception is for visually classified sources in the auto-bent sample.  Here, we find that FR II sources are in richer environments.

\begin{figure}
\begin{center}
\capstart
\includegraphics[scale=0.49]{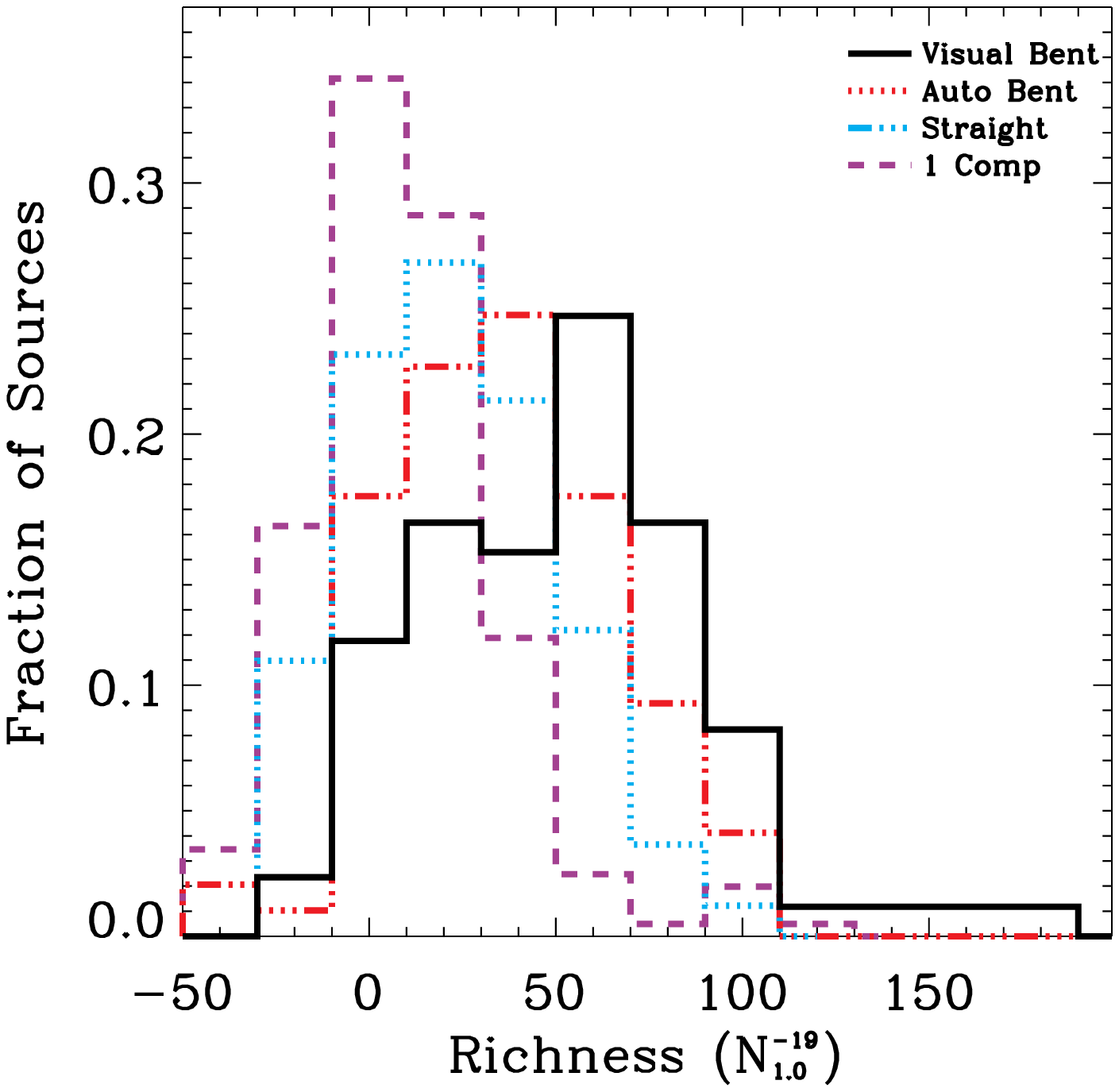}
\includegraphics[scale=0.49]{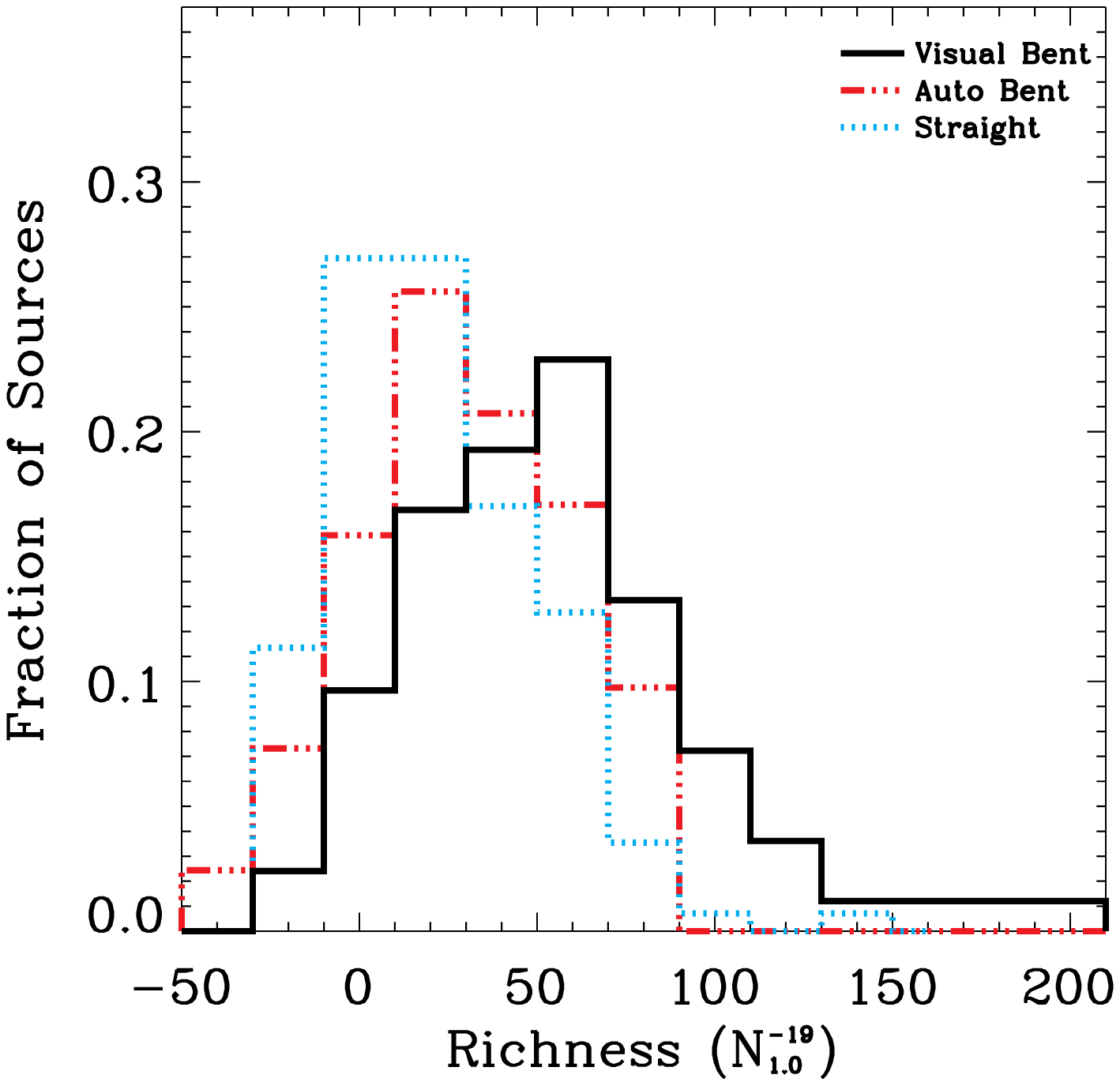}
\caption{The distribution of cluster richness associated with FR I sources (classified using the \citet{ledlow1996} criteria [left panel] and visually [right panel]) for the different samples.  The black solid line represents the visual-bent sample, the red dotted line represents the auto-bent sample, the blue dash-dotted line represents the straight sample, and the purple dashed line represents the single-component sample.  The single-component sample is not included in the right panel because they were not classified visually.  The visual-bent sample is the best sample for selecting rich clusters.  In general, bent sources are more often associated with rich cluster environments than non-bent sources and extended sources are more often associated with rich environments than single-component sources.} \label{fr1hist}
\end{center}
\end{figure}

\begin{figure}
\begin{center}
\capstart
\includegraphics[scale=0.49]{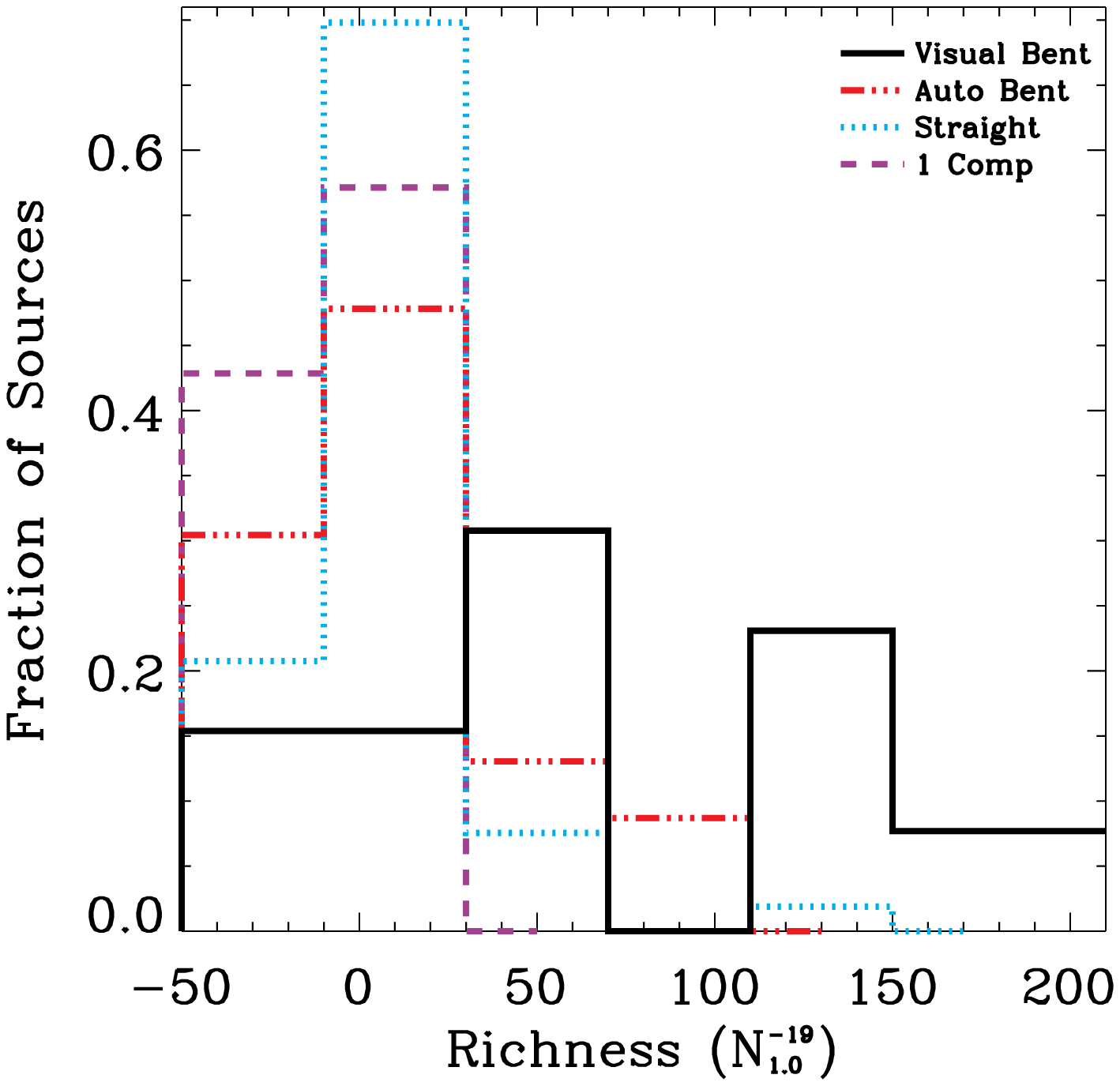}
\includegraphics[scale=0.49]{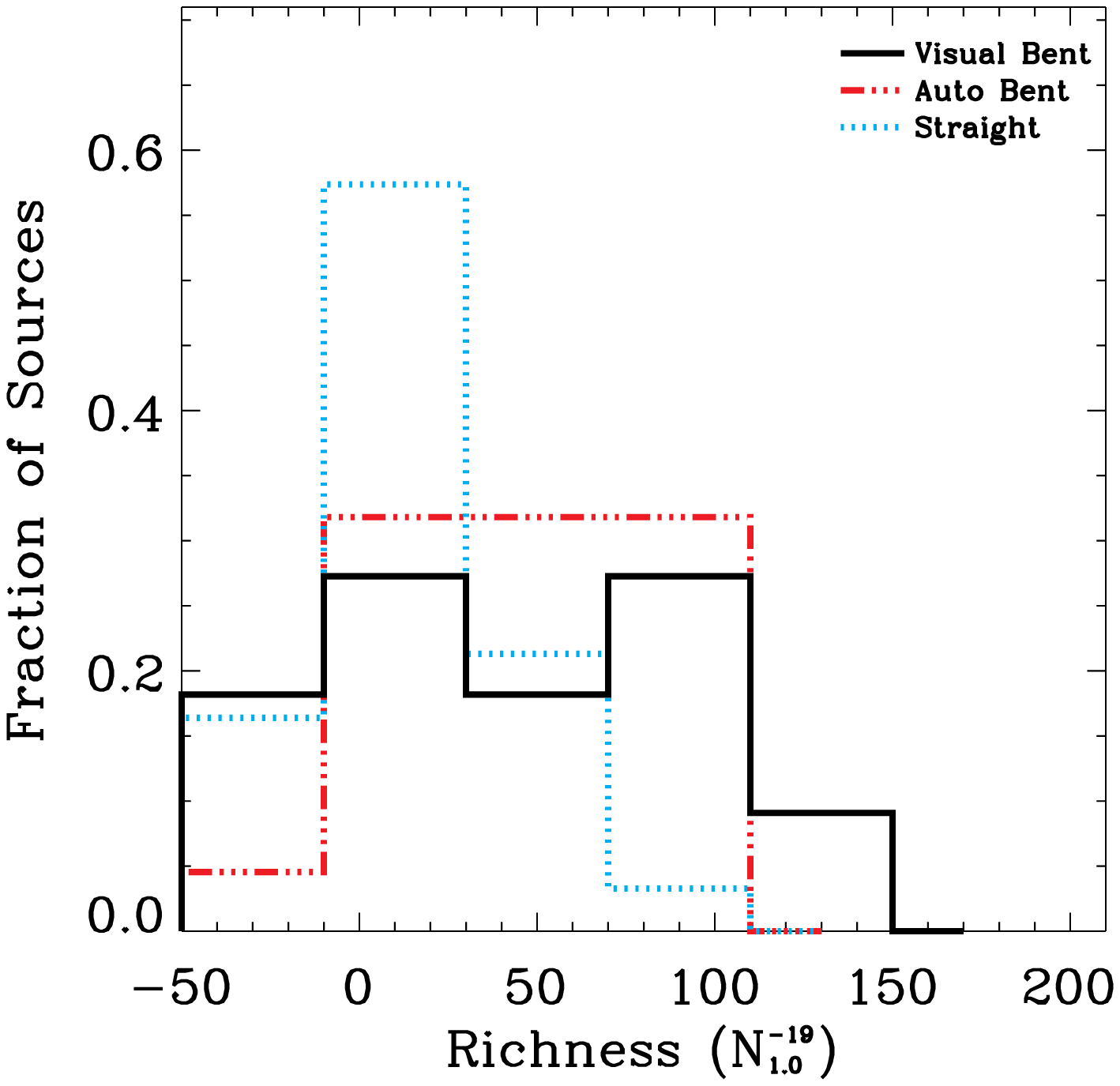}
\caption{The distribution of cluster richness associated with FR II sources (classified using the \citet{ledlow1996} criteria [left panel] and visually [right panel]) for the different samples.  The line styles and colors are the same as in Figure~\ref{fr1hist}.  The single-component sample is not included in the right panel because they were not classified visually.  The visual-bent sample is the best sample for selecting rich clusters.  In general, bent sources are more often associated with rich cluster environments than non-bent sources and extended sources are more often associated with rich environments than single-component sources.} \label{fr2hist}
\end{center}
\end{figure}

Figures~\ref{fr1hist} and \ref{fr2hist} show the fraction of sources with varying richness for the different samples.  Figure~\ref{fr1hist} shows the distribution for the FR I sources.  The left panel shows the distribution of richness for the sources that have been classified as FR I through the classification scheme of \citet{ledlow1996}.  The right panel shows that same distribution, except for sources that have been visually classified as FR I sources.  Figure~\ref{fr2hist} shows the distributions for the FR II sources.  The left panel shows the distribution of cluster richnesses for sources that have been classified as FR II through the \citet{ledlow1996} method, and the right panel shows those sources classified as FR II visually.  These distributions are only for the subset of sources needing no Schechter correction (see \S\ref{schechter_correction}).  We examined the likelihood that the distribution of cluster richness for the different samples could come from the same parent population using the K-S test.  We found that the probability for the visual-bent sample (with an average richness of $57\pm7$ sources) to come from the same parent population as the straight sample (with an average richness of $27\pm6$ sources) is $3.14\times10^{-10}\%$ and for the visual-bent sample compared to the single-component sample (with an average richness of $11\pm6$ sources) that probability is $1.50\times10^{-24}\%$.  The probability between the auto-bent sample (with an average richness of $45\pm7$ sources) and the straight sample is $1.54\times10^{-3}\%$ and for the auto-bent sample compared to the single-component sample that probability is $2.46\times10^{-16}\%$.  If we group both the visual- and auto-bent samples together and compare them to the straight sample, the likelihood that the distribution in cluster richness comes from the same parent population is $1.55\times10^{-9}\%$ and for the single-component sample compared to the bent samples, the likelihood is $1.22\times10^{-28}\%$.  The likelihood of the distribution of cluster richness of FR I sources coming from the same parent population as FR II sources (using all four samples and the \citet{ledlow1996} criteria for determining FR type) is $1.22\times10^{-3}\%$.  That likelihood rises to $1.47\%$ when the FR type is determined visually.  Therefore, the bent-double sources are more often located in cluster environments than the straight sources and the single-component sources.  Further, FR I sources are more likely to be located in cluster environments than FR II sources, although when FR type is determined visually the distribution in cluster richness is more similar.

Table~\ref{table_sources} lists sources from the visual-bent, auto-bent, and straight samples with $N^{-19}_{1.0}$ values of $40$ or higher.  This table is available in its entirety in machine-readable and Virtual Observatory (VO) forms in the online journal. A portion is shown here for guidance regarding its form and content.  Column 1 lists the source name and column 2 lists the sample (V for visual-bent, A for auto-bent, and S for straight) that the source belongs to.  Columns 3 and 4 are the RA and Dec of the source in J2000 coordinates.  Column 5 is the \SDSS\/ de-reddened $r$-band magnitude of the associated optical source and column 6 is the $r$-band absolute magnitude.  Column 7 lists the total $1440$ MHz flux of the radio source in mJy.  Column 8 gives the opening angle of the radio source.  The total power of the radio source, in W Hz$^{-1}$, is given in column 9.  The FR types, determined using the criteria from \citet{ledlow1996} and visually, are in columns 10 and 11, respectively.  Column 12 lists the redshift of the source.  The $N^{-19}_{1.0}$ value is given in column 13 along with the Schechter correction factor, $f_c$, in column 14.  Finally, column 15 lists the Abell cluster (if there is one) located within 3 Mpc (projected) of the radio source.  Note that some radio sources appear twice, if they appear in both the visual-bent and auto-bent samples.  In these fifteen overlap cases we see that the opening angle is often different between the two samples.  This is due to the different methods for determining the center of the radio source.  For the visual-bent sample, the source center was defined as the location of the corresponding optical source if there was one, and the location of the radio core if an optical counterpart was lacking \citep{blanton2000a}.  This is different than the auto-bent sample where the source center was always defined as the location of the radio component opposite the longest side when making a triangle of the three radio sources \citep{proctor2006}.  These differences in the central position account for the differences in opening angle measurement between the visual-bent and auto-bent samples.

An intriguing question remains for the sources with low $N^{-19}_{1.0}$ measurements (i.e. $N^{-19}_{1.0} < 5$).  Sources falling into this category in the visual- and auto-bent samples require some type of explanation as to how the radio lobes were bent.  If they are not in a cluster (at least as optically identified) what could be responsible for the observed bending of the lobes?  One possible explanation is that these radio sources reside in fossil groups \citep{ponman1994,jones2000}.  Fossil groups are thought to be compact galaxy groups within which the member galaxies have merged into one, or few, galaxies near the center-of-mass of the cluster.  These fossil groups can be detected, via their ICM, in the X-ray.  \citet{donghia2005} found that in their simulations $\sim33\%$ of groups actually exist as fossil groups.  Observationally, \citet{vikhlinin1999} and \citet{jones2003} have found that percentage to be between $10\%$-$20\%$.  Further X-ray observations of the sources in our samples with low $N^{-19}_{1.0}$, specifically in the visual- and auto-bent samples, could help to shed light on whether these sources are in fossil groups.

\section{Physical Properties of Radio Selected Galaxy Clusters} \label{physical_properties_of_radio_selected_galaxy_clusters} 
\begin{figure}
\begin{center}
\capstart
\includegraphics{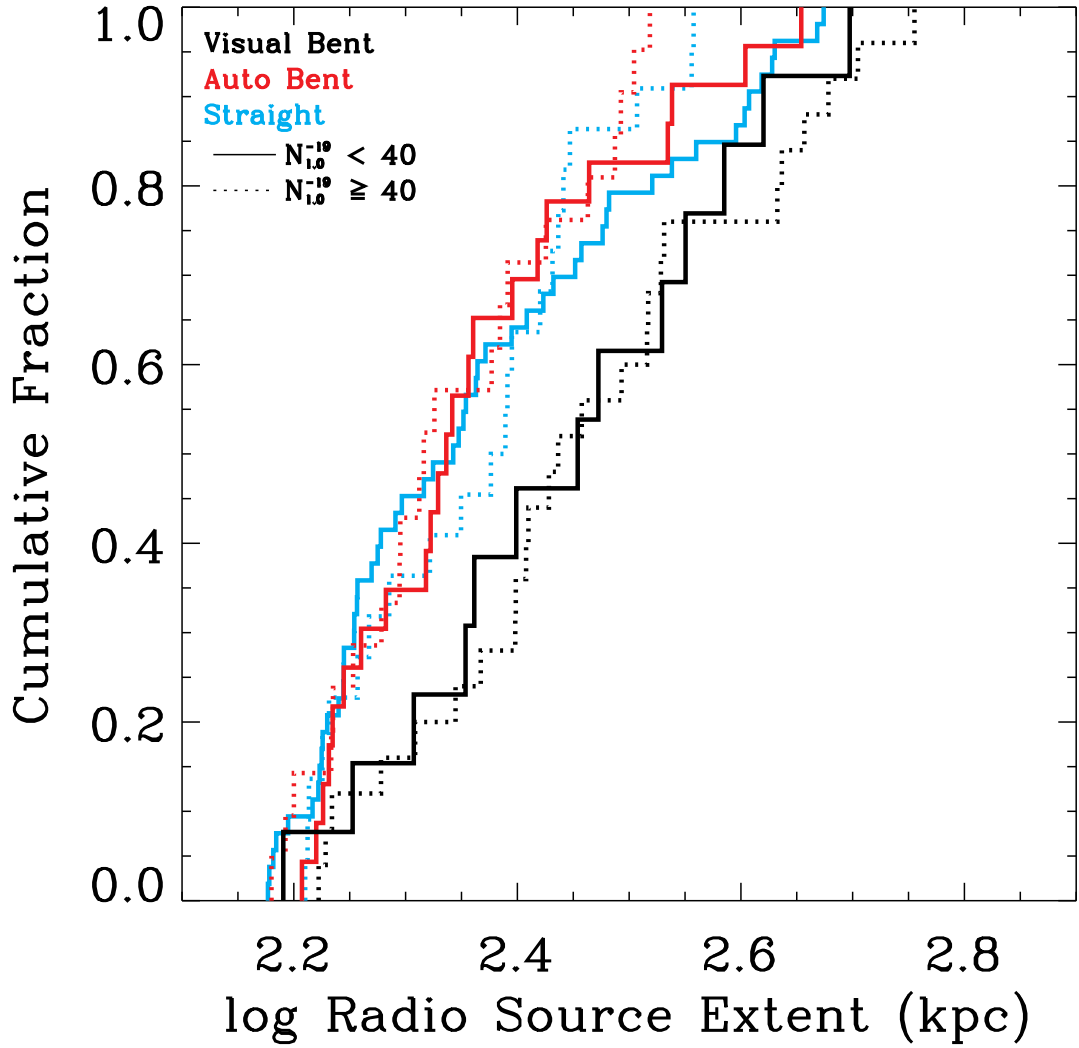}
\caption{The relationship between radio source physical extent and cluster richness.  The line colors and styles are the same as in Figure~\ref{multicumufrac}.  All of the sources in each sample that are contained within the small box from Figure~\ref{zvsize} and have redshifts and magnitudes such that there is no need for a Schechter correction (see \S\ref{schechter_correction}) are included.  There appears to be no difference in the size distributions of sources in clusters versus those not in clusters.} \label{clustercumufrac}
\end{center}
\end{figure}

\index{Physical Properties of Radio Selected Galaxy Clusters}
We have assembled a large collection of radio-selected potential galaxy clusters with known properties such as optical colors, redshift, richness, radio source extent, radio power, bending angle of the radio source and radio morphology.  In Figure~\ref{clustercumufrac}, we plot the radio source size distribution for the different samples, separating the sources located in rich clusters ($N^{-19}_{1.0}\geq40$, short dashed line) from those in poor clusters ($N^{-19}_{1.0} < 40$, solid line).  Only sources contained within the small box in Figure~\ref{zvsize} and that have colors and redshifts such that they need no Schechter correction (see \S\ref{schechter_correction}) are included in the plot.  This includes $13$, $23$ and $53$ sources with $N^{-19}_{1.0} < 40$ and $25$, $21$ and $22$  sources with $N^{-19}_{1.0} \ge 40$ in the visual-bent, auto-bent and straight samples, respectively.

\begin{figure}
\begin{center}
\capstart
\includegraphics[scale=0.49]{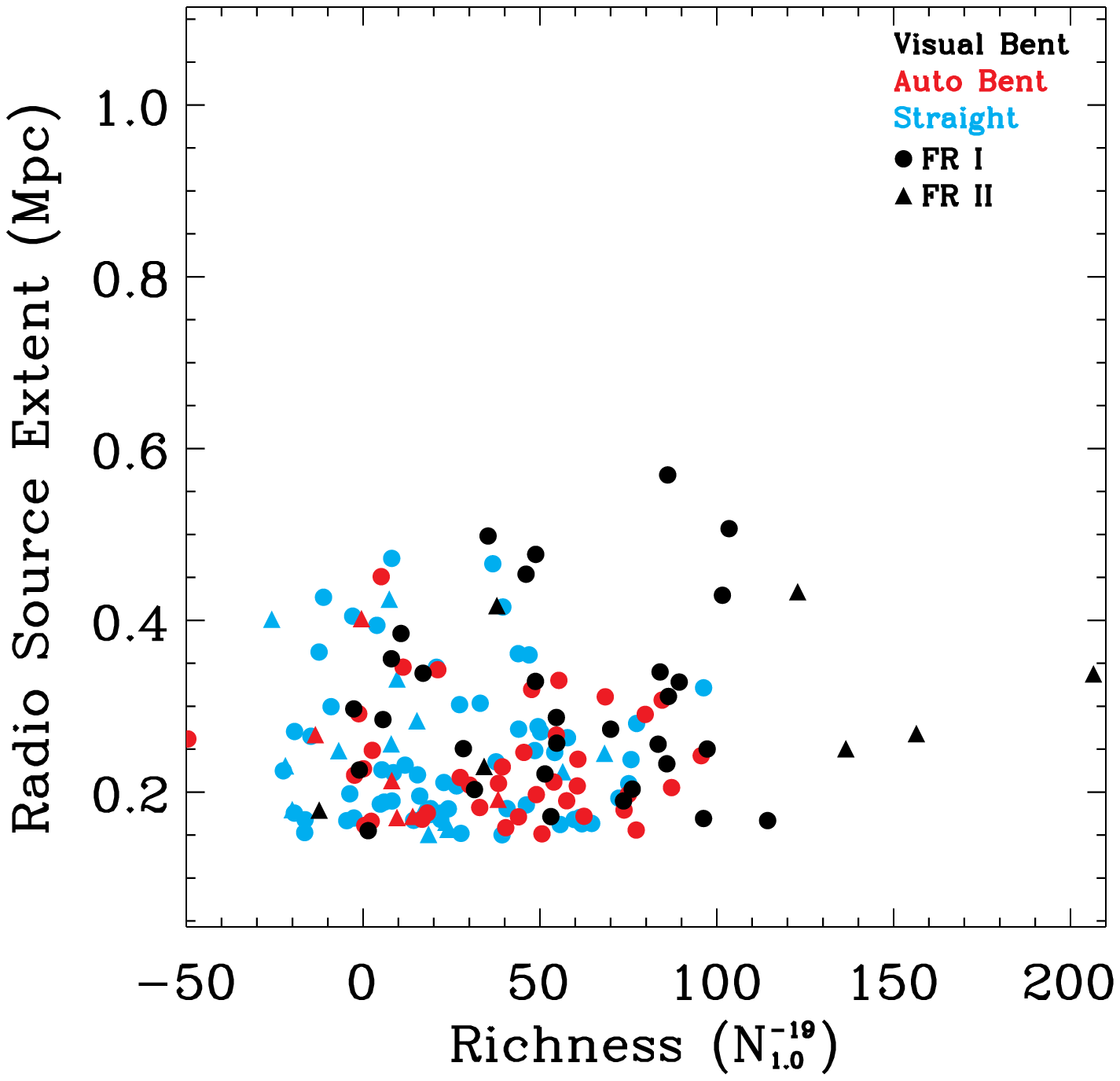}
\includegraphics[scale=0.49]{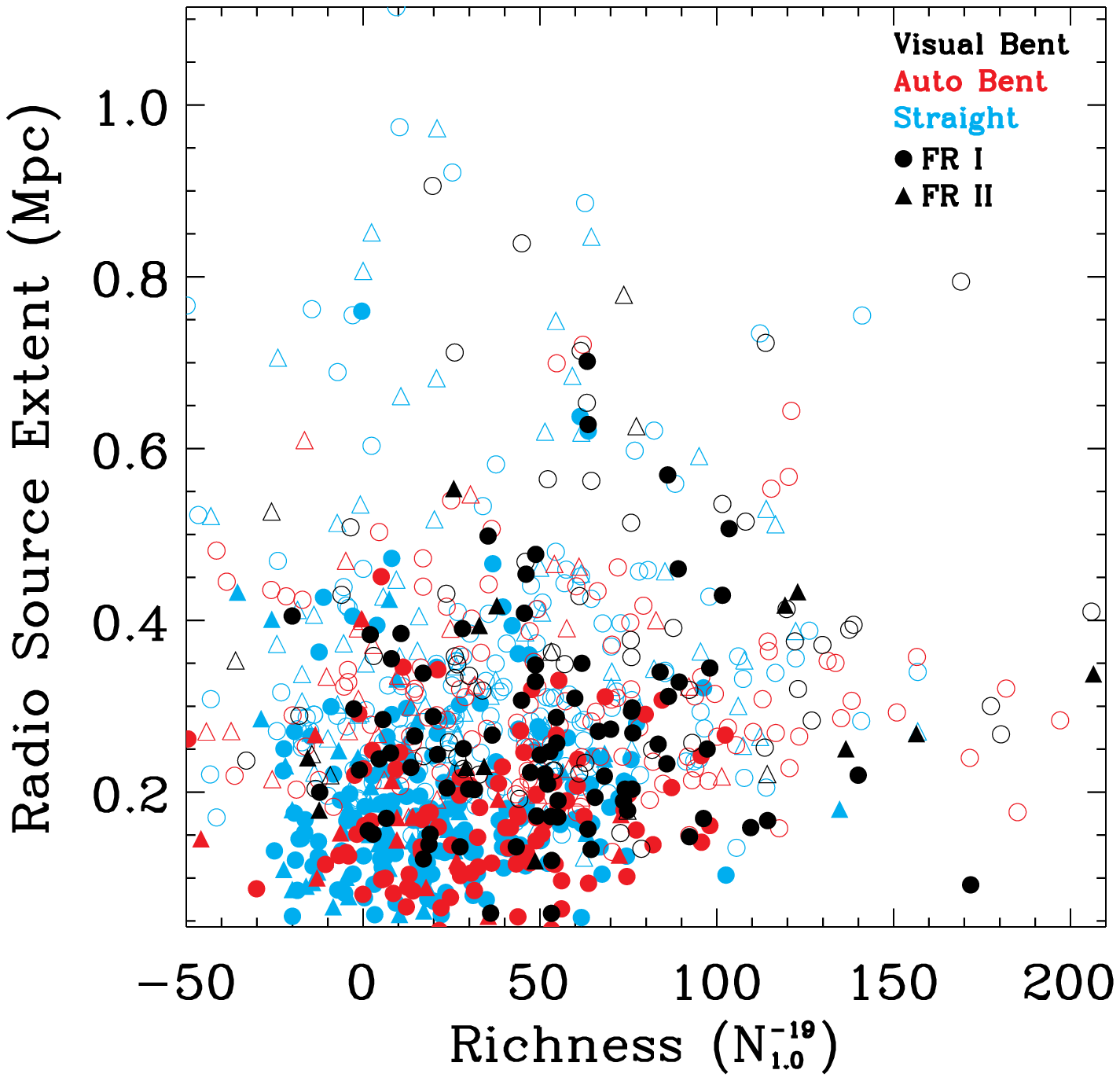}
\caption{The relationship between the richness of a cluster and the physical size of the radio source associated with that cluster.  The left-hand panel plots only those sources in the small box from Figure~\ref{zvsize} and need no Schechter correction (see \S\ref{schechter_correction}).  The right-hand panel shows all sources, including those with Schechter corrections and located outside of the small box.  The symbols are the same as in Figure~\ref{zvcolor}.  The sources with Schechter corrections in the range of $1<f_c\le 3$ are identified as open symbols.  There does not appear to be a relationship between cluster richness and radio source size.} \label{richvsize}
\end{center}
\end{figure}

We see no obvious difference in radio source extent for sources located in clusters versus those not in clusters.  We performed a K-S test on the size distributions of sources in clusters versus those not in clusters and found that the FR I and II sources (determined using the \citet{ledlow1996} criteria) in the visual-bent sample had a $96.2\%$ probability of coming from the same parent population, those in the auto-bent sample had a $85.1\%$ probability of coming from the same parent population, and those in the straight sample had a $72.0\%$ probability of coming from the same parent population.  This confirms what a visual inspection of Figure~\ref{clustercumufrac} implies, there is no difference in physical extent for sources in clusters versus those not in clusters.  Plotting radio source extent against $N^{-19}_{1.0}$ (Figure~\ref{richvsize}) further helps to illustrate this.  We see no obvious trend of radio source extent as a function of the $N^{-19}_{1.0}$ value.

\begin{figure}
\begin{center}
\capstart
\includegraphics[scale=0.49]{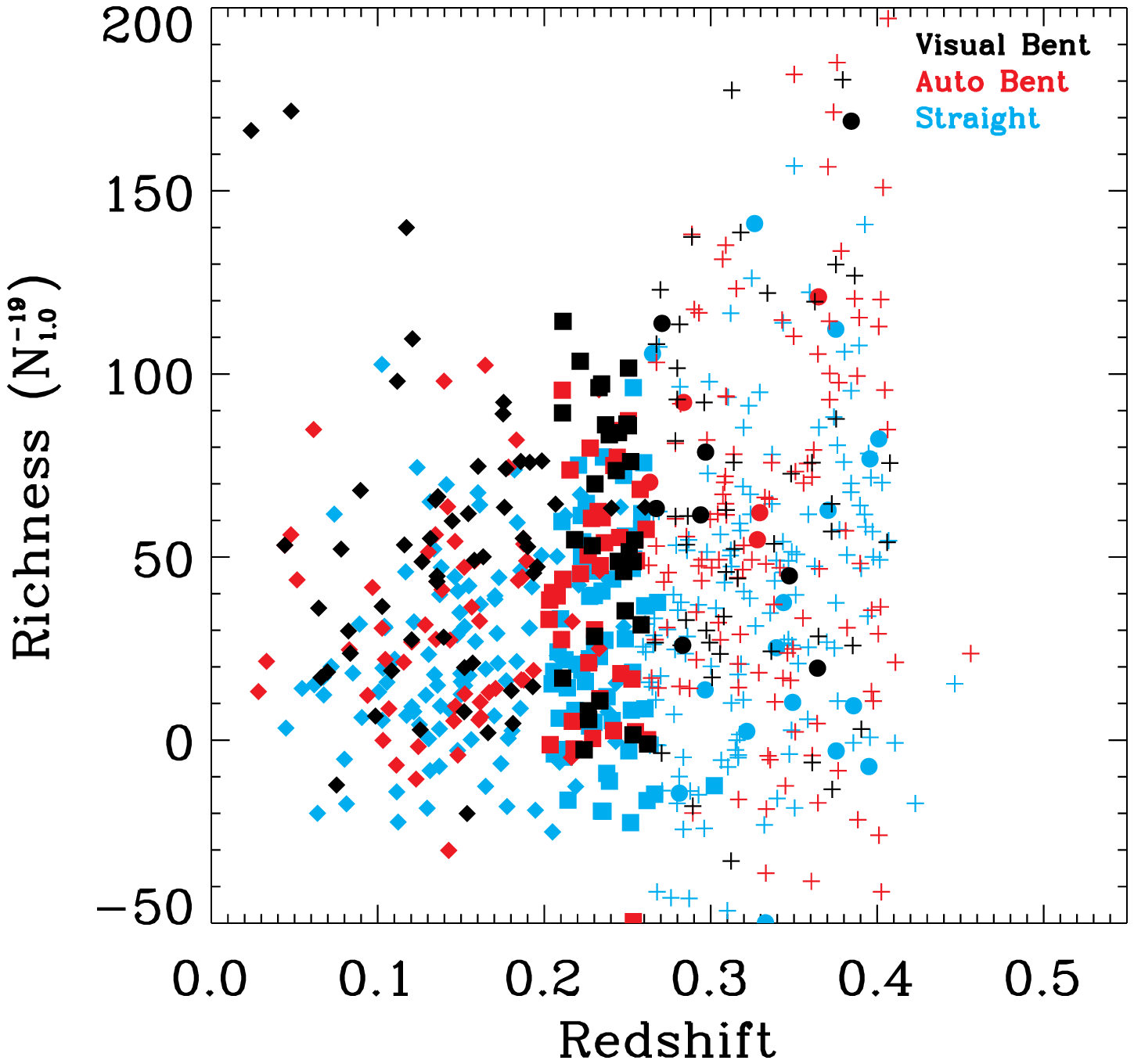}
\includegraphics[scale=0.49]{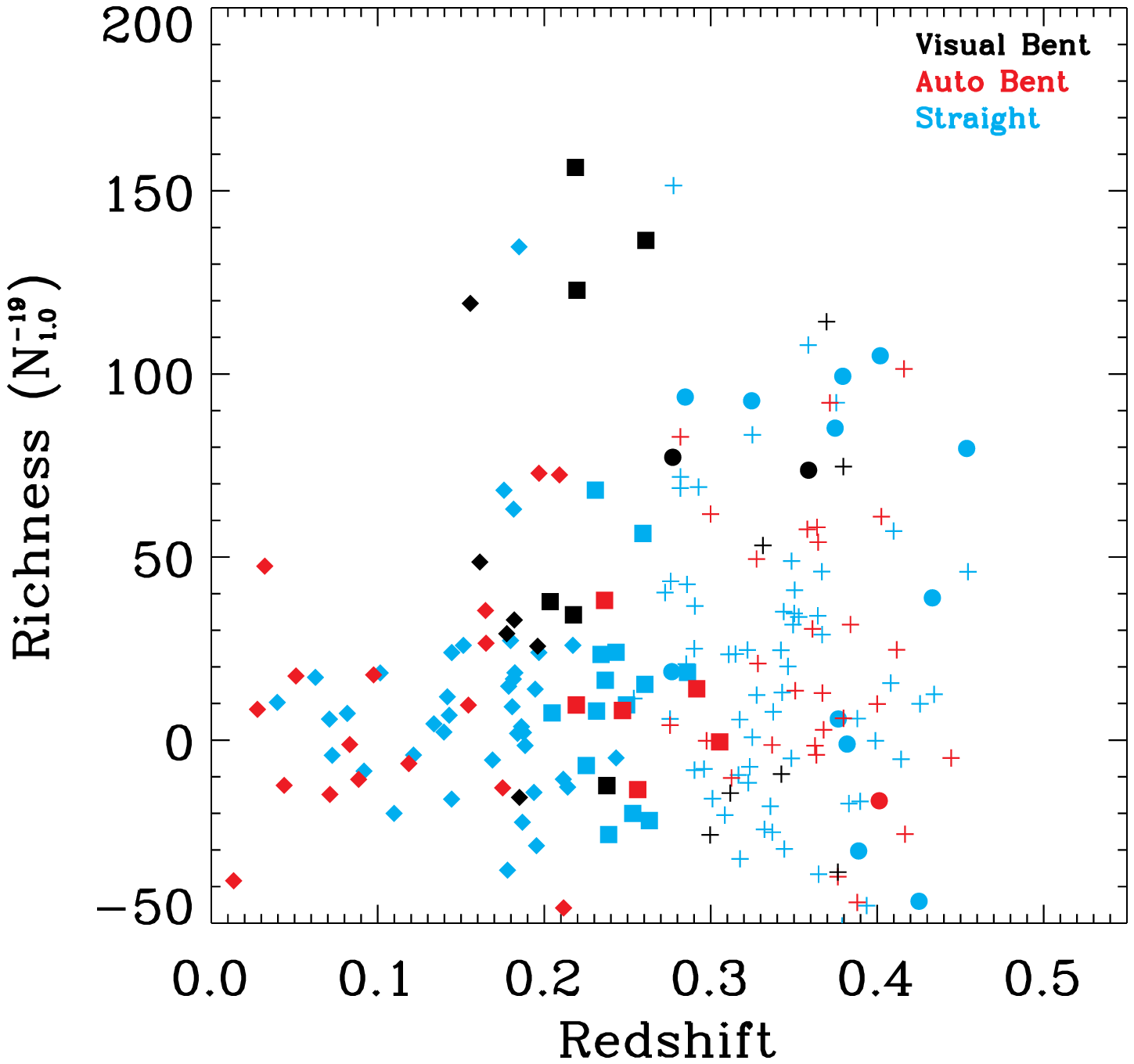}
\caption{Redshift vs. richness.  We have plotted the FR I (left panel) and FR II (right panel) sources separately, using the \citet{ledlow1996} criteria to determine morphology. The filled-diamonds represent sources that are not contained within the small box from Figure~\ref{zvsize} and have no Schechter correction (see \S\ref{schechter_correction}).  The filled-squares represent sources contained within the small box and have no Schechter correction.  The pluses are sources that are contained within the small box but have $1 < f_c \le 3$.  Finally, the filled-circles represent sources that are not within the small box and have $1 < f_c \le 3$.  Black symbols represent the visual-bent sample, red symbols represent the auto-bent sample, and blue symbols represent the straight sample.  Our results show a possible trend of the richest clusters being associated with sources at higher-redshifts.} \label{richvz}
\end{center}
\end{figure}

\begin{figure}
\begin{center}
\capstart
\includegraphics[scale=0.49]{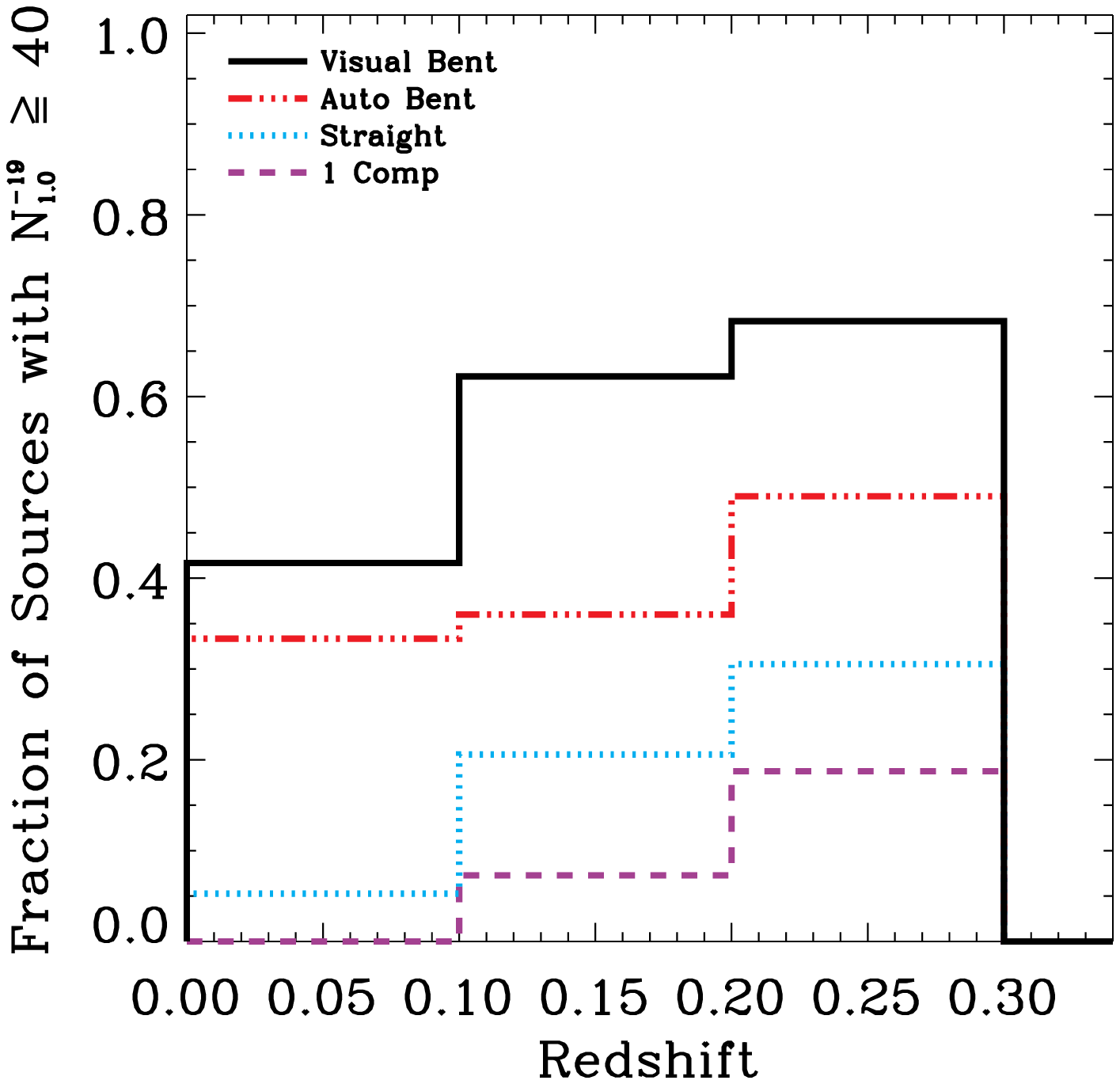}
\includegraphics[scale=0.49]{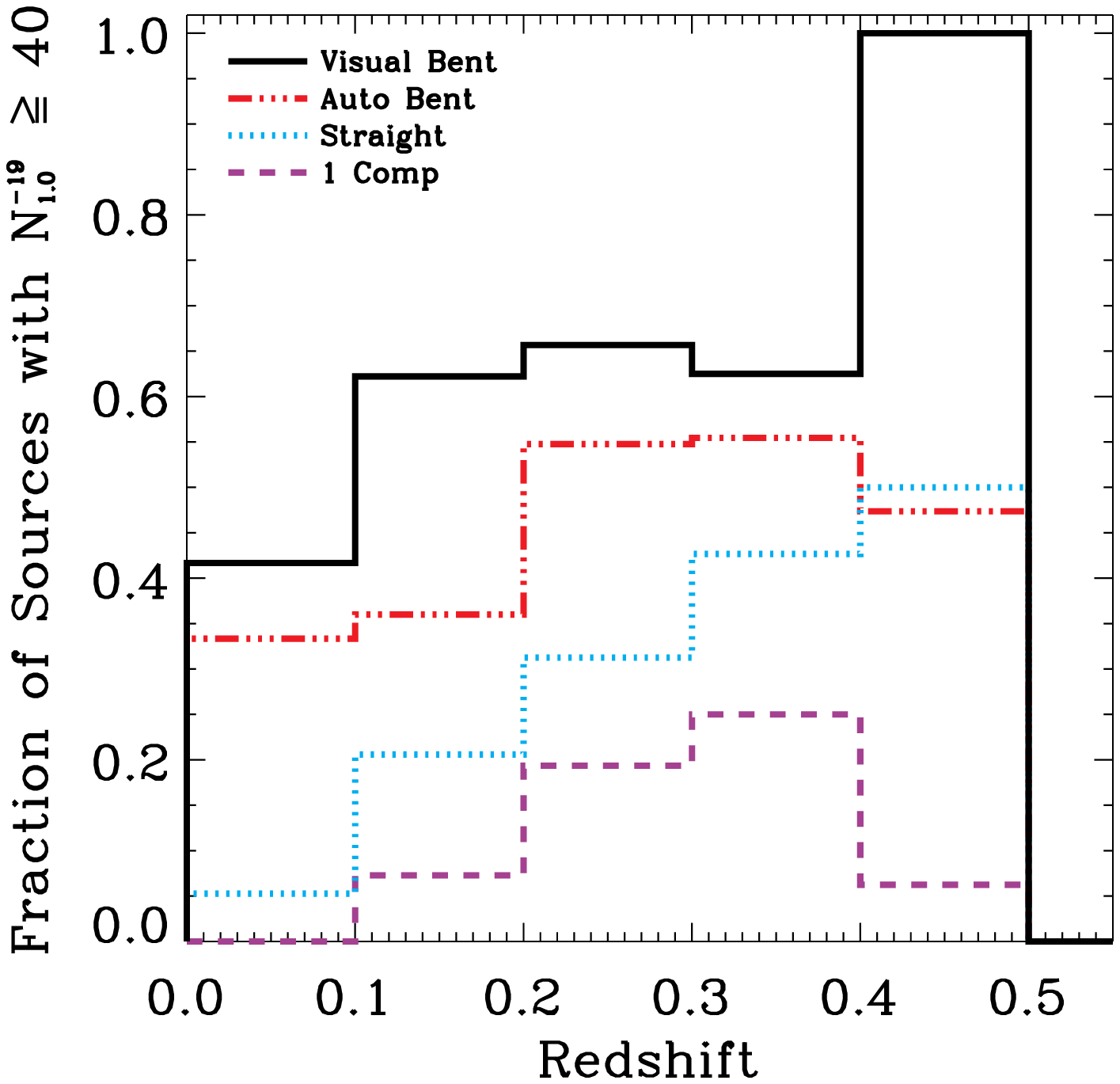}
\caption{The fraction of sources in each redshift bin that are associated with clusters (those having $N^{-19}_{1.0}\ge40$) compared to the total number of radio sources with optical counterparts in each redshift bin.  The line styles and colors are the same as in Figure~\ref{fr1hist}.  The left-hand panel shows the results when only taking into account those sources that have no Schechter correction (see \S\ref{schechter_correction}).  The right-hand panel includes all sources with Schechter corrections up to our limit ($f_c \le 3$) as described earlier.} \label{clustzhist}
\end{center}
\end{figure}

Previous authors (i.e. \citet{allington-smith1993,zirbel1997}) found that in general, nearby FR II sources are more often located in poor environments, while at higher redshifts, they tend to be found in richer environments.  Figure~\ref{richvz} shows the relationship between the redshift and the cluster richness for the FR I (left-panel) and FR II (right-panel) sources in our samples.  Our samples seem to show a similar trend between the richness of the cluster and the redshift of the cluster for both FR I and FR II sources.  Examining the Spearman correlation between the redshift and  $N^{-19}_{1.0}$ for FR I sources (from \citet{ledlow1996}) there is a slight positive correlation of $0.21$, implying that there is a trend between the richness and redshift of a cluster containing an FR I source.  This correlation drops to $0.049$ for FR II sources (again classified as in \citet{ledlow1996}), implying very little to no correlation between the richness and redshift of a cluster containing an FR II source.  If we ignore the FR type we find a positive correlation of $0.15$.  Therefore, searching around radio sources, we find richer environments at higher redshifts (see Figure~\ref{clustzhist}).  The fractions are highest for bent-lobed sources, as seen in Tables~\ref{table_richness} and~\ref{table_richness_visual}.  This could follow from the results of \citet{martini2009}.  If clusters contain a higher number of AGN at higher redshifts ($z\sim1$) then we might expect to find richer environments around AGN when searching at high-redshift as compared to low-redshift.

\begin{figure}
\begin{center}
\capstart
\includegraphics[scale=0.49]{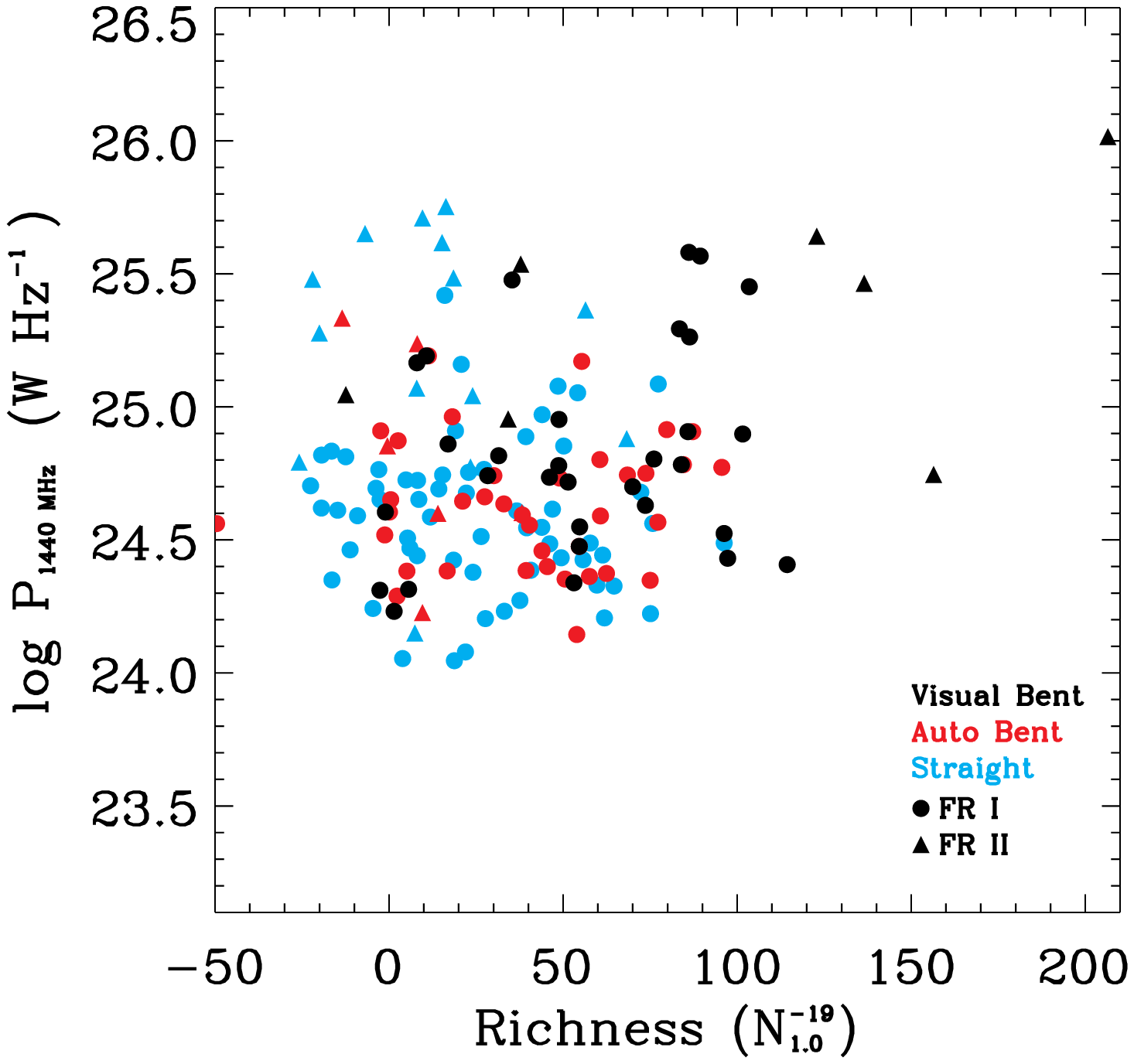}
\includegraphics[scale=0.49]{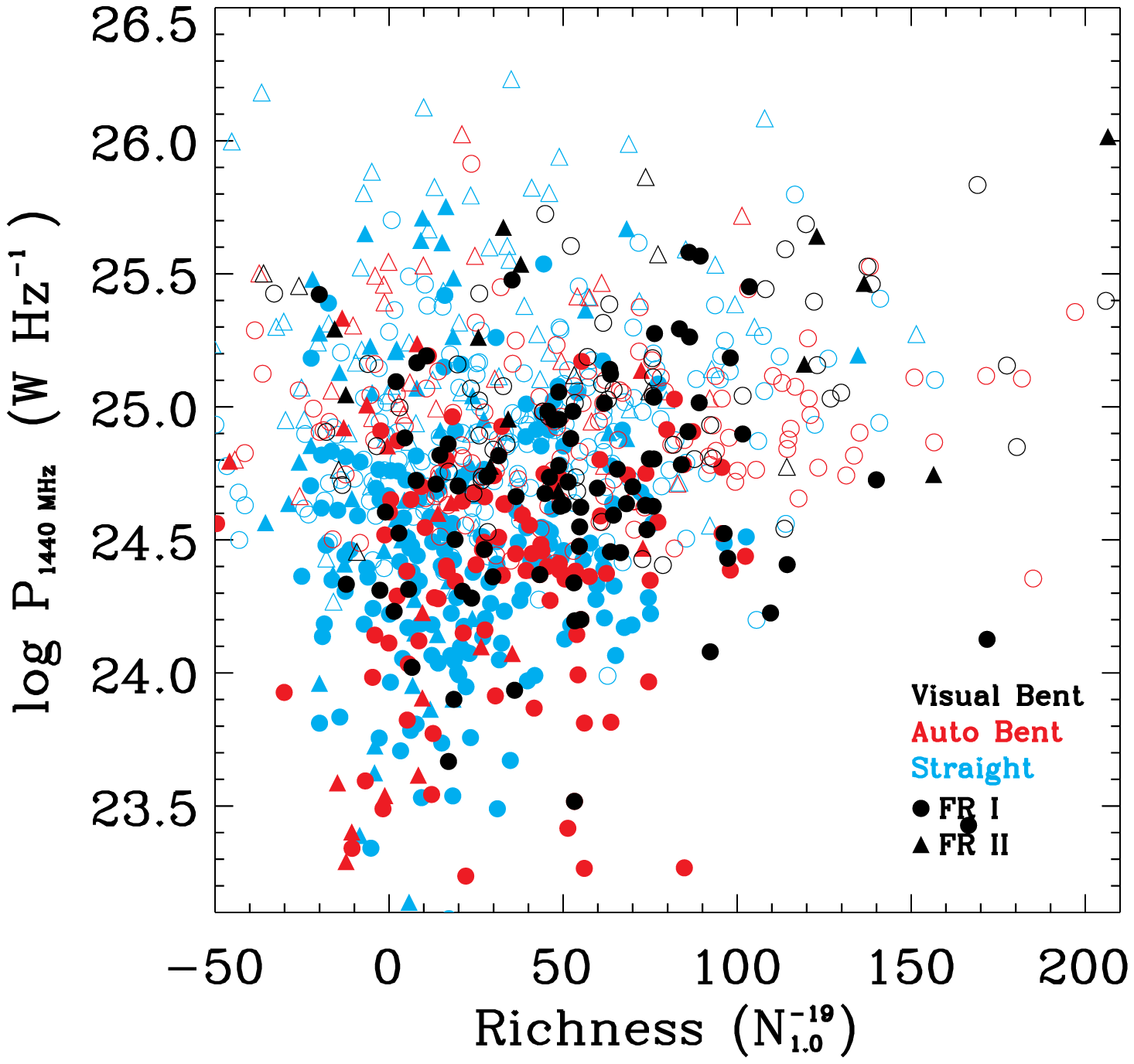}
\caption{Richness vs. $1440$ MHz radio power.  The left-hand panel plots only those sources in the small box from Figure~\ref{zvsize} and need no Schechter correction (see \S\ref{schechter_correction}).  The right-hand panel shows all sources that have Schechter corrections less than $f_c= 3$, including sources outside of the small box.  The symbols are the same as in Figure~\ref{zvcolor}.  The open symbols represent sources that have Schechter corrections in the range $1<f_c\le3$.  It appears that the richest clusters in our samples have a radio power that is consistent with them being located on the FR I/II break.  These are likely to be WATs.} \label{richvpower}
\end{center}
\end{figure}

It has been observed previously \citep{burns1990,blanton2000,blanton2001,mao2009,mao2010} that wide-angle-tail (WAT) sources tend to be located in rich cluster environments.  \citet{burns1990} found evidence that the bent-lobed nature of WATs is a result of a recent cluster-cluster or cluster-subcluster merger.  WATs typically have powers near the classic FR I/II break ($P_{1440\: \rm{MHz}} \sim10^{25}$ W Hz$^{-1}$).  Figure~\ref{richvpower} shows that the richest clusters have radio sources that fall near this break.  The left-hand panel of Figure~\ref{richvpower} shows only the sources contained within the small box in Figure~\ref{zvsize} that need no Schechter correction (see \S\ref{schechter_correction}) and the right-hand panel shows all sources, including ones not contained within the small box, that need no Schechter correction.  Table~\ref{table_richness_wat} gives the breakdown of the number of sources with richnesses above and below N$^{-19}_{1.0}=100$ and the percentage of those sources that have radio powers between $24.75 < \log P_{1440\: \rm{MHz}} < 25.75$, a typical radio power range where WATs are found.  We see that a higher fraction of clusters with  N$^{-19}_{1.0}>100$ have radio sources with powers typical of WATs than less rich clusters.  

We also examined the relationship between the opening angle of the radio source and the richness of the cluster environment.  There was no obvious correlation between the two as might be expected since the angles are affected by viewing angle, ICM density, and galaxy velocity.  \citet{blanton2000a} found similar results, seeing no correlation between the opening angle and the cluster richness.

\section{Correlation with X-ray Catalogs} \label{correlation_with_x-ray_catalogs} \index{Correlation with X-ray Catalogs}
X-ray observations can confirm a source's association with a galaxy cluster.  In addition, X-ray observations can provide information about the temperature and the mass of the galaxy cluster.  We searched the {\it ROSAT} All-Sky Survey (RASS) for possible X-ray counterparts for the sources in our samples.  We used the RASS Faint Source Catalog (RASS-FSC), whose selection criteria included any source in the $0.1$-$2.4$ keV energy band that had a detection likelihood of at least $7$ and contained at least $6$ source photons.  In the {\it ROSAT} All-Sky Survey catalogs, the detection likelihood is defined as $-ln(P)$, where P is the probability that the observed distribution of photons originates from a spurious background fluctuation \citep{voges1999}.

We examined the rate at which each sample had sources with X-ray counterparts, and whether this was related to the radio source being found in an optical cluster.  We searched within a radius of $1.0$ Mpc for any possible X-ray sources in the RASS-FSC associated with all radio sources with an optical counterpart and also more specifically for those radio sources found in a cluster environment with a richness measurement of $N^{-19}_{1.0}\ge40$.  Given the flux limits, sources at high redshift are unlikely to be detected in the RASS-FSC.  To eliminate some of the effects of the flux limit of the RASS-FSC, we have limited our samples to only sources with redshifts less than $z=0.25$ or $z=0.15$ as sources at these redshifts are more likely to be detected by the RASS-FSC.  The results of this search are listed in Table~\ref{table_x-ray}.  Column 2 gives the fraction of radio sources in each sample with optical counterparts and not in rich optical clusters in the \SDSS\/ and a corresponding X-ray source in the RASS-FSC.  Columns 3 and 4 give the same fraction, but for sources with redshifts less than $z=0.25$ and $z=0.15$, respectively.  Column 5 gives the fraction of radio sources in each sample located in optical clusters with $N^{-19}_{1.0}\ge40$ with a corresponding X-ray source in the RASS-FSC.  Columns 6 and 7 give the same fraction, but for sources with redshifts less than $z=0.25$ and $z=0.15$, respectively.

\begin{figure}
\begin{center}
\capstart
\includegraphics[scale=0.49]{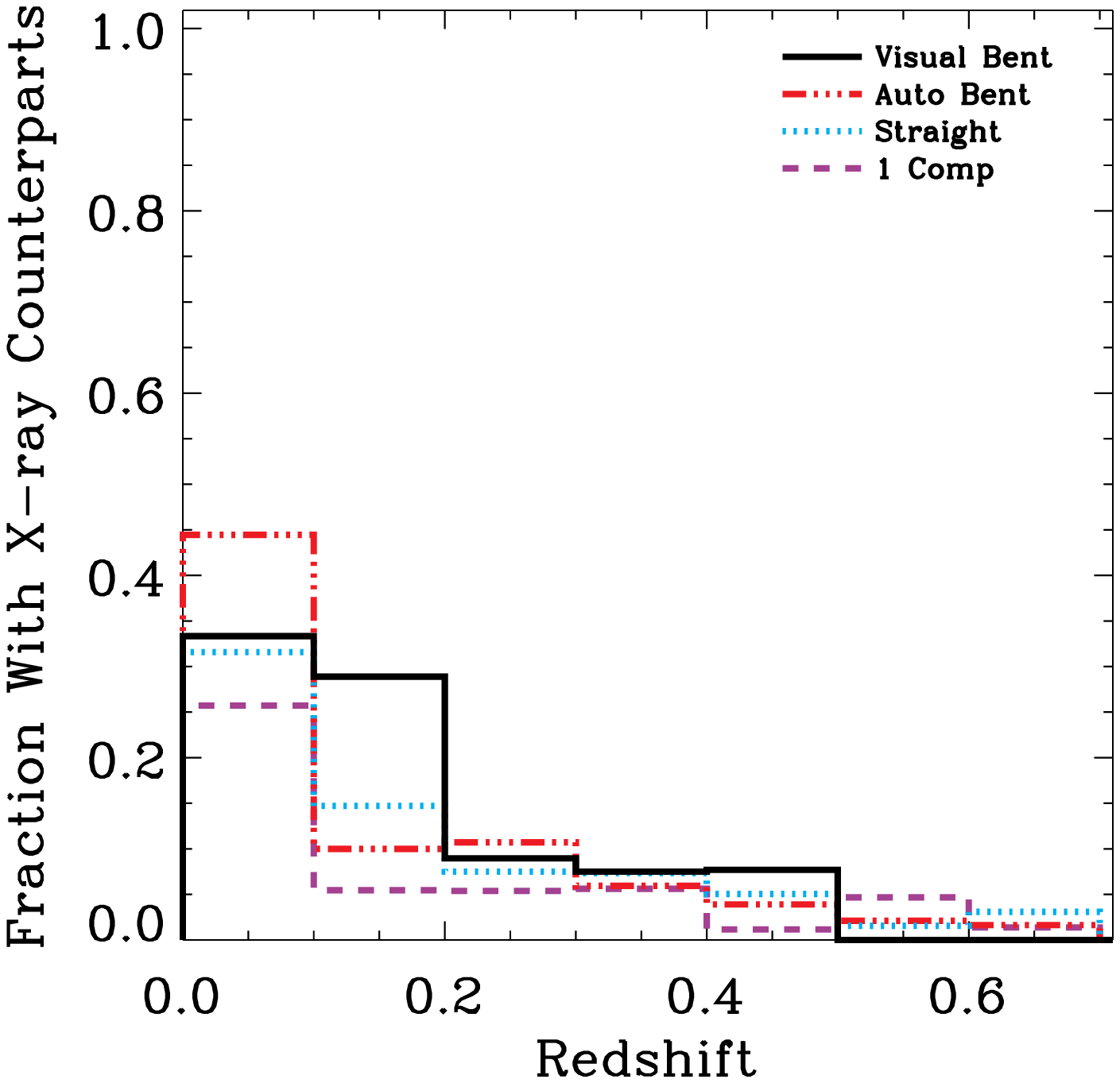}
\includegraphics[scale=0.49]{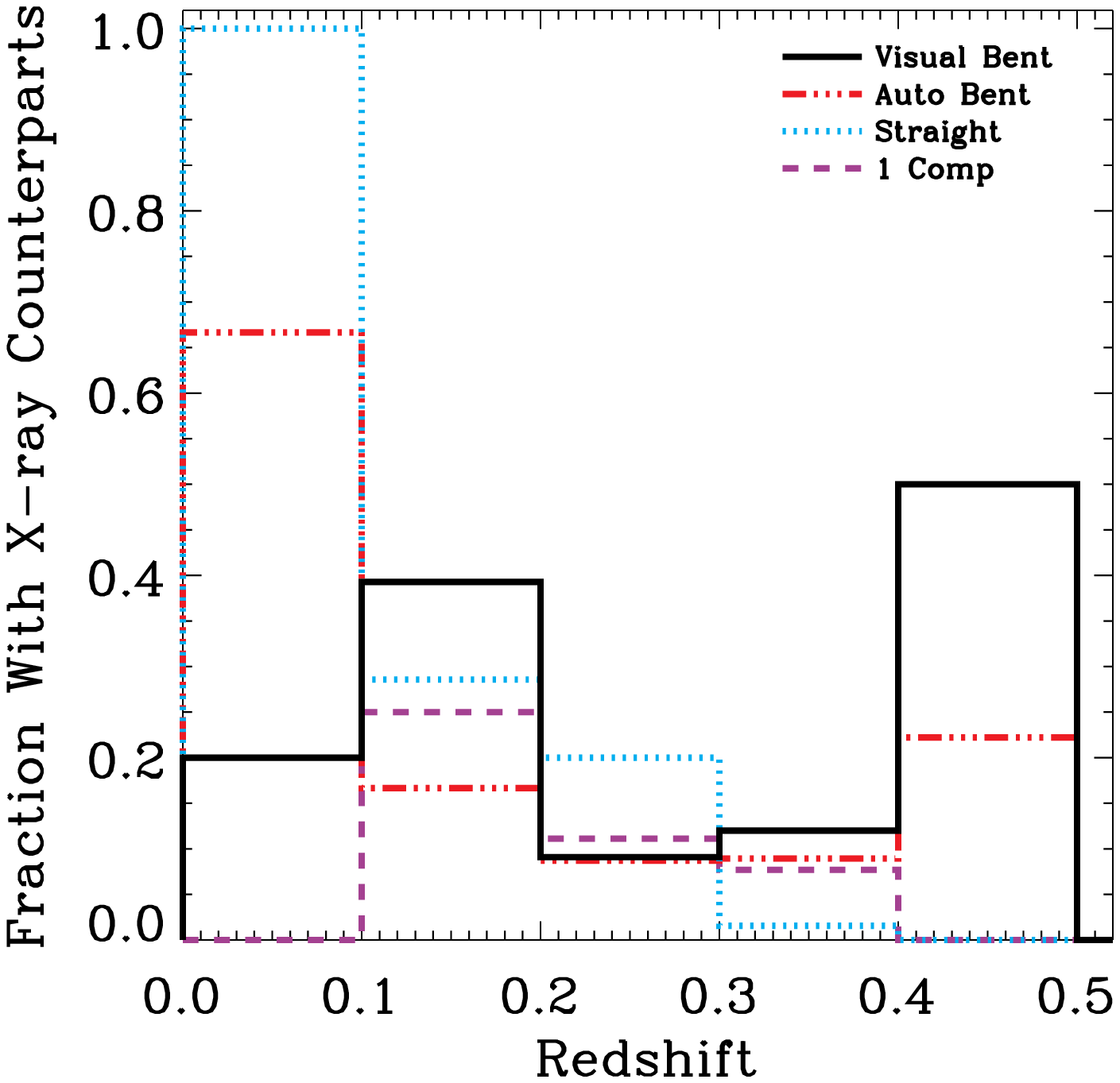}
\caption{The fraction of sources in each redshift bin that have X-ray counterparts in the RASS-FSC within $1.0$ Mpc of the position of the radio source compared to the total number of sources in each redshift bin.  The line styles and colors are the same as in Figure~\ref{fr1hist}.  The left-hand plot shows the distribution for all of the radio sources with optical counterparts not in clusters and the right-hand plot is for only sources with $N^{-19}_{1.0}\ge40$.  We see that the fraction of sources with X-ray counterparts is much higher for those radio sources associated with clusters.  The fraction of all optical sources associated with X-ray sources drops as the redshift increases.  This is likely a result of flux limits of the RASS as well as the angular radius corresponding to 1 Mpc decreasing with redshift resulting in fewer chance matches at high $z$.  X-ray sources associated with radio/optical sources at high redshift are unlikely to be detected.  We see a similar trend for the sources associated with clusters with the exception of the last redshift bin in which the fraction associated with clusters increases.  It is possible that this is due to the small number of sources residing in this redshift bin.} \label{xrayzhist}
\end{center}
\end{figure}

Examination of Table~\ref{table_x-ray} shows that for all of the radio samples, objects associated with optically-detected clusters are more likely to have an X-ray counterpart within a projected distance of $1.0$ Mpc than all sources located in less-dense optical environments.  When we limit these to only sources with redshifts $z\le0.25$ we find that there is a higher percentage of sources with X-ray counterparts (both those not located in optically-detected clusters as well as those within optical clusters), and this fraction is even higher when we limit the sample to sources with $z\le0.15$ (with the exception of the single-component sample).  This is expected given the flux limits of the RASS-FSC.  Lower redshift clusters will be more frequently detected.  In addition, we will detect a larger number of spurious sources at lower redshifts, since our $1.0$ Mpc search region will correspond to a larger angular size on the sky.  Visual-bent sources are more often associated with RASS-FSC X-ray sources than the other samples.  Additionally, multiple-component sources in general are more likely to have an associated RASS-FSC X-ray source than single-component sources.  Figure~\ref{xrayzhist} shows the distribution of redshifts for the optically identified radio sources in our samples that have associated X-ray emission.  

For the sources at the lowest redshifts, using a radius of $1.0$ Mpc within which to search for X-ray counterparts increases the number of chance-coincidences.  We repeated the process of matching X-ray sources to the positions of our radio sources using a fixed maximum angular separation of $5\arcmin$.  The results of this correlation are found in Table~\ref{table_x-ray2}.  The columns are identical to Table~\ref{table_x-ray}.  In general, there are fewer radio sources with RASS-FSC X-ray counterparts located within $5\arcmin$ than within $1.0$ Mpc.  Still, we find that low redshift sources in optical clusters are most likely to have X-ray counterparts.

\section{Summary and Conclusions} \label{conclusions} \index{Conclusions}
We created four samples of radio sources from the \FIRST\/ source catalog based on visual and automatic selection of multi-component, double-lobed, radio sources as well as single-component radio sources.  We find that more than $50\%$ of these radio sources have unique optical counterparts in the \SDSS.  We limited the maximum angular separation between radio and optical source positions such that $95\%$ of the radio/optical sources in each sample (with the exception of the auto-bent sample at $90\%$) are true radio/optical sources and not chance-coincidences as compared to a random distribution of points on the sky.  We created these samples to examine the cluster environments surrounding radio sources, specifically to examine if bent-lobed radio sources are more likely to be associated with galaxy clusters than other types of radio sources.

Using the \SDSS\/ DR7, we searched for optical counterparts around each radio source.  Based on the optical properties in the \SDSS\/ for each of the radio/optical sources, we were able to calculate a redshift, absolute magnitudes in all of the \SDSS\/ wavelength bands, radio power at $1440$ MHz, and the physical extents of the radio sources.  We limited our samples of radio sources with optical counterparts to only those sources that were not too faint, too blue, or too nearby in order to remove sources that we could not properly examine using the \SDSS.  We measured the richess of the source environments using $N^{-19}_{1.0}$, the number of galaxies within 1 Mpc of the radio source with an absolute r-magnitude brighter than $M_r=-19$, corrected for background counts.

We find that, in general, multi-component radio sources are more often associated with galaxy clusters than single-component sources, and bent sources are more often in clusters than straight sources.  The visually-selected, bent sources are most often found to be associated with clusters, followed by (in order) the auto-selected bent sources, the straight sources, and the single-component sources.  In general, we find that FR I sources are found in richer environments than FR II sources, and this is true whether the FR I/II classification is done by visual inspection or by using radio power and optical magnitude criteria.  The difference in environments between FR I and II sources, however, is less clear for some of the samples when this classification is done visually.  Limiting to only those sources not requiring a Schechter correction (our most secure richness measurements) visually-selected, bent-lobed sources are found in clusters or groups $78\%$ of the time, and in rich clusters $62\%$ of the time.  These values are $59\%$ and $41\%$ for the auto-bent sample, $43\%$ and $24\%$ for the straight sample, and $29\%$ and $10\%$ for the single-component sample.

We examined the radio source extents for all of the samples.  In general, we find that the FR II sources are larger than the FR I sources, whether this distinction is made visually or using radio power and optical magnitude criteria.  We do not find an obvious correlation between the richness of the radio source environment and the extent of the radio source.  This implies that there is no preference for radio sources in clusters to have a different characteristic size than radio sources not located in clusters in our samples.

We find that we are selecting richer environments with increasing redshift for all of our samples, for both FR I and II sources up to $z\approx0.5$.  This may be related to clusters having a higher fraction of radio-loud AGN at high redshift as compared to nearby.  When examining the richest clusters in our samples, we find that a large fraction of them host radio sources near the FR I/II break, consistent with them being WAT sources.

We searched for X-ray counterparts around sources in our samples, examining separately objects with optical hosts only and those within clusters.  A number of those sources located in rich cluster environments also have X-ray emission associated with them.  As we restrict the sources to only those at low redshift, we find an even higher fraction with associated X-ray detections, as expected due to the flux limits of the RASS.  We will further examine the X-ray properties of these sources in a future paper.

Since bent, double-lobed radio sources are frequently associated with rich clusters, we can use them as tracers for clusters at high redshifts.  Given the sensitivity of the \FIRST\/ survey, we can detect bent, double-lobed radio galaxies out to $z\approx2$.  The visual-bent sample has $83$ sources that are undetected in the \SDSS\/ to its magnitude limit ($m_r\sim22$).  Given the cluster association rate at low redshifts, we expect $51$ of these sources to be associated with rich clusters at high ($z>0.7$) redshifts.  Using the larger, auto-bent sample, we expect to find 322 rich clusters at $z>0.7$ (with small overlap between the two samples).  We are currently in the process of following up on these sources using deep optical and near-infrared imaging.  This cluster-finding technique is complimentary to other high-redshift cluster surveys and may select clusters with a wider range of masses than other searches.  It may preferentially select clusters that are merging or still in the process of forming.  In addition, it gives us the opportunity to examine AGN feedback at high redshift.

\acknowledgements
We are very grateful to Deanne Proctor for supplying us with the auto-bent- and straight-lobed radio samples and commenting on an earlier draft.  We would also like to thank the anonymous referee for insightful comments.

This work was partially supported by NASA through the Astrophysics Data Analysis Program, grant number NNX10AC98G.

The FIRST project has been supported by grants from the National Geographic Society, the National Science Foundation, NASA, NATO, the Institute of Geophysics and Planetary Physics, Columbia University, and Sun Microsystems.

Funding for the SDSS DR7 has been provided by the Alfred P. Sloan Foundation, the Participating Institutions, the National Aeronautics and Space Administration, the National Science Foundation, the US Department of Energy, the Japanese Monbukagakusho, and the Max Planck Society.  The SDSS website is \url{http://www.sdss.org}.  The SDSS is managed by the Astrophysical Research Consortium for the Participating Institutions.  The Participating Institutions are the University of Chicago, Fermilab, the Institute for Advanced Study, the Japan Participation Group, the Johns Hopkins University, Los Alamos National Laboratory, the Max Planck Institute for Astronomy, the Max Planck Institute for Astrophysics, New Mexico State University, Princeton University, the US Naval Observatory, and the University of Washington.

This research has made use of the NASA/IPAC Extragalactic Database (NED) which is operated by the Jet Propulsion Laboratory, California Institute of Technology, under contract with the National Aeronautics and Space Administration.

\bibliography{bibdeskbiblio}

\newpage

\begin{deluxetable}{lrrrrrrrrrr}
\tabletypesize{\scriptsize}
\tablewidth{0pt}
\tablecolumns{11}
\tablecaption{Samples}
\tablehead{\colhead{Sample Name}&
		  \colhead{Total}&
		  \colhead{Footprint}&
		  \colhead{Unique}&
		  \colhead{Radius}&
		  \colhead{$95\%$}&
		  \colhead{Complete}&
		  \colhead{First Cut}&
		  \colhead{Second Cut}&
		  \colhead{FR I}&
		  \colhead{FR II}
		  \cr
		  \colhead{(1)}&
		  \colhead{(2)}&
		  \colhead{(3)}&
		  \colhead{(4)}&
		  \colhead{(5)}&
		  \colhead{(6)}&
		  \colhead{(7)}&
		  \colhead{(8)}&
		  \colhead{(9)}&
		  \colhead{(10)}&
		  \colhead{(11)}}
\startdata
Visual-Bent & $384$ & $355$ & $291$ & $6\farcs8$ & $272$ & $97.7\%$ & $236$ & $166$ & $143$ & $23$\\
Auto-Bent & $1546$ & $1384$ & $910$ & $4\farcs2$ & $599$\tablenotemark{a} & $83.6\%$ & $519$ & $272$ & $219$ & $53$\\
Straight & $3232$ & $2929$ & $1806$ & $3\farcs2$ & $1288$ & $95.0\%$ & $840$ & $449$ & $322$ & $127$\\
\enddata
\tablecomments{Col. (1): name of the sample; col. (2): total number of radio sources; col. (3): number of radio sources within the \SDSS\/ footprint; col. (4): number of radio sources with a unique optical counterpart in the \SDSS\/ catalog within $10\arcsec$; col. (5): the maximum angular separation between a radio source and an optical source corresponding to our $95\%$ confidence criteria, as described in \S\ref{correlating_first_sources_with_sdss_sources}; col. (6): number of radio/optical counterparts with separations less than the maximum angular separation; col. (7): percentage of radio/optical counterparts with a separation less than the maximum angular separation; col. (8): the number of sources remaining after color selection, as described in \S\ref{identifying_possible_clusters}; col. (9): the number of sources remaining after redshift and apparent magnitude selection, as described in \S\ref{schechter_correction} and \S\ref{nearby_clusters}; col. (10): number of sources classified as FR I based on the criteria of \citet{ledlow1996} after making the first and second cuts; col. (11): number of sources classified as FR II based on the criteria of \citet{ledlow1996} after making the first and second cuts.}
\tablenotetext{a}{The Auto-bent sample has a confidence rate of $90\%$.}
\label{sample_breakdown}
\end{deluxetable}

\begin{deluxetable}{lrrrrrrrr}
\tabletypesize{\footnotesize}
\tablewidth{0pt}
\tablecolumns{9}
\tablecaption{Results of Visual FR I/II Classification}
\tablehead{\colhead{Sample}&
		  \colhead{Total}&
		  \colhead{FR I}&
		  \colhead{FR II}&
		  \colhead{FR I?}&
		  \colhead{FR II?}&
		  \colhead{Other}&
		  \colhead{FR I\tablenotemark{1}}&
		  \colhead{FR II\tablenotemark{2}}
		  \cr
		  \colhead{(1)}&
		  \colhead{(2)}&
		  \colhead{(3)}&
		  \colhead{(4)}&
		  \colhead{(5)}&
		  \colhead{(6)}&
		  \colhead{(7)}&
		  \colhead{(8)}&
		  \colhead{(9)}}
\startdata
Visual-Bent & $272$ & $200$ & $48$ & $15$ & $8$ & $1$ & $134$ & $22$\\
Auto-Bent & $599$ & $369$ & $137$ & $49$ & $22$ & $22$ & $178$ & $56$\\
Straight & $1288$ & $659$ & $498$ & $63$ & $36$ & $32$ & $272$ & $138$
\enddata
\tablenotetext{1}{The number of visually-classified FR I sources remaining after redshift and apparent magnitude selection, as described in \S\ref{schechter_correction} and \S\ref{nearby_clusters}.}
\tablenotetext{2}{The number of visually-classified FR II sources remaining after redshift and apparent magnitude selection, as described in \S\ref{schechter_correction} and \S\ref{nearby_clusters}.}
\label{fr_class}
\end{deluxetable}

\begin{deluxetable}{l@{\extracolsep{5pt}}rrrrrrrr}
\tabletypesize{\footnotesize}
\tablewidth{0pt}
\tablecolumns{9}
\tablecaption{Sources Contained in Boxes (Figure~\ref{zvsize})}
\tablehead{& \multicolumn{4}{c}{Small Box} & \multicolumn{4}{c}{Large Box} \cr
		  \cline{2-5} \cline{6-9}
		  \colhead{Sample}&
		  \colhead{FR I\tablenotemark{1}}&
		  \colhead{FR II\tablenotemark{1}}&
		  \colhead{FR I\tablenotemark{2}}&
		  \colhead{FR II\tablenotemark{2}}&
		  \colhead{FR I\tablenotemark{1}}&
		  \colhead{FR II\tablenotemark{1}}&
		  \colhead{FR I\tablenotemark{2}}&
		  \colhead{FR II\tablenotemark{2}}
		  \cr
		  \colhead{(1)}&
		  \colhead{(2)}&
		  \colhead{(3)}&
		  \colhead{(4)}&
		  \colhead{(5)}&
		  \colhead{(6)}&
		  \colhead{(7)}&
		  \colhead{(8)}&
		  \colhead{(9)}}
\startdata
Visual-Bent & 96 & 24 & 88 & 21 & 19 & 19 & 25 & 9\\
Auto-Bent & 200 & 51 & 164 & 56 & 112 & 42 & 106 & 26\\
Straight & 261 & 107 & 224 & 110 & 115 & 101 & 125 & 77
\enddata
\tablenotetext{1}{FR type determined using the \citet{ledlow1996} criteria.}
\tablenotetext{2}{FR type determined visually.}
\label{table_size_distribution}
\end{deluxetable}

\begin{deluxetable}{l@{\extracolsep{5pt}}rrrrrr}
\tabletypesize{\footnotesize}
\tablewidth{0pt}
\tablecolumns{7}
\tablecaption{Cluster Richness, $N^{-19}_{1.0}$}
\tablehead{ & \multicolumn{3}{c}{N$^{-19}_{1.0} \ge 20$} & \multicolumn{3}{c}{N$^{-19}_{1.0} \ge 40$} \cr
		  \cline{2-4} \cline{5-7}
		  \colhead{Sample}&
		  \colhead{FR I\&II}&
		  \colhead{FR I}&
		  \colhead{FR II}&
		  \colhead{FR I\&II}&
		  \colhead{FR I}&
		  \colhead{FR II}
		  \cr
		  \colhead{(1)}&
		  \colhead{(2)}&
		  \colhead{(3)}&
		  \colhead{(4)}&
		  \colhead{(5)}&
		  \colhead{(6)}&
		  \colhead{(7)}}
\startdata
Visual-Bent\tablenotemark{a} & $78\%$(98) & $76\%$(85) & $85\%$(13) & $62\%$(98) & $65\%$(85) & $46\%$(13)\\
Visual-Bent\tablenotemark{b} & $78\%$(166) & $80\%$(143) & $70\%$(23) & $63\%$(166) & $65\%$(143) & $48\%$(23)\\
Auto-Bent\tablenotemark{a} & $59\%$(120) & $67\%$(97) & $26\%$(23) & $41\%$(120) & $47\%$(97) & $13\%$(23)\\
Auto-Bent\tablenotemark{b} & $66\%$(272) & $73\%$(219) & $38\%$(53) & $50\%$(272) & $56\%$(219) & $25\%$(53)\\
Straight\tablenotemark{a} & $43\%$(217) & $49\%$(164) & $23\%$(53) & $24\%$(217) & $28\%$(164) & $9\%$(53)\\
Straight\tablenotemark{b} & $52\%$(449) & $58\%$(322) & $39\%$(127) & $32\%$(449) & $37\%$(322) & $21\%$(127)\\
Single-Component\tablenotemark{a} & $29\%$(209) & $29\%$(202) & $14\%$(7) & $10\%$(209) & $10\%$(202) & $0\%$(7)\\
Single-Component\tablenotemark{b} & $35\%$(358) & $36\%$(335) & $22\%$(23) & $15\%$(358) & $15\%$(335) & $13\%$(23)
\enddata
\tablecomments{The numbers in parentheses correspond to the number of sources in each sample and sub-sample from which the percentages are calculated.}
\tablenotetext{a}{Only sources needing no Schechter correction}
\tablenotetext{b}{Schechter corrected sources included}
\label{table_richness}
\end{deluxetable}

\begin{deluxetable}{l@{\extracolsep{5pt}}rrrr}
\tabletypesize{\footnotesize}
\tablewidth{0pt}
\tablecolumns{5}
\tablecaption{Cluster Richness, $N^{-19}_{1.0}$ with visual FR classifications}
\tablehead{ & \multicolumn{2}{c}{N$^{-19}_{1.0} \ge 20$} & \multicolumn{2}{c}{N$^{-19}_{1.0} \ge 40$} \cr
		  \cline{2-3} \cline{4-5}
		  \colhead{Sample}&
		  \colhead{(FR I)}&
		  \colhead{(FR II)}&
		  \colhead{(FR I)}&
		  \colhead{(FR II)}
		  \cr
		  \colhead{(1)}&
		  \colhead{(2)}&
		  \colhead{(3)}&
		  \colhead{(4)}&
		  \colhead{(5)}}
\startdata
Visual-Bent\tablenotemark{a} & $80\%$(83) & $64\%$(11) & $63\%$(83) & $55\%$(11)\\
Visual-Bent\tablenotemark{b} & $79\%$(134) & $73\%$(22) & $63\%$(134) & $59\%$(22)\\
Auto-Bent\tablenotemark{a} & $59\%$(82) & $77\%$(22) & $39\%$(82) & $59\%$(22)\\
Auto-Bent\tablenotemark{b} & $66\%$(178) & $70\%$(56) & $48\%$(178) & $59\%$(56)\\
Straight\tablenotemark{a} & $45\%$(141) & $38\%$(61) & $26\%$(141) & $20\%$(61)\\
Straight\tablenotemark{b} & $53\%$(272) & $50\%$(138) & $32\%$(272) & $31\%$(138)
\enddata
\tablecomments{The numbers in parentheses correspond to the number of sources in each sample and sub-sample from which the percentages are calculated.}
\tablenotetext{a}{Only sources needing no Schechter correction}
\tablenotetext{b}{Schechter corrected sources included}
\label{table_richness_visual}
\end{deluxetable}

\clearpage
\LongTables

\begin{deluxetable}{lcccccrrccccrcc}
\tabletypesize{\tiny}
\tablewidth{0pt}
\tablecolumns{15}
\tablecaption{All Extended sources with N$^{-19}_{1.0} \ge 40$}
\tablehead{\colhead{Source Name}&
		  \colhead{Smpl\tablenotemark{1}}&
		  \colhead{$\alpha$}&
		  \colhead{$\delta$}&
		  \colhead{$m_r$}&
		  \colhead{$M_r$}&
		  \colhead{S}&
		  \colhead{Angle}&
		  \colhead{$P_{1440\: \rm MHz}$}&
		  \colhead{FR\tablenotemark{2}}&
		  \colhead{FR\tablenotemark{3}}&
		  \colhead{$z$}&
		  \colhead{N$^{-19}_{1.0}$}&
		  \colhead{$f_c$}&
		  \colhead{Abell}
		  \cr
		   & & & &
		  \colhead{(mag)}&
		  \colhead{(mag)}& 
		  \colhead{(mJy)}&
		  \colhead{(deg)}&
		  \colhead{(W Hz$^{-1}$)}&
		   & & & & &
		  \cr
		  \colhead{(1)}&
		  \colhead{(2)}&
		  \colhead{(3)}&
		  \colhead{(4)}&
		  \colhead{(5)}&
		  \colhead{(6)}&
		  \colhead{(7)}&
		  \colhead{(8)}&
		  \colhead{(9)}&
		  \colhead{(10)}&
		  \colhead{(11)}&
		  \colhead{(12)}&
		  \colhead{(13)}&
		  \colhead{(14)}&
		  \colhead{(15)}}
%\rotate
\startdata
J002859.3$+$002000 & S & 00:28:59.38 & $+00$:20:00.6 & 17.12 & $-23.40$ & 19.8 & 169.5 & 2.77e+24 & I & I & 0.222\tablenotemark{a} & 61 & 1.00 & \nodata\\
J005625.7$-$011543 & V & 00:56:25.70 & $-01$:15:43.0 & 13.36 & $-24.48$ & 515.3 & 85.0 & 7.60e+24 & I & I & 0.078\tablenotemark{b} & 52 & 1.00 & A0119\\
J005702.0$-$005230 & A & 00:57:02.07 & $-00$:52:30.7 & 13.41 & $-23.11$ & 72.5 & 147.2 & 3.29e+23 & I & I & 0.044\tablenotemark{a} & 53 & 1.00 & A0119\\
J005702.1$-$005231 & V & 00:57:02.10 & $-00$:52:31.0 & 13.41 & $-23.11$ & 72.5 & 147.0 & 3.29e+23 & I & I & 0.044\tablenotemark{a} & 53 & 1.00 & A0119\\
J010242.5$-$005032 & V & 01:02:42.50 & $-00$:50:32.0 & 17.46 & $-23.54$ & 66.5 & 87.0 & 1.33e+25 & I & I & 0.261\tablenotemark{b} & 64 & 1.00 & \nodata\\
J010456.5$+$000423 & V & 01:04:56.50 & $+00$:04:23.0 & 18.26 & $-23.26$ & 83.4 & 8.0 & 2.50e+25 & I & I & 0.312\tablenotemark{b} & 206 & 1.37 & \nodata\\
J011403.1$-$011058 & V & 01:14:03.10 & $-01$:10:58.0 & 16.90 & $-23.15$ & 16.5 & 139.0 & 1.59e+24 & I & I & 0.187\tablenotemark{a} & 55 & 1.00 & \nodata\\
J011403.1$-$011057 & A & 01:14:03.15 & $-01$:10:58.0 & 16.90 & $-23.15$ & 16.5 & 138.9 & 1.59e+24 & I & I & 0.187\tablenotemark{a} & 55 & 1.00 & \nodata\\
\enddata
\tablecomments{A question mark in Col. 11 represents a dubious visual FR I/II classification.  Three successive question marks represent an inability to visually classify the radio source as either FR I or II.}
\tablenotetext{1}{The sample each source belongs to:  V = visual bent sample, A = auto bent sample, and S = straight sample.}
\tablenotetext{2}{FR type determined following \citet{ledlow1996}.}
\tablenotetext{3}{FR type determined visually.}
\tablenotetext{a}{The redshift comes from a spectroscopic measurement.}
\tablenotetext{b}{The redshift comes from a photometric measurement.}
\tablenotetext{c}{The difference in redshift between the identified Abell cluster (from NED) and the radio cluster is greater than $1500$ km s$^{-1}$.}
\tablenotetext{d}{The Abell cluster does not have a confirmed redshift in NED.}
\label{table_sources}
\end{deluxetable}

\clearpage

\begin{deluxetable}{l@{\extracolsep{5pt}}crcc}
\tabletypesize{\footnotesize}
\tablewidth{0pt}
\tablecolumns{5}
\tablecaption{Richness distribution of sources with WAT-like radio powers}
\tablehead{ & \multicolumn{2}{c}{N$^{-19}_{1.0}<100$} & \multicolumn{2}{c}{N$^{-19}_{1.0}>100$}\cr
		  \cline{2-3} \cline{4-5}
		  \colhead{Sample}&
		  \colhead{\#}&
		  \colhead{\%}&
		  \colhead{\#}&
		  \colhead{\%}
		  \cr
		  \colhead{(1)}&
		  \colhead{(2)}&
		  \colhead{(3)}&
		  \colhead{(4)}&
		  \colhead{(5)}}
\startdata
Visual-Bent & $138$ & $55.8\%$ & $28$ & $67.9\%$\\
Auto-Bent & $242$ & $44.6\%$ & $30$ & $86.7\%$\\
Straight & $430$ & $46.5\%$ & $19$ & $68.4\%$\\
Single-Component & $351$ & $3.7\%$ & $7$ & $42.9\%$
\enddata
\tablecomments{Col. (1): name of the sample; col. (2): number of sources with a cluster richness measurement less than $100$; col. (3): percentage of sources with N$^{-19}_{1.0}<100$ and $24.75 < \log P_{1440\: \rm{MHz}} < 25.75$; col. (4): number of sources with a cluster richness measurement greater than $100$; col. (5): percentage of sources with N$^{-19}_{1.0}>100$ and $24.75 < \log P_{1440\: \rm{MHz}} < 25.75$.}
\label{table_richness_wat}
\end{deluxetable}

\begin{deluxetable}{l@{\extracolsep{5pt}}rrrrrr}
\tabletypesize{\scriptsize}
\tablewidth{0pt}
\tablecolumns{7}
\tablecaption{RASS-FSC Matches (within 1.0 Mpc)}
\tablehead{ & \multicolumn{3}{c}{N$^{-19}_{1.0}<40$} & \multicolumn{3}{c}{N$^{-19}_{1.0}\ge40$} \cr
		   \cline{2-4} \cline{5-7}
		  \colhead{Sample}&
		  \colhead{All}&
		  \colhead{$z\le0.25$}&
		  \colhead{$z\le0.15$}&
		  \colhead{All}&
		  \colhead{$z\le0.25$}&
		  \colhead{$z\le0.15$}
		  \cr
		  \colhead{(1)}&
		  \colhead{(2)}&
		  \colhead{(3)}&
		  \colhead{(4)}&
		  \colhead{(5)}&
		  \colhead{(6)}&
		  \colhead{(7)}}
\startdata
Visual-Bent & $11.3\%$(62) & $14.7\%$(34) & $25.0\%$(12) & $19.2\%$(104) & $28.3\%$(53) & $43.8\%$(16)\\
Auto-Bent & $8.8\%$(137) & $15.6\%$(64) & $14.0\%$(29) & $13.3\%$(135) & $22.7\%$(44) & $38.5\%$(13)\\
Straight & $9.2\%$(304) & $11.2\%$(152) & $16.0\%$(58) & $11.7\%$(145) & $20.5\%$(44) & $30.0\%$(10)\\
Single-Component & $6.2\%$(305) & $7.4\%$(175) & $13.0\%$(84) & $11.3\%$(53) & $21.1\%$(19) & $0.0\%$(2)
\enddata
\tablecomments{The numbers in parentheses correspond to the number of sources in each sample and sub-sample from which the percentages are calculated.}
\label{table_x-ray}
\end{deluxetable}

\begin{deluxetable}{l@{\extracolsep{5pt}}rrrrrr}
\tabletypesize{\scriptsize}
\tablewidth{0pt}
\tablecolumns{7}
\tablecaption{RASS-FSC Matches (within $5\farcm0$)}
\tablehead{ & \multicolumn{3}{c}{N$^{-19}_{1.0}<40$} & \multicolumn{3}{c}{N$^{-19}_{1.0}\ge40$} \cr
		   \cline{2-4} \cline{5-7}
		  \colhead{Sample}&
		  \colhead{All}&
		  \colhead{$z\le0.25$}&
		  \colhead{$z\le0.15$}&
		  \colhead{All}&
		  \colhead{$z\le0.25$}&
		  \colhead{$z\le0.15$}
		  \cr
		  \colhead{(1)}&
		  \colhead{(2)}&
		  \colhead{(3)}&
		  \colhead{(4)}&
		  \colhead{(5)}&
		  \colhead{(6)}&
		  \colhead{(7)}}
\startdata
Visual-Bent & $6.5\%$(62) & $5.9\%$(34) & $8.0\%$(12) & $20.2\%$(104) & $26.4\%$(53) & $31.3\%$(16)\\
Auto-Bent & $7.3\%$(137) & $7.8\%$(64) & $0.0\%$(29) & $14.8\%$(135) & $20.5\%$(44) & $30.8\%$(13)\\
Straight & $9.2\%$(304) & $8.6\%$(152) & $9.0\%$(58) & $12.4\%$(145) & $20.5\%$(44) & $30.0\%$(10)\\
Single-Component & $5.6\%$(305) & $3.4\%$(175) & $5.0\%$(84) & $13.2\%$(53) & $15.8\%$(19) & $0.0\%$(2)
\enddata
\tablecomments{The numbers in parentheses correspond to the number of sources in each sample and sub-sample from which the percentages are calculated.}
\label{table_x-ray2}
\end{deluxetable}

\clearpage

\clearpage

\clearpage

\clearpage

\clearpage

\clearpage

\clearpage

\clearpage

\clearpage

\clearpage

\clearpage

\clearpage

\clearpage

\clearpage

\clearpage

\clearpage

\clearpage

\clearpage

\clearpage

\clearpage

\clearpage

\clearpage

\clearpage

\clearpage

\end{document}